\documentclass[12pt,preprint]{aastex}

\slugcomment{{\sc Astrophysical Journal Supplement, in press}}

\shorttitle{Subaru/XMM-Newton Deep Survey. IV.}
\shortauthors{Ouchi et al.}

\begin{document}


\title{The Subaru/XMM-Newton Deep Survey (SXDS). IV.\\ 
Evolution of Ly$\alpha$ Emitters from $z=3.1$ to $5.7$ 
in the 1 deg$^2$ Field: \\
Luminosity Functions and AGN\altaffilmark{1}}

\author{Masami Ouchi        \altaffilmark{2,3,4,5},
        Kazuhiro Shimasaku  \altaffilmark{6},
	Masayuki Akiyama    \altaffilmark{7},
	Chris Simpson       \altaffilmark{8},\\
	Tomoki Saito        \altaffilmark{9},
	Yoshihiro Ueda      \altaffilmark{10},
	Hisanori Furusawa   \altaffilmark{7},
	Kazuhiro Sekiguchi  \altaffilmark{11},\\
	Toru Yamada         \altaffilmark{12},
	Tadayuki Kodama     \altaffilmark{11},
	Nobunari Kashikawa  \altaffilmark{11},\\
	Sadanori Okamura    \altaffilmark{6},
	Masanori Iye        \altaffilmark{11},
	Tadafumi Takata     \altaffilmark{11},\\
	Michitoshi Yoshida  \altaffilmark{13}, and
	Makiko Yoshida      \altaffilmark{6}
	}

\altaffiltext{1}{Based on data collected at 
        Subaru Telescope, which is operated by 
        the National Astronomical Observatory of Japan.}
\altaffiltext{2}{Space Telescope Science Institute,
        3700 San Martin Drive, Baltimore, MD 21218, USA}
\altaffiltext{3}{Observatories of the Carnegie Institution of Washington,
        813 Santa Barbara St., Pasadena, CA 91101, USA; ouchi \_at\_ ociw.edu.}
\altaffiltext{4}{Hubble Fellow}
\altaffiltext{5}{Carnegie Fellow}
\altaffiltext{6}{Department of Astronomy, School of Science,
        University of Tokyo, Tokyo 113-0033, Japan}
\altaffiltext{7}{Subaru Telescope, National Astronomical Observatory, 
        650 N.A'ohoku Place, Hilo, HI 96720, USA}
\altaffiltext{8}{Astrophysics Research Institute, 
        Liverpool John Moores University, Twelve Quays House, 
        Egerton Wharf, Birkenhead CH41 1LD, UK}
\altaffiltext{9}{Physics Department, Graduate School of Science, 
        Ehime University, 2-5 Bunkyou, Matuyama, 790-8577, Japan}
\altaffiltext{10}{Department of Astronomy, Kyoto University, Kyoto 606-8502, Japan}
\altaffiltext{11}{National Astronomical Observatory, 
        Tokyo 181-8588, Japan}
\altaffiltext{12}{Astronomical Institute, Graduate School of Science, 
        Tohoku University, Aramaki, Aoba, Sendai 980-8578, Japan}
\altaffiltext{13}{Okayama Astrophysical Observatory,
    National Astronomical Observatory, Kamogata, Okayama 719-0232, Japan}

\begin{abstract}
We present luminosity functions (LFs) and 
various properties of Ly$\alpha$ emitters (LAEs) 
at $z=3.1$, $3.7$, and $5.7$, 
in a 1 deg$^2$ sky of the Subaru/$XMM-Newton$ Deep Survey (SXDS) Field. 
We obtain a photometric sample of 858 LAE candidates 
based on deep Subaru/Suprime-Cam imaging data, and a spectroscopic 
sample of 84 confirmed LAEs 
from Subaru/FOCAS and VLT/VIMOS spectroscopy 
in a survey volume of $\sim 10^{6}$ Mpc$^{3}$ with 
a limiting Ly$\alpha$ luminosity of $\sim 3\times 10^{42}$ erg s$^{-1}$.
We derive the LFs of Ly$\alpha$ and UV-continuum ($\simeq 1500$ \AA)
for each redshift, taking into account the statistical error and the field-to-field 
variation. We find that the {\it apparent} Ly$\alpha$ LF
shows no significant evolution between $z=3.1$ and 5.7 within 
factors of 1.8 and 2.7 in $L^*$ and $\phi^*$, respectively.
On the other hand, the UV LF of LAEs increases from $z=3.1$ to 5.7,
indicating that galaxies 
with Ly$\alpha$ emission are more common at earlier epochs.
We identify six LAEs with AGN activities from our spectra 
combined with VLA, Spitzer, and XMM-Newton data.
Among the photometrically selected LAEs at $z=3.1$ and 3.7, 
only $\simeq 1$ \% show AGN activities, while the brightest LAEs 
with $\log L({\rm Ly\alpha})\gtrsim 43.4-43.6$ erg s$^{-1}$
appear to always host AGNs. 
Our LAEs are bluer in UV-continuum color than dropout galaxies, 
suggesting lower extinction and/or younger stellar populations.
Our stacking analyses provide upper limits to the 
radio luminosity and the $f_{\rm HeII}/f_{\rm Ly\alpha}$ line fraction,
and constrain the hidden star formation (+low-luminosity AGN) and 
the primordial population in LAEs.
\end{abstract}


\section{Introduction}

Ly$\alpha$ emitters (LAEs) provide clues to evolution
and formation of galaxies. Most of LAEs are known as 
star-forming galaxies at $z\simeq 2-7$ with a faint-UV continuum, but 
a prominent Ly$\alpha$ emission line, which are
produced by star-forming activities
with a typical star-formation rate of $\sim 1-10M_\odot$yr$^{-1}$ 
(e.g., \citealt{cowie1998,ouchi2003,gawiser2006,pirzkal2006}).
LAEs are believed to have a negligible fraction of AGN 
activities, since an X-ray detection has been reported
for only one LAE at $z\sim 3$ (\citealt{gawiser2006}, 
see also \citealt{wang2004}).
The strong Ly$\alpha$ emission and a blue UV continuum color
imply a young metal-poor star-forming galaxies \citep{malhotra2002,
finkelstein2007}. Recent infrared observations and population synthesis models
indicate that typical stellar mass of LAEs is as small as 
$\sim 10^8-10^9 M_\odot$ and that the stellar age is 
as young as $\lesssim 10$ Myr
(\citealt{pirzkal2006}; see also \citealt{chary2005,gawiser2006,lai2006,nilsson2007}).
LAEs show a compact UV morphology with a size smaller than 1 kpc
\citep{pascarelle1996,pirzkal2006}.
The fraction of LAEs (or galaxies with a Ly$\alpha$ emission line)
increases at faint UV magnitudes \citep{ouchi2003,ando2006,shimasaku2006}.
These pieces of evidence suggest that LAE samples include 
a young low-mass population at high-$z$.

More interestingly, some of LAEs have a Ly$\alpha$ emission line with a
large rest-frame equivalent width (EW) of $EW_0>240$\AA\ 
that cannot be explained by
a star-formation with a Salpeter IMF 
\citep{malhotra2002,dawson2004,shimasaku2006}. 
It is reported that about 10-40\% of spectroscopically-identified LAEs
have an $EW_0$ of $>240$\AA\ at $z=4.5$ \citep{dawson2004} 
and $z=5.7$ \citep{shimasaku2006}.
The population synthesis models indicate that
such large EW objects would have a top heavy IMF, 
a very poor metallicity and/or a very young age$<10^7$ yr 
(\citealt{charlot1993,malhotra2002,schaerer2003}, see also 
\citealt{kudritzki2000}). 
Large-EW objects often show a spatially-extended Ly$\alpha$
envelope whose projected size can be as large as
$\gtrsim 100$ kpc (\citealt{steidel2000}).
These objects are referred to as Ly$\alpha$ blobs.
Ly$\alpha$ blobs are originally discovered in
an overdensity region of Lyman break galaxies 
(LBGs) at $z\sim 3$ (\citealt{steidel2000,matsuda2004}, 
see also \citealt{dey2005}). However, recent deep surveys
also find Ly$\alpha$ blobs in blank fields 
\citep{nilsson2006,saito2006}.
Some of these Ly$\alpha$ blobs
show evidence of AGN activities \citep{dey2005}, 
while others are not \citep{matsuda2004,saito2006}.
Theoretical studies predict that
these large-EW objects with an extended Ly$\alpha$ profile
are candidates of cooling clouds (e.g. \citealt{yang2006})
and population III galaxies (e.g. \citealt{schaerer2003}),
which are at the very beginning stage of galaxy formation.
Thus, LAEs are keys to understanding the early stage
of galaxy formation with less-massive (or dwarf) population some of which
include young galaxies even at the very beginning stage of radiating clouds 
and population III.
The studies of LAEs complement the recent work of high-$z$ massive 
and/or old galaxies (e.g. \citealt{vandokkum2006,kriek2006,daddi2007}).

LAEs are usually identified by 
a redshifted Ly$\alpha$ emission line that falls in a passband
of narrow-band filter (e.g. \citealt{hu2002,kodaira2003}),
tuned at $z\sim3$ up to $z\sim 9$ (e.g. \citealt{iye2006,willis2006}).
Recent blank-field slitless, multi-silt, and IFU spectroscopy
also searched for LAEs 
\citep{martin2004,tran2004,kurk2004,vanbreukelen2005,martin2006}. 
Blind spectroscopy in cluster regions
yielded candidates of gravitationally 
lensed LAEs that enable us to study
intrinsically faint LAEs \citep{ellis2001,santos2004a,stark2007}.
The numbers of photometrically and spectroscopically 
identified LAEs are now over several hundreds
(e.g. \citealt{hu2002,ouchi2003,malhotra2004,taniguchi2005, 
shimasaku2006,kashikawa2006,venemans2007}), and the numbers
are increasing by recent and on-going wide-field projects such as 
WFILAS \citep{westra2005,westra2006}, 
MUSYC \citep{gawiser2006,gronwall2007},
Hawaii (see \citealt{hu2006}), and
SSA22 surveys (Nakamura et al. in preparation).

In spite of increasing observational data of LAEs,
evolution of LAEs is not clearly understood. 
Most of observational results agree that
the LF of Ly$\alpha$ luminosity shows no evolution 
at $z=3-6$ within the errors of measurements.
(e.g. \citealt{rhoads2001,ouchi2003,hu2004,yamada2005,vanbreukelen2005,
shimasaku2006,tapken2006,murayama2007}; c.f. \citealt{maier2003}). 
Although these measurements qualitatively agree,
the constraints on the LF evolution are weak,
due to the small statistics and field variations.
Indeed, the presence of large-scale structures of LAEs has been reported 
\citep{shimasaku2003,ouchi2005a} at these redshifts,
which produce a significant source gradient even within a
$0.2$ deg$^2$ field \citep{ouchi2003,hu2004}. Note that
the narrow-band LAE studies observe a thin slice ($\Delta z=0.1$) of 
the Universe whose survey volume is about $1/10$ of that of dropout
or LBG surveys ($\Delta z = 1$) at a given search area 
(see e.g. \citealt{ouchi2004b}). 
We need to obtain statistical properties of LAEs including clustering
with a good accuracy. Moreover, we should investigate
rare populations of LAEs with a large EW or AGN, 
which are essential for understanding galaxy formation
and its relation to the formation of massive black holes.

Here we carry out a systematic wide-field narrow-band survey for
LAEs at $z=3.1-6.6$ in the wide-field (1 deg $^2$)
Subaru/XMM-Newton Deep Field (SXDS) that is covered by 
deep Subaru optical images ($i'[5\sigma]=27.0$; Furusawa in preparation) 
and the near-infrared images of UKIDSS/Ultra Deep Survey
(UKIDSS/UDS; \citealt{lawrence2006}) as well as VLA \citep{simpson2006a},
XMM-Newton (Ueda et al. in preparation), and Spitzer \citep{lonsdale2003}.
Except for the data of $z=6.6$ LAEs, 
we have completed narrow-band imaging for $z=3.1$,
$3.7$ and $5.7$ LAEs. In this paper, we show
properties of LAEs at $z=3.1-5.7$.
Note that the first results from the $z=5.7$ data 
have been presented in \citet{ouchi2005a}.
Our systematic survey will not only put strong
constraints on properties of LAEs at $z=3.1-5.7$, 
but also give the useful baselines to be compared with higher
redshift LAEs for studying reionization of the Universe
\citep{malhotra2004,kashikawa2006}.

We show photometric and spectroscopic data, and make the spectroscopic
samples in Section \ref{sec:observations}.
We define the photometric samples of LAEs in Section \ref{sec:photometric}.
The AGNs in our LAE samples are presented 
in Section \ref{sec:multi-wavelength_properties}.
We derive LFs of LAEs and investigate evolution of the LF
in Section \ref{sec:luminosity_functions}. 
In Section \ref{sec:properties_of_lya_lines}, we address the properties
of LAEs, such as EW, UV-continuum slope, star-formation rate, and AGN activity.
We discuss our results and summarize them in Sections \ref{sec:discussion}
and \ref{sec:conclusions}, respectively.
We will present clustering and stellar population
of these LAEs in two companion papers of 
Ouchi et al. (in preparation) and Ono et al. (in preparation).

Throughout this paper, magnitudes are in the AB system
\citep{oke1974,fukugita1995}. EW is presented in rest frame 
(i.e. $EW_0$).
The values for the cosmological parameters adopted in this paper
are: 
$(h, \Omega_m,\Omega_\Lambda,n)=
(0.7,0.3,0.7,1.0)$.

\section{Observations and Data Reduction}
\label{sec:observations}

\subsection{Imaging Data}
\label{sec:observations_observations_imaging}
\begin{figure}
\epsscale{0.95}
\plotone{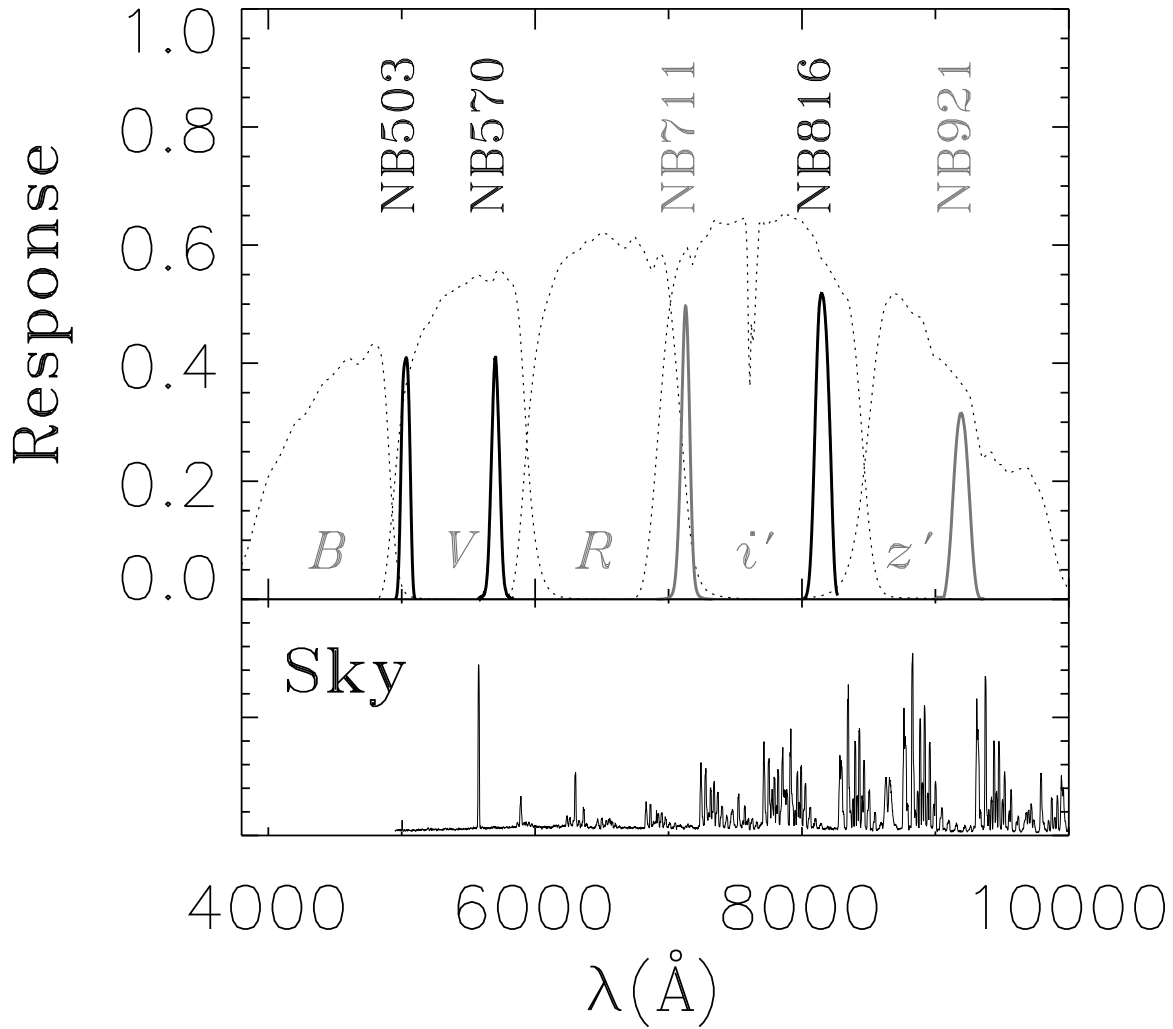}
\caption{
$Top\ panel:$ Response of the narrow-band filters,
NB503, NB570, and NB816 (black solid lines), of this work, 
together with that of broad-band filters of $B$,$V$,$R$,$i'$, and $z'$ (dotted lines) 
and two other narrow-band filters of
$NB711$ \citep{ouchi2003,shimasaku2003} 
and $NB921$ (gray solid lines). $NB921$ filter is used for
our on-going observations in the SXDS field.
These response curves include atmospheric absorption, 
quantum efficiency, and transmittance of optical elements of 
the telescope and instrument.
$Bottom\ panel:$ The spectrum of night-sky emission.
Note that the passbands of the narrow-bands do not include
a strong sky emission.
\label{fig:plot_sed_SXDSfilter_paper}}
\end{figure}

We carried out narrow-band imaging with 
Subaru/Suprime-Cam (\citealt{miyazaki2002}; see also \citealt{iye2004})
in 7 nights during 2003 September 28-30, October 22-24, and October 26.
We summarize details of these observations as well as image qualities
in Table \ref{tab:obs}.
We obtained narrow-band images with three bands,
$NB503$ ($\lambda_c=5029$\AA, $\Delta \lambda=74$\AA), 
$NB570$ ($\lambda_c=5703$\AA, $\Delta \lambda=69$\AA),
and $NB816$ ($\lambda_c=8150$\AA, $\Delta \lambda=120$\AA).
We show the response curves of
our narrow-band and broad-band filters 
in Figure \ref{fig:plot_sed_SXDSfilter_paper}.
These response curves include atmospheric absorption, 
quantum efficiency, and transmittance of optical elements of 
the telescope and instrument.
We have chosen these three narrow bands to identify LAEs
at $z\simeq 3.1$, $3.7$, and $5.7$.
Note that there are no strong OH sky lines within
the passbands of these narrow-band filters 
(Figure \ref{fig:plot_sed_SXDSfilter_paper}).
Since Suprime-Cam has a field-of-view (FoV) of 0.255 deg$^2$,
the 1 deg$^2$ area of SXDS is covered with 5 pointings
of Suprime-Cam. These 5 pointings are referred to as 
SXDS-C (Center), SXDS-N (North), SXDS-S (South), SXDS-E (East), and
SXDS-W (West) whose central coordinates are the same as 
the archival broad-band images of the SXDS project 
(see Table 1 of Furusawa et al. in preparation). 
The on-source exposure times of
$NB503$, $NB570$, and $NB816$ filters are typically
$1.2-1.5$, $1.2-1.5$, and $4.0-5.7$ hours per pointing, respectively. 
The observational condition was photometric
and clear throughout the first 6 nights, and especially good in
three nights for $NB816$ data in 2003 September 28-30.
The seeing size was typically $0''.4-0''.6$ and $0''.5-1''.0$
for 2003 September 28-30, and October 22-24 and 26, respectively.
In addition to these narrow-band data,
we use archival data of very deep broad-band ($B$, $V$, $R$,
$i'$ and $z'$) images of the SXDS project.
The narrow-band data are reduced with 
the Suprime-Cam Deep field REDuction package (SDFRED;
\citealt{yagi2002,ouchi2004a}). With the standard
parameter sets of SDFRED, we perform bias subtraction,
flat-fielding, distortion+atmospheric-dispersion 
corrections, sky subtraction, image alignments, and stacking. 
Before stacking, we mask out bad data areas
such as dead pixels and satellite trails. Cosmic rays
are removed in the process of stacking with the rejected-mean 
algorithm. 
The $5\sigma$ sky noise of the reduced narrow-band images
are $NB503\simeq 25.3$, 
$NB570\simeq 24.9$, and $NB816\simeq 26.0$
in a $2''.0$-diameter circular aperture
(see \citealt{ouchi2005a} for the $NB816$ data). 
Typical PSF sizes of images after reduction
are $\simeq 0''.8$ in FWHM.

We do not use, either contaminated areas with halos of bright 
stars and CCD blooming or low signal-to-noise (S/N) regions 
located around the edge of 
the FoV, which are caused by dithering.
After we reject these bad areas, 
the effective total areas are 0.983, 0.965, and 1.034 deg$^2$
for $NB503$, $NB570$, and $NB816$ images, respectively.
These effective areas correspond to the survey volumes of 
$7.0\times 10^{5}$ Mpc$^{3}$ ($z=3.1$), 
$6.1\times 10^{5}$ Mpc$^{3}$ ($z=3.7$), and
$9.2\times 10^{5}$ Mpc$^{3}$ ($z=5.7$),
if we assume a simple top-hat selection function of LAEs
whose redshift distribution is defined by the FWHM of narrow-band
filters.

During the observations, we took images of spectrophotometric 
standard stars of GD50 and GD71 in $NB503$ and $NB816$ bands
and GD248, G93-48, and GD50 in $NB570$ band
\citep{oke1990,bohlin1995}.
These standard stars were observed $1-4$ time(s), when the
sky was thought to be photometric.
We calculate photometric zero-points 
from photometry of standard stars.
We check these photometric zero points 
with the narrow- and broad-band images
by comparing with colors of stellar objects in our field
and 175 Galactic stars calculated from
spectra given in \citet{gunn1983}. 
We find that the colors of stellar objects in our data
are consistent with those of \citeauthor{gunn1983}'s stars
within $0.01-0.05$ magnitude.
The photometric-zero points thus obtained are regarded
as more accurate than 0.05 mag in the entire 1 deg$^2$ field.

Our narrow-band images are registered to match the coordinates
of SXDS broad-band images based on hundreds of stellar objects 
commonly detected in both narrow-
and broad-band images. The astrometry of our objects is the same as 
those of SXDS version 1.0 catalog (Furusawa et al. in preparation).
The errors in the {\it relative} positions of objects are $\sim 0''.04$
in r.m.s. The r.m.s accuracy of the {\it absolute} positions is 
estimated in Furusawa et al. (in preparation) to be $\sim 0''.2$.
After the registration, we homogenize the PSF sizes of broad and
narrow-band images, referring to these stellar objects. The PSF sizes
of narrow-band images match to those of broad-band images with
an accuracy of $\delta{\rm FWHM} \simeq 0''.01$.

\subsection{Photometric Catalogs}
\label{sec:observations_data_photometriccatalogs}

Source detection and photometry are performed using
SExtractor 
\citep{bertin1996}.
We measure both MAG\_AUTO of SExtractor and
$2''.0$-diameter aperture magnitudes.
We adopt MAG\_AUTO as total magnitudes, while 
we use a $2''.0$-diameter aperture magnitude 
to measure colors of objects in order to obtain
colors of faint objects with a good signal-to-noise ratio.
We make $NB503$-, $NB570$-, and $NB816$-detection catalogs,
and limit to $NB503<25.3$, $NB570<24.7$, and $NB816<26.0$,
respectively, that correspond to about $5\sigma$ detection limits
on a $2''.0$-diameter aperture.
Our $NB503$-, $NB570$-, and $NB816$-detection
catalogs contain 98,907, 64,362, and 278,458, respectively.
We correct the magnitudes of objects
for Galactic extinction of $E(B-V)=0.020$ \citep{schlegel1998}.

\subsection{Spectroscopic Data}
\label{sec:observations_observations_spectroscopy}
We carried out spectroscopic follow-up observations for our LAE candidates
with Faint Object Camera and Spectrograph
(FOCAS; \citealt{kashikawa2002}) of Subaru 
and Visible Multi-Object Spectrograph (VIMOS; \citealt{lefevre2003})
of VLT.

The FOCAS observations were carried out in the MOS mode on 
2003 December 20, 23, and 25, 
2004 October 17, November 9, 11, and
2005 November 2-3. The sky was clear during these observations,
except on 2005 November 3.
Since these observations were conducted 
under the SXDS project and the Subaru open use programs of
Akiyama et al., Yamada et al., and Sekiguchi et al.,
the slits of our objects shared the eight MOS masks with those of 
the other targets.
The on-source exposure time ranges typically from 7200 to 10800 seconds.
These spectroscopic observations were 
made with SY47 order-cut
filter and 300 lines mm$^{-1}$ grating 
having the blaze wavelengths of 5500\AA\ (300B) 
with a slit width of $0''.8$.
The spectral range and resolution are $\lambda=4900-9400$\AA\ and 
$\lambda/\Delta\lambda \simeq 500-1000$, respectively.

The VIMOS spectroscopy was conducted in the programs of
Simpson et al. (in preparation) and \citet{saito2007}.
The Simpson et al.'s observations were made 
between UT 2004 Dec 17 and UT 2006 Jan 2. The medium
resolution grating was used with the GG475 order sorting filter, which
provides a spectral resolution $\lambda/\Delta\lambda \approx 580$
over the wavelength range 4800\,\AA--1\,$\mu$m. Each mask was observed
with $1\times2700 + 2\times1350$\,second exposures, often on separate
nights. Data reduction is described in detail in Simpson et al.\ (in
preparation) but broadly followed the standard pipeline
method. 
The Saito et al.'s observations were carried out in 2004 November 6 to 9.
The HR-Orange grism and the GG435 order-cut filter were used
for MOS masks with a slit width of $1''.0$.
The spectral resolution of these observations is $R\approx 2160$\AA,
which is $>2-4$ times higher than 
the other observations of our spectroscopy.
The effective on-source integration time was 4-7 hours. 
Data reduction and details of observations
are described in \citet{saito2007}.

\subsection{Spectroscopic Catalogs and Samples}
\label{sec:observations_data_spectroscopiccatalogs}

We took spectra for objects with colors similar to LAEs
which show a narrow-band excess and a continuum break 
in blue band.
We observed 128 and 29 objects at $z=3.1/3.7$ and $5.7$, respectively.
Since we share the MOS masks with targets from the other projects, we had tight
constraints on our target selection for spectroscopy. 
Thus, we include unreliable LAE candidates with colors
similar to foreground objects and very faint narrow-band magnitudes, 
some of which potentially have a bright Ly$\alpha$
emission line redshifted to the edge of passband of our narrow band.
As a result, we have identified line emitters from 60\% of our targets.
Although the success rate of identification is not high by the constraints
of our spectroscopy, this target selection allows us to obtain spectra for objects
in wide color and magnitude ranges.

\begin{figure}
\epsscale{0.80}
\plotone{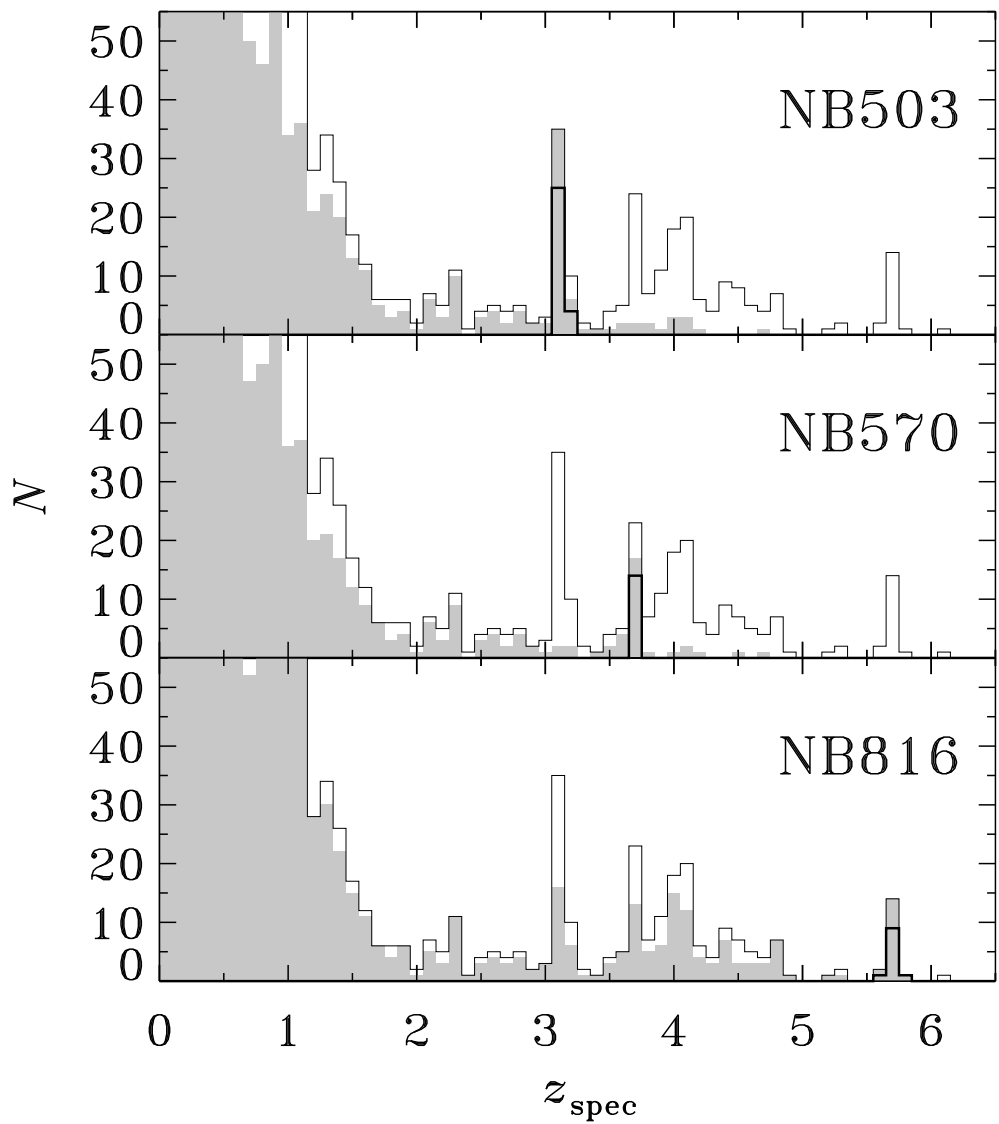}
\caption{Redshift distribution of our objects with a spectrum.
Thin solid lines in all three panels show the distribution
of our LAEs and objects of the SXDS spectroscopic catalog.
The shaded histograms indicate the sources detected at a $5 \sigma$ level 
with $NB503$ (top), $NB570$ (middle), and $NB816$ (bottom).
The thick solid lines in top, middle, and bottom panels
represent our spectroscopic LAEs that meet the color criteria of 
equations (\ref{eq:laeselection_nb503}), (\ref{eq:laeselection_nb570}),
and (\ref{eq:laeselection_nb816}), respectively.
\label{fig:nz_zall}}
\end{figure}

In addition to these spectra for our LAEs,
we use the SXDS version 1.0 spectroscopic data
taken with Subaru/FOCAS and AAO/2dF. 
The combination of these and our data
provide 3,233 spectroscopic redshifts in our field (Akiyama et al. in preparation).
We refer this catalog to estimate the contamination rates of 
our LAE sample in Section \ref{sec:photometric_lae_completeness}.
Figure \ref{fig:nz_zall} presents
the redshift distribution of our sources from the combined catalogs.

\begin{figure}
\epsscale{0.80}
\plotone{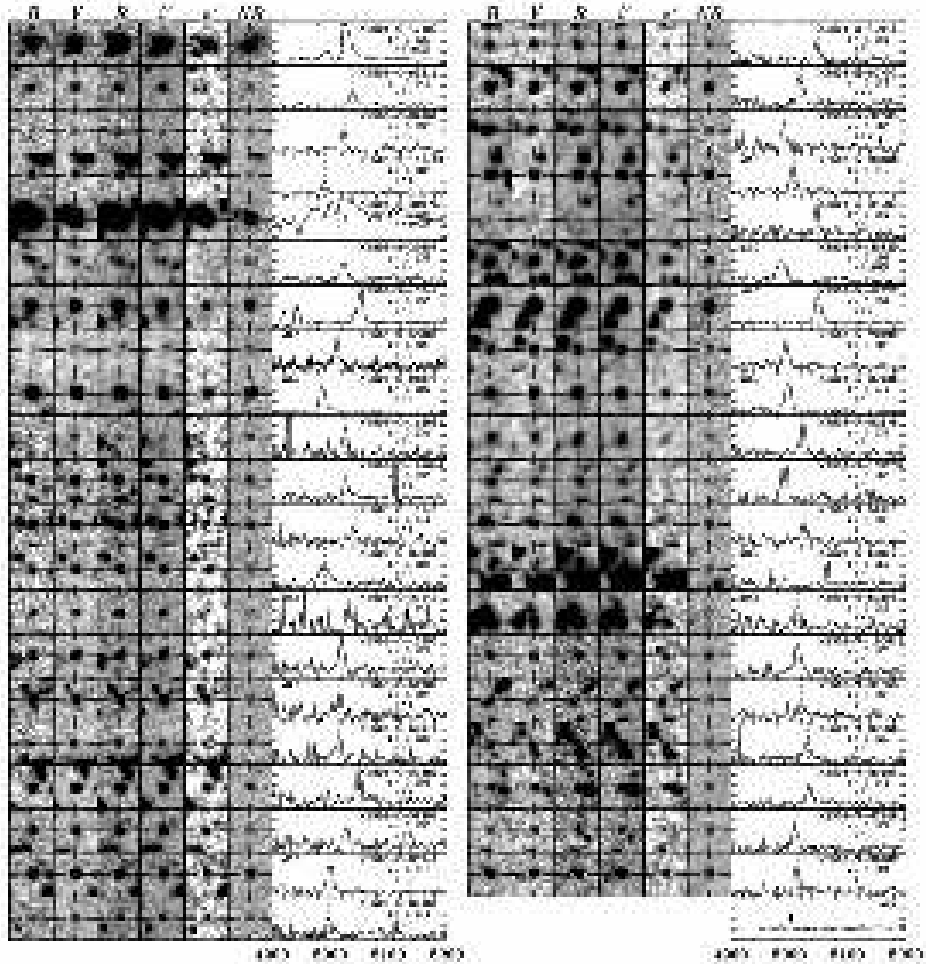}
\caption{
Spectra and snapshots of our $z=3.1$ LAEs.
Each object has a spectrum in right and
snapshots of $B$, $V$, $R$, $i'$, $z'$,
and $NB503$ images in left. Each snapshot
is presented in a $6''\times 6''$ box. In the panel of
spectrum, the tick of $y$ axis is marked 
in $2.5\times 10^{-18}$ erg s$^{-1}$ cm$^{-2}$ \AA$^{-1}$; 
for panels in which a factor is shown in the upper left
corner, multiply the scale by this factor
to obtain a correct scale. The wavelength (in $x$ axis) 
is in unit of \AA.
The object name and redshift (+AGN classification, if any)
are presented in the right corner of each spectrum panel.
'P' ('N') in parentheses indicates that the object
is (or is not) selected with the color criteria and included 
in our photometric sample (Section \ref{sec:photometric_lae_definitions}).
The vertical dotted lines mean the center of the emission line.
The right bottom panel shows a typical spectrum of the sky background 
with an arbitrary normalization.
\label{fig:image_spec_all_nb503}}
\end{figure}

\begin{figure}
\epsscale{1.0}
\plotone{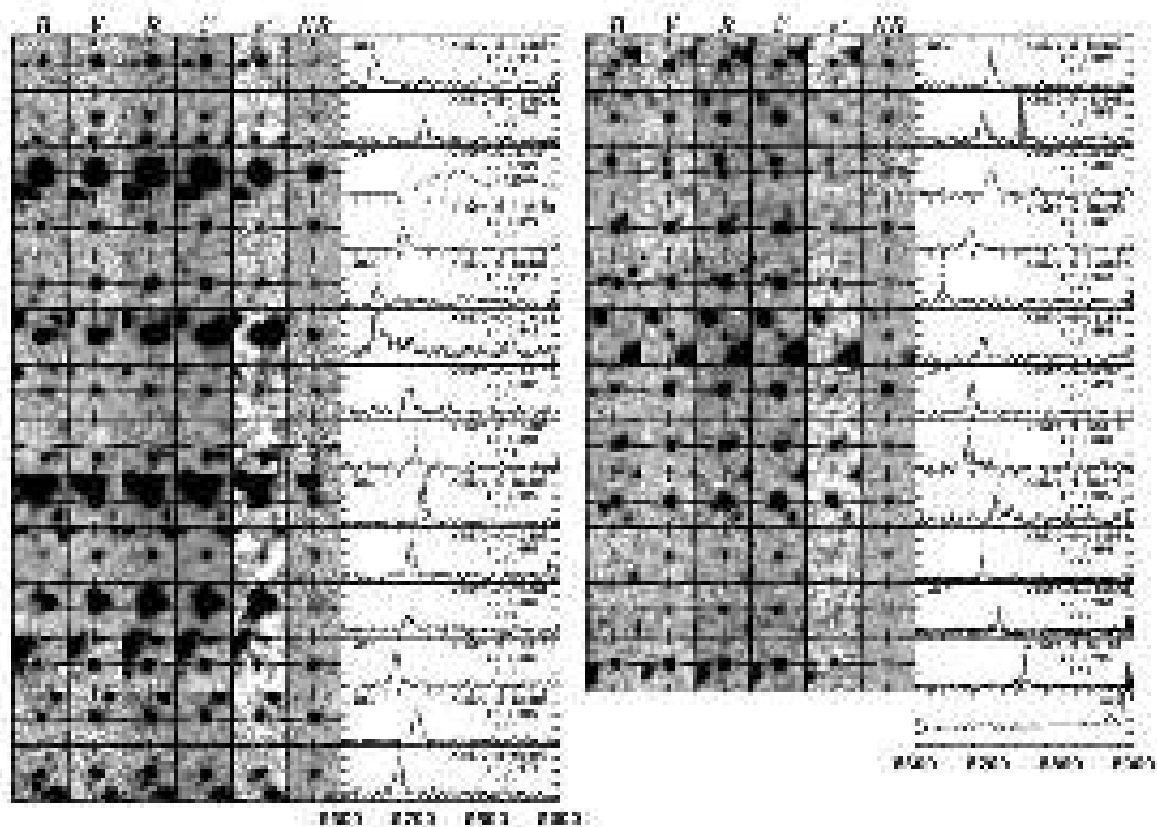}
\caption{
Same as Figure \ref{fig:image_spec_all_nb503},
but for our $z=3.7$ LAEs.
\label{fig:image_spec_all_nb570}}
\end{figure}

\begin{figure}
\epsscale{0.95}
\plotone{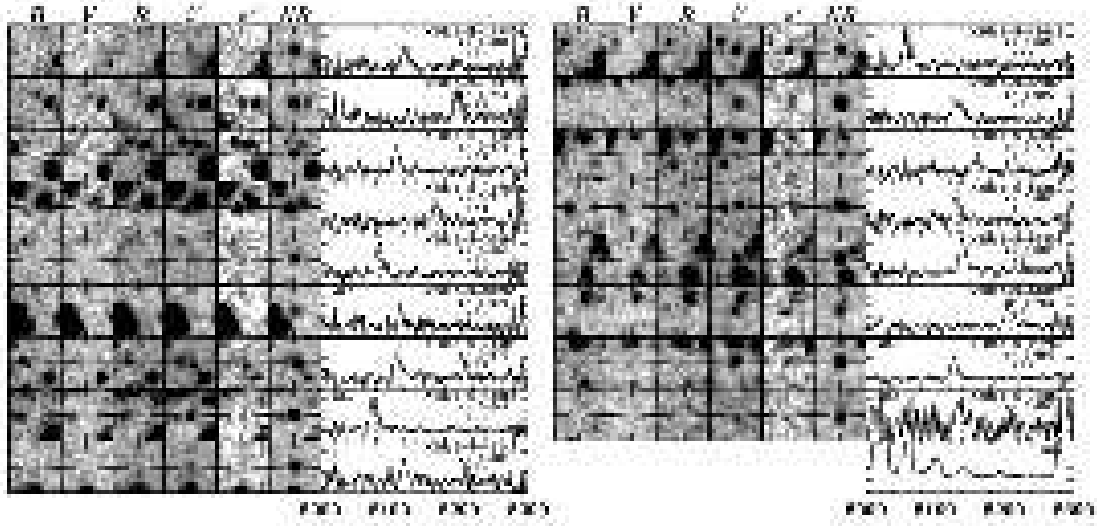}
\caption{
Same as Figure \ref{fig:image_spec_all_nb503},
but for our $z=5.7$ LAEs. The ticks of $y$ axis in a spectrum panel
are marked in $0.5\times 10^{-18}$ erg s$^{-1}$ cm$^{-2}$ \AA$^{-1}$.
\label{fig:image_spec_all_nb816}}
\end{figure}

\begin{figure}
\epsscale{1.0}
\plotone{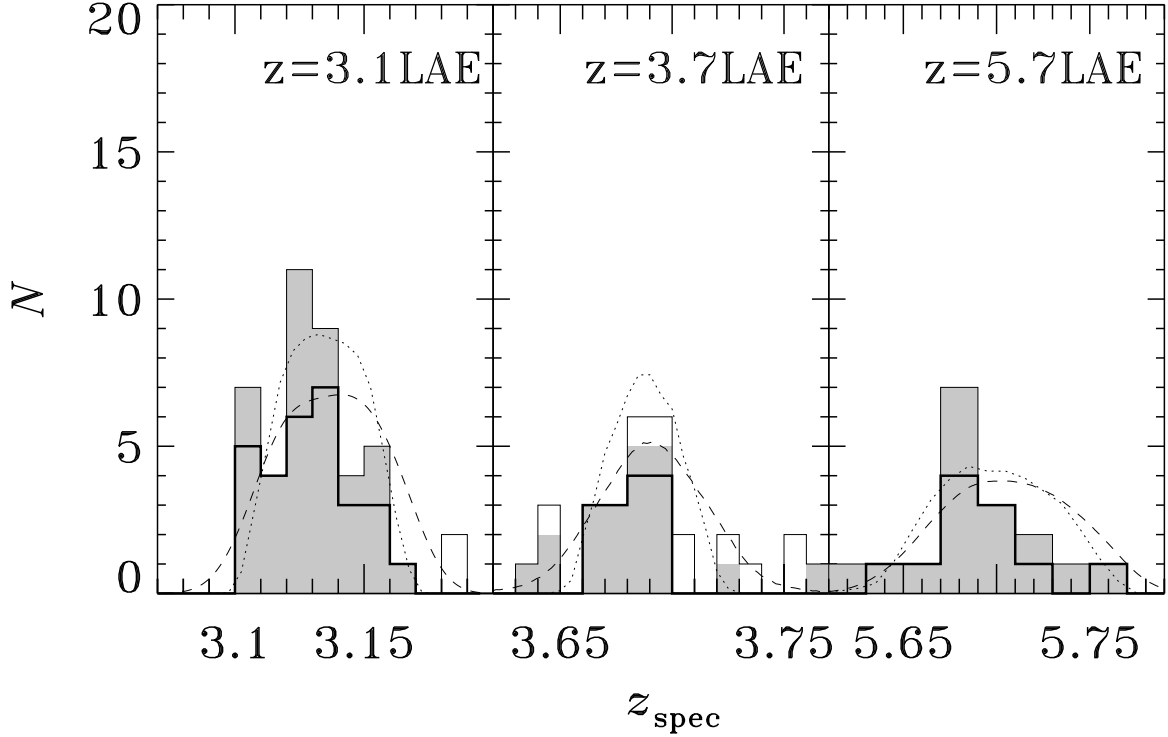}
\caption{The redshift distribution of spectroscopically identified 
LAEs at $z=3.1$ (left), $3.7$(middle), and $5.7$ (right).
Thin solid lines indicate all the spectroscopically identified
objects. The shaded histograms present the sources at the $5 \sigma$ level
in $NB503$ (left), $NB570$ (middle), and $NB816$ (right) bands.
The thick solid lines are histograms for LAEs selected by
the color criteria of equations 
(\ref{eq:laeselection_nb503}) (left), (\ref{eq:laeselection_nb570}) (middle), 
and (\ref{eq:laeselection_nb816}) (right).
The dashed lines represent selection functions of LAEs that
are simply calculated from the response curves of narrow-band filters.
The dotted lines are selection functions from the results of 
our simulations in Section \ref{sec:ly_alpha_luminosity_functions}.
Both of the selection functions are normalized by the numbers of
identified LAEs.
\label{fig:nz_zlae}}
\end{figure}

We have investigated our spectra with an emission line at the passbands
of our narrow bands. We carefully check the possibilities of low-$z$
[{\sc Oii}], [{\sc Oiii}], and H$\alpha$ emitters.
For $z=3.1$ LAE candidates with an emission at $\sim 5000$\AA,
we look at the spectrum at the wavelengths of 
an H$\alpha$ line of a $z=0.004$ [{\sc Oiii}] emitter
and H$\alpha$/[{\sc Oiii}] lines of a $z=0.3$ [{\sc Oii}] emitter.
Similarly, the possibilities of a $z=0.1$ [{\sc Oiii}] and a $z=0.5$ 
[{\sc Oii}] emitters are examined for $z=3.7$ LAE candidates.
For $z=5.7$ LAE candidates, our spectra allow us to identify
a $z=0.2$ H$\alpha$ emitter and a $z=0.6$ [{\sc Oiii}] emitter with corresponding 
[{\sc Oii}]/[{\sc Oiii}] and [{\sc Oii}] lines, respectively.
However, no nebular emission lines enter in the wavelengths of our optical spectra
for a $z=1.2$ [{\sc Oii}] emitter. 
Because our spectral resolution
is not high enough to clearly identify an [{\sc Oii}] doublet, we examine
the possibility of an [{\sc Oii}] emitter by the detection of
blue continuum in $B$ or $V$ bands whose $2\sigma$ upper limits
reach very deep magnitudes of $B=28.7$ and $V=27.7$.
In this way, we discriminate low-$z$ emitters from high-$z$ LAEs.
We also visually inspect images and spectra, and remove 
spurious objects.
By these analyses, we have identified 41, 26, and 17 LAEs at $z=3.1$, $3.7$, $5.7$.
We refer to these 84 LAEs as the spectroscopic samples of LAEs.
In Figures \ref{fig:image_spec_all_nb503}-\ref{fig:image_spec_all_nb816}
we show spectra of $z=3.1-5.7$ LAEs, 
together with snapshots of broad- and narrow-band images.
Table \ref{tab:laes_with_redshifts} summarizes the properties of LAEs 
with a spectrum. Figure \ref{fig:nz_zlae} plots
the redshift distribution of these 
spectroscopically-identified LAEs at $z=3.1$, $3.7$, and $5.7$.

\section{Photometric Samples of Ly$\alpha$ Emitters at $z=3.1-5.7$}
\label{sec:photometric}

\subsection{Definitions of $z=3.1$, $3.7$, and $5.7$ Ly$\alpha$ Emitters}
\label{sec:photometric_lae_definitions}

\begin{figure}
\epsscale{1.1}
\plotone{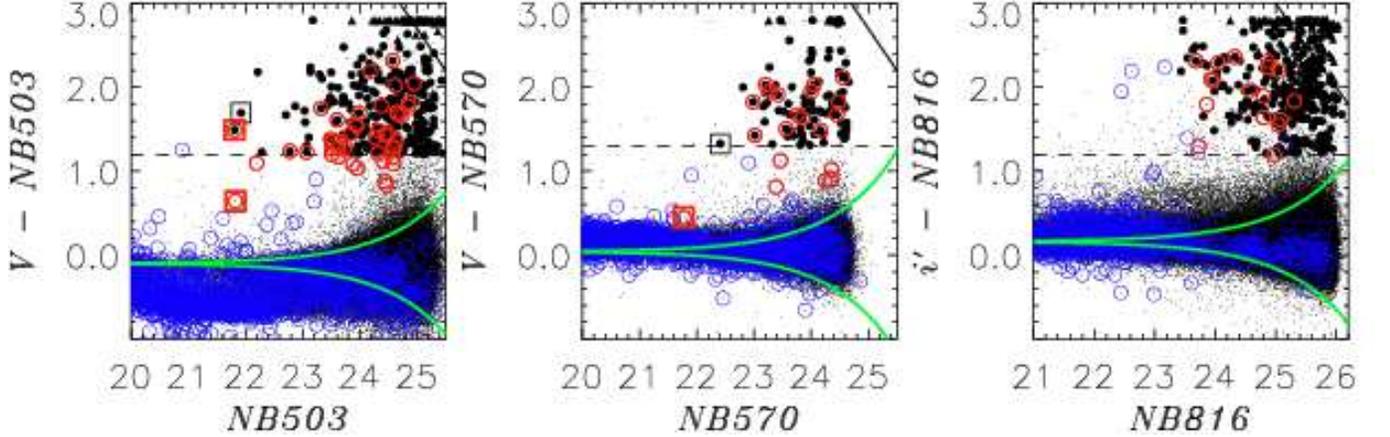}
\caption{
Color magnitude diagrams of $NB503$ (left),
$NB570$ (middle), and $NB816$ (right).
The black dots plot colors of all the
detected objects. The photometrically-selected LAEs
are presented with filled circles and triangles.
The LAEs of triangles have a broad-band 
($V$ or $i'$) magnitude
fainter than that of the $2 \sigma$ level,
and show the lower limits of the narrow-band
excess color. The red and blue open circles
denote spectroscopically-identified objects
in the redshift range of LAEs and interlopers, respectively.
We define the redshift ranges of LAEs as
$z=3.09-3.18$, $3.63-3.75$, and $5.60-5.78$
in the diagrams of $NB503$, $NB570$, and $NB816$, respectively
(see Figure \ref{fig:nz_zlae} for these redshift ranges).
The red and black squares mark AGNs from our multi-wavelength data
(Table \ref{tab:multi-wavelength_prop})
with and without a spectroscopic redshift, respectively.
For a display purpose, we place symbols at 
the narrow- v.s. broad-band color of $2.8$
for objects with a color redder than $2.8$.
The green lines indicate $3\sigma$ errors of
the colors of narrow v.s. broad bands
for a source with a color of $V-NB503=-0.10$,
$V-NB570=0.04$, and $i'-NB816=0.16$, which correspond
to the average color of all objects.
The solid and dashed lines represent the $2\sigma$ limit
of a broad band, and the color cut of narrow-band excess.
Note that the narrow-band magnitudes, i.e. 
$NB503$, $NB570$, and $NB816$, are total magnitudes, while
the colors of $V-NB503$, $V-NB570$, and $i'-NB816$
are defined with a $2''$-diameter aperture.
\label{fig:cm_all}}
\end{figure}

\begin{figure}
\epsscale{1.1}
\plotone{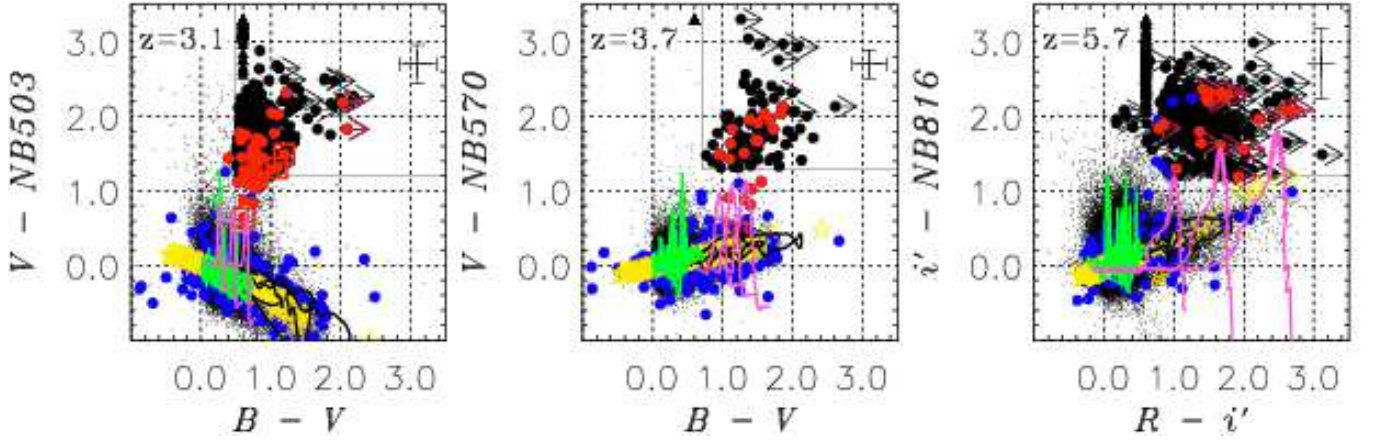}
\caption{
Two color diagrams of
continuum and narrow-band excess.
{\it Left panel:} 
The diagram of $B-V$ v.s. $V-NB503$ for our $z=3.1$ LAEs.
The black dots present colors of all the detected objects.
The photometrically selected LAEs
are plotted with filled circles and triangles.
The LAEs of triangles have an off-band
($V$) magnitude fainter than $2 \sigma$ level,
and show the lower limits of $V-NB503$. 
The red and blue circles/triangles
indicate spectroscopically identified objects
in and out of the LAE redshift range, respectively.
We take the LAE redshift range same as
that of Figure \ref{fig:cm_all}.
The red squares mark spectroscopically identified AGNs
found in our multi-wavelength data
(Table \ref{tab:multi-wavelength_prop}).
For a display purpose, we place symbols at 
$V-NB503=3.3$ for the objects with 
a color of $V-NB503>3.3$.
The objects with no continuum ($V$-band) detection
below a $2\sigma$ level are plotted at a color of $B-V=0.6$.
The error bars at the upper-right corner indicate
$1\sigma$ errors for a typical LAE whose brightness 
is the median magnitudes of our photometrically selected LAEs.
The solid line shows the color criteria of eq. (\ref{eq:laeselection_nb503})
for sources with a detection in the broad band ($V$).
The colors are defined with a $2''$-diameter aperture.
The curves present tracks of model galaxies 
at different redshifts.
Pink lines indicate model-LAE SEDs which are
composite spectra of a 0.03 Gyr single burst
model galaxy \citealp{bruzual2003}
and a Lyman $\alpha$ emission
($EW_{\rm rest}$(Ly$\alpha$)=22 \AA);
from the left to right, three different amplitudes
of IGM absorption are applied:
$0.5\tau_{\rm eff}$, $\tau_{\rm eff}$, and
$1.5\tau_{\rm eff}$, where $\tau_{\rm eff}$ is
the \citeauthor{madau1995}'s \citeyearpar{madau1995}
original median opacity.
The narrow-band excess in each of the peaks in
the red lines
indicates the Lyman $\alpha$ emission
of LAEs at $z=3.1$.
Green lines show 6 templates of
nearby starburst galaxies \citep{kinney1996} up to $z=2$,
which are 6 classes of
starburst galaxies with different dust extinction
($E(B-V)=0.0-0.7$). The narrow-band excess
peaks in the green line
correspond to the emission lines of
[{\sc Oiii}]($z=0.004$), H$\beta$($z=0.03$),
or [{\sc Oii}]($z=0.3$).
Black lines show colors of typical
elliptical, spiral, and irregular galaxies \citep{coleman1980}
which are redshifted from $z=0$ to $z=2$. Yellow star marks show
175 Galactic stars given by \citet{gunn1983}.
{\it Middle panel:} 
Same as the left panel but for $z=3.7$ LAEs.
{\it Right panel:} 
Same as the left panel but for $z=5.7$ LAEs.
We do not show the error bar of a $R-i'$ color at the upper-right corner,
because 
a typical LAE has no $R$-band flux above a $1\sigma$ noise level.
\label{fig:cc_all}}
\end{figure}

We plot color-magnitude diagrams in Figure \ref{fig:cm_all}
for objects in our photometric and spectroscopic catalogs.
Each panel of Figure \ref{fig:cm_all} shows
narrow-band excess color and narrow-band magnitude
for $NB503$, $NB570$, and $NB816$.
Figure \ref{fig:cc_all} presents two-color diagrams 
based on the $NB503$-, $NB570$-, 
and $NB816$-detection catalogs, together with 
colors of model galaxies and Galactic stars.
Colors of model objects indicate that LAEs can be isolated
from low-$z$ galaxies and Galactic stars
by their narrow-band excess of Ly$\alpha$ emission
and red continuum colors.

We compare colors of galaxies in our LAE samples with 
those of the 3,233 spectroscopically-identified objects
which include LAEs and foreground/background interlopers.
As expected, spectroscopically-identified LAEs
are located in the upper-right part of the
two-color diagrams.

Based on 
these color diagrams,
we define the selection criteria of three LAE samples:
{\footnotesize
\begin{eqnarray}
\label{eq:laeselection_nb503}
V-NB503>1.2\ \&\
\left[(V<V_{2\sigma}\ \&\ B-V>0.5)\ {\rm or}\ (V\ge V_{2\sigma}\ \&\ B-V_{2\sigma}>0.5)\right]
\ \ \ \ {\rm for\ {\it z=3.1}\ LAEs},\ \ \\
\label{eq:laeselection_nb570}
V-NB570>1.3\ \&\
\left[(V<V_{2\sigma}\ \&\ B-V>0.7)\ {\rm or}\ (V\ge V_{2\sigma}\ \&\ B-V_{2\sigma}>0.7)\right]
\ \ \ \ {\rm for\ {\it z=3.7}\ LAEs},\ \ \\
\label{eq:laeselection_nb816}
i'-NB816>1.2\ \&\
B>B_{2\sigma}\ \&\
V>V_{2\sigma}\ \&\
\left[(R<R_{2\sigma}\ \&\ R-i'>1.0)\ {\rm or}\ (R\ge R_{2\sigma})\right]
\ \ \ \ {\rm for\ {\it z=5.7}\ LAEs},\ \ 
\end{eqnarray}
}
where
$B_{2\sigma}$, $V_{2\sigma}$, and $R_{2\sigma}$ are
$2\sigma$ limiting magnitudes of $B$, $V$, and $R$ images,
i.e. $B=28.7$, $V=27.7$, and $R=28.0$, respectively. 
With these criteria, we include the objects with no detectable 
continuum in $V$ (for $z=3.1$ and 3.7) and $i'$ (for $z=5.7$). 
Thus, these photometric samples are complete to the narrow-band
magnitude limits, which include all the Ly$\alpha$ emitting objects 
at these redshifts either with or without a detectable-UV continuum.
The narrow-band excess colors, $V-NB503>1.2$, $V-NB570>1.3$,
and $i'-NB816>1.2$, isolate an emission line object at each redshift
with a flat continuum ($f_\nu=$const) whose rest-frame EW is greater 
than $\simeq 45$\AA.
Note that the actual EW limits for our LAE samples are different from
$\simeq 45$\AA, since LAEs do not have a flat continuum, but a continuum break,
i.e. Gunn-Peterson trough, between blue and red sides of Ly$\alpha$ emission. 
Thus, the narrow- and broad-band colors do not provide
a clear limit of EWs.
We estimate the limits of rest-frame EWs 
to be approximately $\sim 64$\AA, $\sim 44$\AA, 
$\sim 27$\AA\ for our $z=3.1$, $3.7$, and $5.7$ LAE samples
with models of our Monte-Carlo simulations described 
in Section \ref{sec:ly_alpha_luminosity_functions}.

We apply these selection criteria to our photometric catalogs.
We use $NB503$-, $NB570$-, and $NB816$-detection catalogs 
and find 356, 101, and 401 LAEs
at $z=3.1$, $3.7$, and $5.7$, respectively. 
In this way, we obtain LAE samples composed of
858 photometrically-identified LAEs in total.
We refer to these 858 LAEs as the photometric samples of LAEs,
which are summarized in Table \ref{tab:sample}.
Note that these photometric samples are purely composed of 
objects satisfying the selection criteria of 
eq. (\ref{eq:laeselection_nb503}), 
(\ref{eq:laeselection_nb570}), or
(\ref{eq:laeselection_nb816}).
Our photometric samples do not include spectroscopically-identified
LAEs escaping from the photometric criteria.

Figures \ref{fig:number_density_nb503}-\ref{fig:number_density_nb816}
show surface densities of LAEs.
The red circles are the average surface densities.
The black points with 5 different symbols
indicate the surface densities in 5 sub-fields
($\simeq 0.2$ deg$^2$) of Suprime-Cam, i.e., SXDS-C, -N, -S, -E, -W.
The detection completeness is corrected with the simulations
described in Section \ref{sec:photometric_lae_completeness}.
For these completeness-corrected points,
the difference of surface densities among the five $\simeq 0.2$ deg$^2$ fields
are negligible for all narrow-band detected objects,
but are as large as a factor of, say, $\simeq 2-5$, for LAEs. 
These large differences probably come from
field-to-field variation as well as Poisson errors.
We evaluate the field-to-field variation in our survey area, $\sigma_{g}$, 
with

\begin{eqnarray}
\label{eq:field_variation}
\sigma_{g} & = & 
\sigma_{g:1FoV} (\sigma_{DM}/\sigma_{DM:1FoV})\\
\sigma_{g:1FoV}^2 & = &
[\left<(\Sigma_{g:1FoV}-\bar{\Sigma}_{g})^2\right>-\bar{\Sigma}_{g}]/
\bar{\Sigma}_{g}^2
\end{eqnarray}
where $\sigma_{DM}$ and $\sigma_{DM:1FoV}$ are
the rms fluctuation of dark matter in all the survey volume of 1 deg$^2$
and Suprime-Cam's 1-FoV volume ($\simeq 0.2$ deg$^2$), respectively.
We calculate the fluctuations of dark matter with the power spectrum,
adopting the transfer function given by \citet{bardeen1986}.
$\sigma_{g:1FoV}$ is the fluctuation of number density of LAEs for the 1 FoV.
$\Sigma_{g:1FoV}$ and $\bar{\Sigma}_{g}$ are LAE's surface densities
of the 1 FoV and the average of survey area.

Since these estimates of field-to-field variation are based on
a large but single contiguous field, it is important to check
whether our field is located at the sky of
an overdense or underdense region.
In fact, a large-scale overdensity or underdensity of 
Ly$\alpha$ sources could also be produced by 
an inhomogeneous distribution of Ly$\alpha$ absorbers
(i.e. neutral hydrogen) along the line of sight.

In Figure \ref{fig:number_density_nb816} 
we compare the surface densities of our LAEs with those on 
completely different sky areas but selected with the same $NB816$ filter 
by \citet{shimasaku2006} and \citet{hu2004}. 
We find that the surface densities of our $z=5.7$ LAEs are 
consistent with those of \citet{shimasaku2006}
and \citet{hu2004} within the scatters and Poisson errors
of our 5 sub-fields. Moreover, Figure \ref{fig:number_density_nb816} shows 
that Shimasaku et al.'s and Hu et al.'s measurements scatter around
our average surface-density curve.
Thus, we conclude that our single contiguous 1 deg$^2$ field has no signature
of overdensity or underdensity, and that our 5 sub-fields
represent the average field-to-field variation.

In the same manner as the calculations of surface densities,
Figure \ref{fig:color_hist_all} plots
color distributions with the uncertainties of field variance
for our LAE samples. In Section \ref{sec:ly_alpha_luminosity_functions},
we compare these color distributions (as well
as the surface densities) with those reproduced by our Monte-Carlo simulations,
and determine the best-estimates of luminosity function by
$\chi^2$ fitting.

\subsection{Completeness and Contamination of the Samples}
\label{sec:photometric_lae_completeness}

 First, we estimate the detection completeness of each narrow-band
images, $f_{\rm det}$, as a function of narrow-band magnitude. 
We distribute 7875 artificial objects
that mimic LAEs 
on our original 1 deg$^2$ images after adding photon noise,
and detect them in the same manner as
for the detection of our LAE catalogs with SExtractor. 
We repeat this process 10 times, and compute the detection fraction.
We find that the detection completeness is typically 
$\gtrsim 80-90$\% for relatively luminous sources 
which are 0.75 magnitude brighter than
the magnitude cuts of our samples 
(i.e. $NB503\le 24.55$, $NB570\le 23.95$, and $NB816\le 25.25$).
The detection completeness is $\gtrsim 50-60$\%
even for the faintest magnitude bin of our samples
(i.e. $NB503=24.8-25.3$, $NB570=24.2-24.7$, and $NB816=25.5-26.0$).

 Second, we estimate the contamination 
of our photometric LAE samples, which we need to
consider in calculating the LF.
We use our spectroscopic catalog of 3,233 objects 
(Section \ref{sec:observations_data_spectroscopiccatalogs}).
We define the contamination fraction, $f_{\rm cont}$, 
with
\begin{eqnarray}
f_{\rm cont} & = & 1 - N_{\rm LAE}^{\rm in} / N_{\rm all}^{\rm in}
\label{eq:contamination_completeness}
\end{eqnarray}
where 
$N_{\rm LAE}^{\rm in}$ and $N_{\rm all}^{\rm in}$ are
the numbers of spectroscopically-identified LAEs
and all spectroscopic objects, respectively, lying in our color criteria 
(i.e. eqs \ref{eq:laeselection_nb503}-\ref{eq:laeselection_nb816}).

Since we have spectroscopic objects with no identification 
(see Section \ref{sec:observations_data_spectroscopiccatalogs}),
we calculate $f_{\rm cont}$ 
for the following two extreme cases.
If we omit the unclear objects, we find 
$N_{\rm LAE}^{\rm in} / N_{\rm all}^{\rm in} =  (29/29, 14/14, 11/11)$ 
for $z=(3.1, 3.7, 5.7)$. Note that we see no obvious interlopers 
in our photometric samples.
In case that all the no-identification objects are interlopers,
$N_{\rm LAE}^{\rm in} / N_{\rm all}^{\rm in} =  (7/8, 6/7, 3/4)$, 
where we take magnitude cuts of $(NB503,NB570,NB816)=(24.0,24.0,24.5)$ 
that are bright enough to be completely identified by our spectroscopy.
Thus, the contamination rates are taken within the ranges of 
$f_{\rm cont}=(0-13\%,0-14\%,0-25\%)$
for $z=(3.1, 3.7, 5.7)$ photometric LAE samples, respectively.
Since these contamination rates are negligibly small,
we do not correct them for 
contamination. It is notable that statistical errors from contamination
cannot be as large as $\sim 30$\%.

\section{AGN and Multi-Wavelength Detection in our Samples}
\label{sec:multi-wavelength_properties}
We identify AGNs in LAEs with our spectra and
SXDS multi-wavelength images of
radio, sub-mm, mid-infrared, and X-ray,
and investigate AGN activities in out LAEs.
We summarize the properties of these AGNs
in Table \ref{tab:multi-wavelength_prop}.
In this section, we formally define
the positional uncertainty of optically-identified 
LAEs, $\sigma_{\rm opt}=FWHM/2.35$,
with a FWHM of narrow-band morphology.

\subsection{Spectra with AGN signatures}
\label{sec:agn_spectra}
We find 3 and 1 objects at $z=3.1$ and $3.7$
whose spectrum shows high-ionization lines,
such as N{\sc v}, C{\sc iv}, He{\sc ii}, and C{\sc iii}],
originated from AGN activities. Figure \ref{fig:disp_agn}
plots spectra of these four objects. All of these
AGNs show broad lines in these high-ionization lines
with a line width of $v_{\rm FWHM} \gtrsim 1000$km $^{-1}$,
although the line width for
one of these object, NB503-N-35820, has a large uncertainty.
Radio bright objects, i.e. radio galaxies, are also included
in our sample. The properties of these AGNs are presented
in Table \ref{tab:multi-wavelength_prop}.

We also find that NB503-N-42377 shows a marginal 
($\sim 2\sigma$) {\sc Civ} emission. Although
a typical high-$z$ AGN has a line ratio of 
Ly$\alpha/${\sc Civ}$=5-10$ (e.g. \citealt{mccarthy1993}), NB503-N-42377 has
Ly$\alpha/${\sc Civ}$\gtrsim 25$. With the facts of 
the marginal {\sc Civ} detection and the large flux ratio of Ly$\alpha/${\sc Civ},
we rule out an AGN being the dominant power source of Ly$\alpha$
for NB503-N-42377, and do not classify this object as an AGN.

\subsection{X-ray Detection}
\label{sec:x-ray_properties}
We use a deep XMM-Newton EPIC image of SXDS, and identify
an X-ray counterpart of LAEs with the 
combined X-ray catalog of SXDS ver 5.0 (Ueda et al. in prep.).
This X-ray catalog is made of sources with
a detection likelihood greater than 7
in either of 6 bands, 0.3-0.5, 0.5-2, 2-4.5, 
4.5-10, 0.5-4.5, and 2-10 keV (see Ueda et al. in prep). 
We calculate a combined positional error
by taking
a sum of square errors of optical and X-ray positions.
Although there are 3 X-ray counterpart candidates
with a $<2\sigma$ level of the combined error, 
we find by our visual inspection that
one of the candidate, NB503-N-87126, clearly 
appears to be confused by the neighboring foreground objects.
This X-ray counterpart is regarded as
a foreground object.
We thus identify 2 X-ray counterparts;
1 and 1 counterparts at $z=3.1$ and $3.7$,
respectively (Table \ref{tab:multi-wavelength_prop}).
No counterparts are found for $z=5.7$ LAEs.

\subsection{Infrared Detection}
\label{sec:infrared_properties}
The Spitzer MIPS $24\mu$m-band image is investigated
for infrared counterparts of LAEs. We use
relatively shallow data of the Spitzer Wide-area InfraRed Extragalactic
(SWIRE; \citealt{lonsdale2003}) survey.
The SWIRE MIPS data cover almost entire SXDS field
of a 0.9 deg$^2$, and miss only 13, 6, and 15 LAEs
\footnote{
All of X-ray and radio emitting LAEs listed 
in Table \ref{tab:multi-wavelength_prop} 
are included in the MIPS area.
}
at $z=3.1$, 3.7, and 5.7, respectively,
which are located at the edge of 
our SXDS-W field (see Figure 1 of Morokuma et al. 2007 submitted).
The depth of MIPS data is $\sigma=48\mu$Jy (Shupe et al. in preparation;
see \citealt{ivison2007}). We run SExtractor to detect and 
measure source fluxes with a $12''$-diameter aperture. We apply 
the aperture correction of a factor of 1.698 to the aperture fluxes
(see \verb+http://ssc.spitzer.caltech.edu/mips/apercorr/+).
We have cross matched our LAEs with MIPS sources detected at
a $>5 \sigma$ level, i.e. $>240\mu$Jy.
We find 3 (and 0) MIPS counterparts at $z=3.1$ (at $z=3.7$ and $5.7$)
within the $3 \sigma$ positional allowance of our LAEs
(Table \ref{tab:multi-wavelength_prop}).

\subsection{Sub-mm Detection}
\label{sec:sub-mm_properties}
We compare the positions of LAEs with
SCUBA $850\mu$m sources of
Submm HAlf-Degree Extragalactic Survey (SHADES;
\citealt{coppin2006}). The rms noise level is $\simeq 2.2$mJy.
The SHADES image covers
an $\sim 360$ arcmin$^2$ area at the center of
SXDS-C, which corresponds to about 10\% of our 1 deg$^2$ SXDS field.
In this SHADES region, we have about 16, 6, and 3 LAEs 
at $z=3.1$, $3.7$, and $5.7$, respectively.
These LAEs include one X-ray emitting LAE of
NB503-C-49497 
at $z=3.128$ (see Table \ref{tab:multi-wavelength_prop}).
The beam size of the sub-mm image is as large as $FWHM=14.7$ arcsec.
Following \citet{ivison2007}, we search counterparts with 
a positional error circle of radius 8 arcsec.
We find no submm counterparts in our LAEs (including the X-ray emitting
LAE) located at the SHADES area.
We extend our search radius up to 12.5 arcsec for completeness.
We find that a $z=5.7$ LAE of NB816-C-90169 falls on
the large search radius of a submm source of Sxdf850.5
\citep{coppin2006}. However, on our optical images
this submm source is very likely centered
at a foreground galaxy of SXDS-iC-086385 with $i'=21.11$.
Thus, we conclude that none of our LAEs 
has a sub-mm emission down to $\simeq 3-5$mJy
in the SHADS area.

\subsection{Radio Detection}
\label{sec:radio_properties}
We investigate radio properties of LAEs
with deep VLA 1.4-GHz radio image of the SXDS
\citep{simpson2006a}. This radio image reaches
an rms noise level of 12$\mu$Jy beam$^{-1}$.
We search a radio counterpart of LAEs 
in a 3x3 pixel box of the VLA image 
(1 pixel=1.25 arcsec). A radio source with a signal-to-noise
ratio of $>5$ are identified as a radio counterpart
in the same manner as \citet{simpson2006b}.
We then visually inspect LAEs with radio contours,
and find that two LAEs are detected in the radio image
\footnote{
We find that a radio source is located around
NB503-S-29853. Since the flux center of this radio source
is obviously placed outside of NB503-S-29853, 
this radio source is likely associated
with the other source(s).
}
(Table \ref{tab:multi-wavelength_prop}).

\subsection{LAEs Hosting AGN}
\label{sec:lae_hosting_agn}
As summarized in Table \ref{tab:multi-wavelength_prop},
we find that 4 $z=3.1$ and 1 $z=3.7$ LAEs
have a detection in the multi-wavelength data.
Figure \ref{fig:multiband_sed} presents the spectral
energy distribution (SED) of these AGN-LAEs, 
together with those of the average radio quiet/loud quasars \citep{elvis1994}. 
We also plot a UV-optical spectrum of \citet{telfer2002,richards2003}
that is normalized to the average SED of \citep{elvis1994}.
We refer to these average SEDs+spectrum as QSO templates.

Since the detection limits of infrared and X-ray (and radio)
bands are as large as $\nu L_\nu \gtrsim 10^{44}$ $(10^{41})$ erg s$^{-1}$, 
LAEs with any multi-wavelength counterparts likely host AGN. 
In fact, we have 3 spectroscopically-identified LAEs with
multi-wavelength counterparts (Table \ref{tab:multi-wavelength_prop}).
The spectroscopic classifications of these LAEs indicate 
that all of these LAEs have AGN.
Thus, we classify all of 5($=4+1$) LAEs with multi-wavelength detection
as AGN. In addition to these 5 AGNs identified by multi-wavelength data,
we have 4 spectroscopically-identified AGNs. Since 3 AGNs have
both a multi-wavelength detection and spectroscopic identification,
the number of our LAEs with an AGN (hereafter AGN-LAEs)
is 6 in total: 4 and 2 at $z=3.1$ and $3.7$, respectively.
No AGN-LAEs are found
at $z=5.7$. Note that 3 $z=3.1$ and 1 $z=3.7$ AGN-LAEs out of 6 AGN-LAEs
satisfy our color criteria of photometric samples 
(eq. \ref{eq:laeselection_nb503} or \ref{eq:laeselection_nb570};
see Table \ref{tab:multi-wavelength_prop}). 
These numbers (i.e. 3 and 1) are sufficiently smaller than
those of entire photometric samples (356 and 101 for $z=3.1$ and $3.7$).
Since the AGN fraction in our photometric sample is negligibly small,
\footnote{
We discuss the AGN fraction of our photometric samples in
Section \ref{sec:evolution_of_agn_fraction}. We address
star-forming activities of LAEs with the multi-wavelength data
in Section \ref{sec:sfr_multiwavelength}.
}
AGN-LAEs do not largely contribute 
to the statistical properties of LAEs
in the following sections.

\section{Luminosity Functions}
\label{sec:luminosity_functions}

\subsection{Ly$\alpha$ Luminosity Functions}
\label{sec:ly_alpha_luminosity_functions}

 We derive luminosity functions (LFs) of LAEs at $z=3.1$, $3.7$,
and $5.7$ from our photometric samples.
First, we calculate LFs with a simple 
classical method that was taken by most of previous
studies (e.g. \citealt{ouchi2003,ajiki2003,hu2004,malhotra2004}. 
We obtain the number densities of LAEs in each magnitude bin
by simply dividing the observed number counts of LAEs
in a given narrow band by the effective survey volume
defined as the FWHM of the bandpass times the area of the
survey. Here, we calculate the Ly$\alpha$ luminosity of each object 
with the response curves of narrow and broad bands
by subtracting the continuum emission measured from the continuum
magnitude from the total luminosity in the narrow band, 
assuming that Ly$\alpha$ enters in the central wavelength of
the narrow band.
In this calculation, we use total magnitude of narrow-band images.
The continuum emission is estimated by the narrow-band excess
color (i.e. color of narrow-band and broad-band)
defined with a $2''$ aperture, so as to keep high
signal-to-noise ratios and to avoid the effects of source confusion
on broad-band images with high source density.
To check the accuracy of this calculation,
Figure \ref{fig:comp_flya_all} compares 
Ly$\alpha$ fluxes measured from
our images and spectra for our spectroscopic samples.
Figure \ref{fig:comp_flya_all} shows that 
both measurements agree well within error bars
for most of LAEs. These values mostly scatter
within the range of difference of a factor of 2.

Figures \ref{fig:lumifun_full_diff_nb503}-\ref{fig:lumifun_full_diff_nb816} 
present the LFs from this classical method. Note that we corrected the 
detection-completeness by weighting with $f_{\rm det}$
measured in Section \ref{sec:photometric_lae_completeness}.
To check the field-to-field variation and the accuracy of our results, 
we plot the estimates of LFs from the entire 1 deg$^2$ field (filled circles),
together with those from the five $\simeq 0.2$ deg$^2$ fields (open symbols).
In Figures \ref{fig:lumifun_full_diff_nb503}-\ref{fig:lumifun_full_diff_nb570}, 
we calculate the errors of
field-to-field variations with eq. (\ref{eq:field_variation}),
and include these errors in the error bars of LFs for 
the entire 1 deg$^2$ field. We find that the field-to-field variations
are as large as a factor of $\simeq 2-5$ among the five 0.2 deg $^2$ fields,
although the typical scatters of the 0.2 deg $^2$ results 
are {\it not} far beyond the errors of Poisson statistics.
Then we obtain the best-fit Schechter function \citep{schechter1976}
defined by
\begin{equation}
\phi(L)dL=\phi^*(L/L^*)^\alpha \exp(-L/L^*)d(L/L^*).
\label{eq:schechter_l}
\end{equation}

The classical method is accurate when the narrow-band filter 
has an ideal boxcar shape. However, since the shapes of actual filters 
used in LAE surveys are rather close to a triangle, 
the classical method potentially suffers from the following two 
uncertainties.
(a) The narrow-band magnitude of LAEs of 
a fixed Ly$\alpha$ luminosity varies largely as a function 
of redshift. Thus Ly$\alpha$ luminosity may be over- or under-estimated 
for some LAEs. (b) The selection function of LAEs in terms of EW
also changes with redshift; the minimum EW value corresponding 
to a given (fixed) narrow-band excess, such as $i'- NB816$ for $z=5.7$ LAEs, 
becomes larger when the redshift of the object goes away 
from the redshift corresponding to the center of bandpass. 

\begin{figure}
\epsscale{0.60}
\plotone{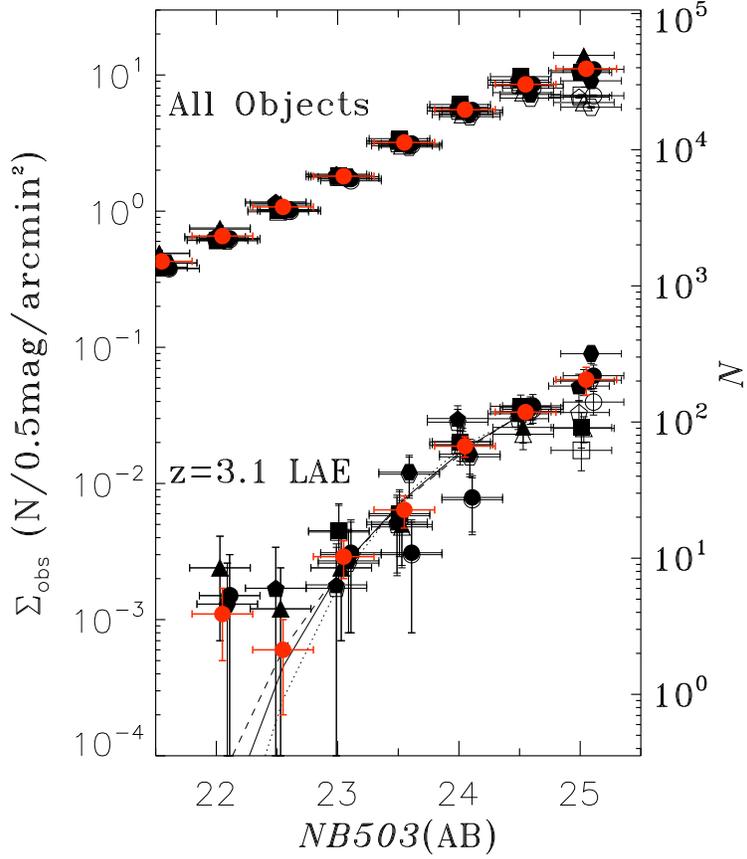}
\caption{
Surface densities of objects detected in
the $NB503$ data. The lower and upper points indicate
surface densities of our $z=3.1$ LAEs and
all the objects detected in the narrow band, respectively.
Black circles, hexagons, triangles,
squares, and pentagons plot the surface densities
of a $\sim 0.2$ deg field, SXDS-C, SXDS-N,
SXDS-S, SXDS-E, and SXDS-W, respectively.
We distinguish between 
the raw and completeness-corrected
surface densities with the open and filled symbols,
respectively. The red filled circles represent the
surface density averaged over our $\simeq 1$ deg$^2$ 
survey field. The errors are given by Poisson statistics
for black symbols, while the errors of red symbols
are made by the geometric mean of Poisson errors
and field-to-field variation of eq. (\ref{eq:field_variation}).
For the presentation purpose,
we slightly shift all the black points along the abscissa.
The exact magnitude
of a black point is the same as a magnitude of neighboring
red point. The vertical axis on the right side
indicates the numbers of objects, i.e. $N/(0.5 {\rm mag})$,
identified in our $\simeq 1$ deg$^2$ survey area.
The solid, dotted, and dashed lines present
the surface densities of LAEs reproduced by our
simulations for the cases of $\alpha=-1.5$, $-1.0$,
and $-2.0$, respectively 
(see Section \ref{sec:ly_alpha_luminosity_functions} for more details).
\label{fig:number_density_nb503}}
\end{figure}

\begin{figure}
\epsscale{0.95}
\plotone{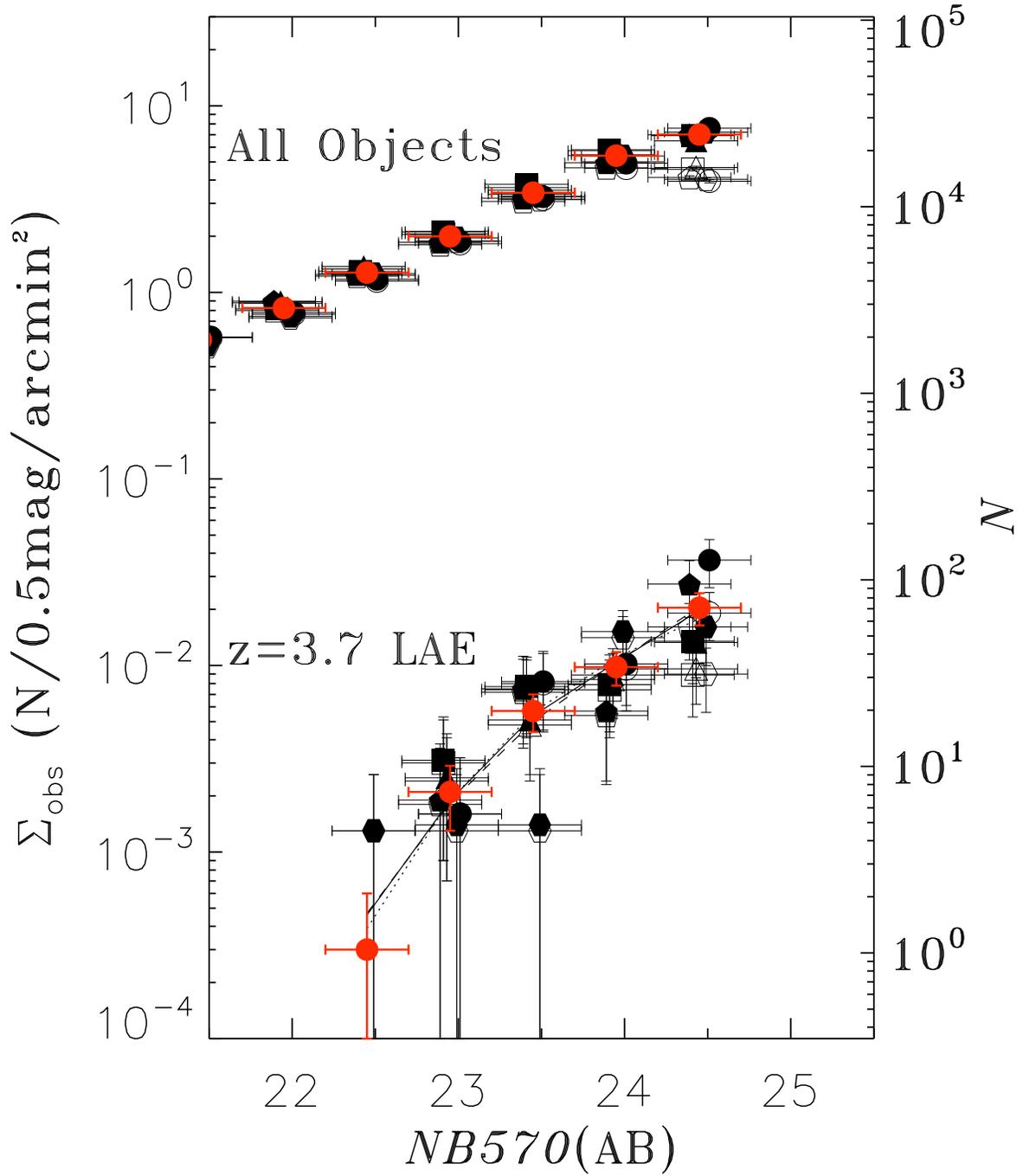}
\caption{
Same as Figure \ref{fig:number_density_nb503},
but for $NB570$ data.
\label{fig:number_density_nb570}}
\end{figure}

\begin{figure}
\epsscale{0.90}
\plotone{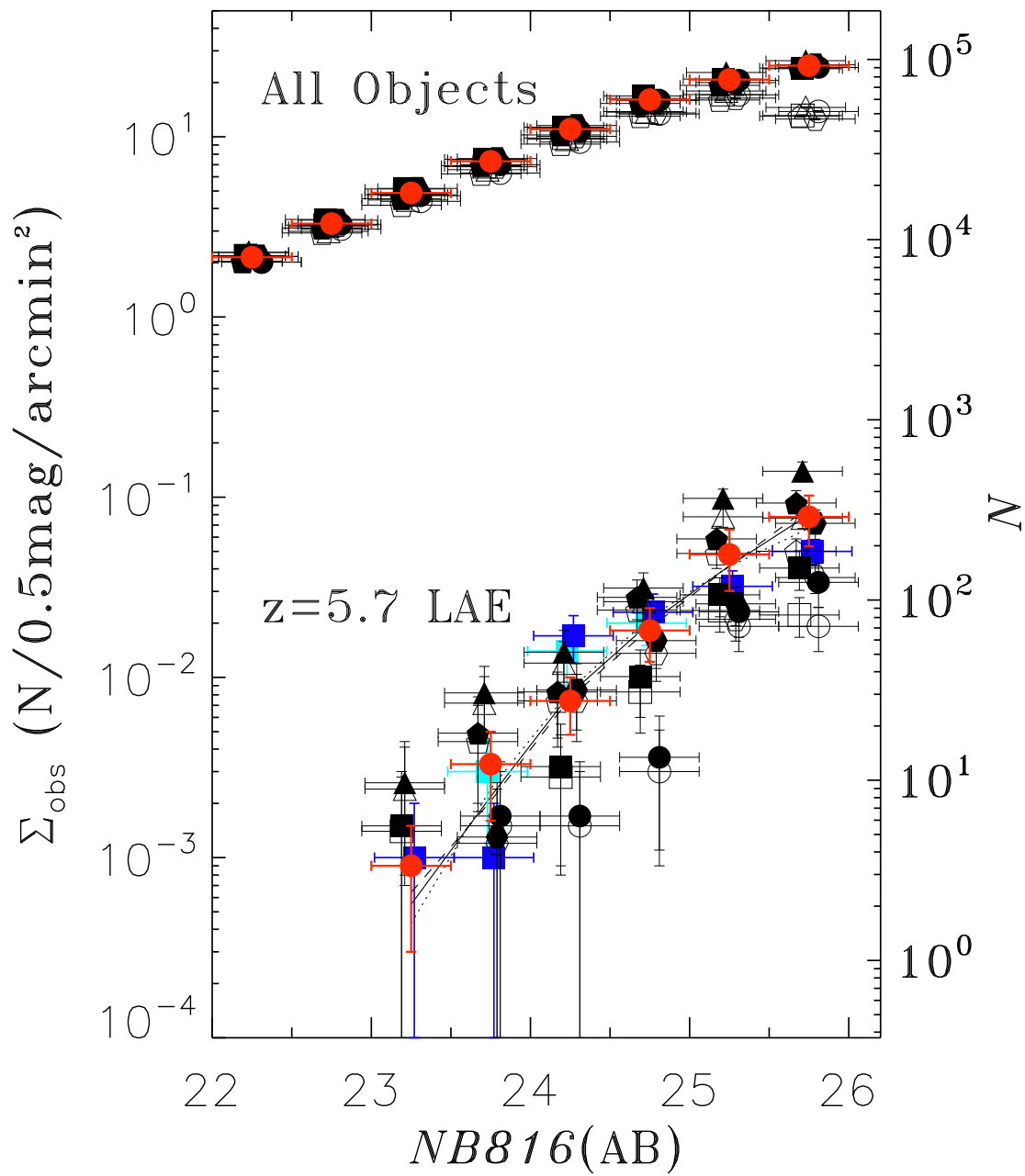}
\caption{
Same as Figure \ref{fig:number_density_nb503},
but for $NB816$ data. The blue and cyan squares
are the surface densities of $z=5.7$ LAEs
detected with the same $NB816$ filter but in different sky areas
by \citet{shimasaku2006}
and \citet{hu2004}, respectively.
\label{fig:number_density_nb816}}
\end{figure}

\begin{figure}
\epsscale{1.0}
\plotone{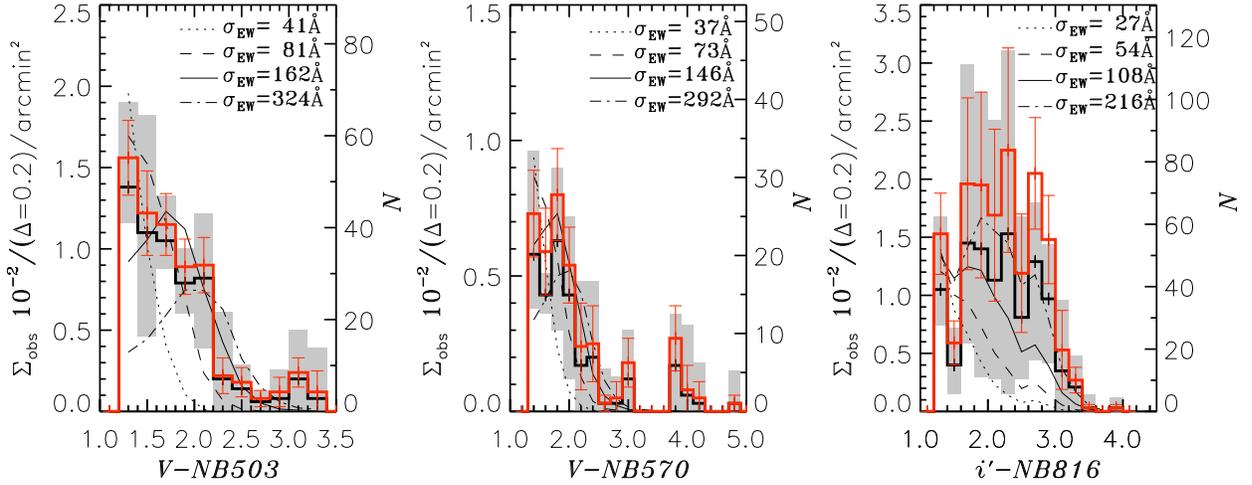}
\caption{
Color histograms of $V-NB503$ (left), $V-NB570$ (middle)
and $i'-NB816$ (right) for 
the photometric sample of $z=3.1$, $3.7$, and $5.7$ LAEs,
respectively.
The black-solid histogram presents the average
of all the LAEs in the $\simeq 1$ deg$^2$ field. 
The shaded region shows the
variations of histograms of five $\sim 0.2$ deg fields, 
SXDS-C, SXDS-N, SXDS-S, SXDS-E, and SXDS-W.
The red histogram is the same as the black-solid histogram,
but with a completeness correction. The error bars of the red histogram
are given by
the geometric mean of Poisson errors
and field-to-field variation of eq. (\ref{eq:field_variation}).
The right-hand vertical axis on each panel
indicates the numbers of LAEs for
the $\simeq 1$ deg$^2$ histogram.
We plot color distribution reproduced by our simulations of fixed $\alpha=-1.5$
with dotted, dashed, solid, and dot-dashed lines,
whose $\sigma_{\rm EW}$ values are presented in the legends.
\label{fig:color_hist_all}}
\end{figure}

\begin{figure}
\epsscale{0.90}
\plotone{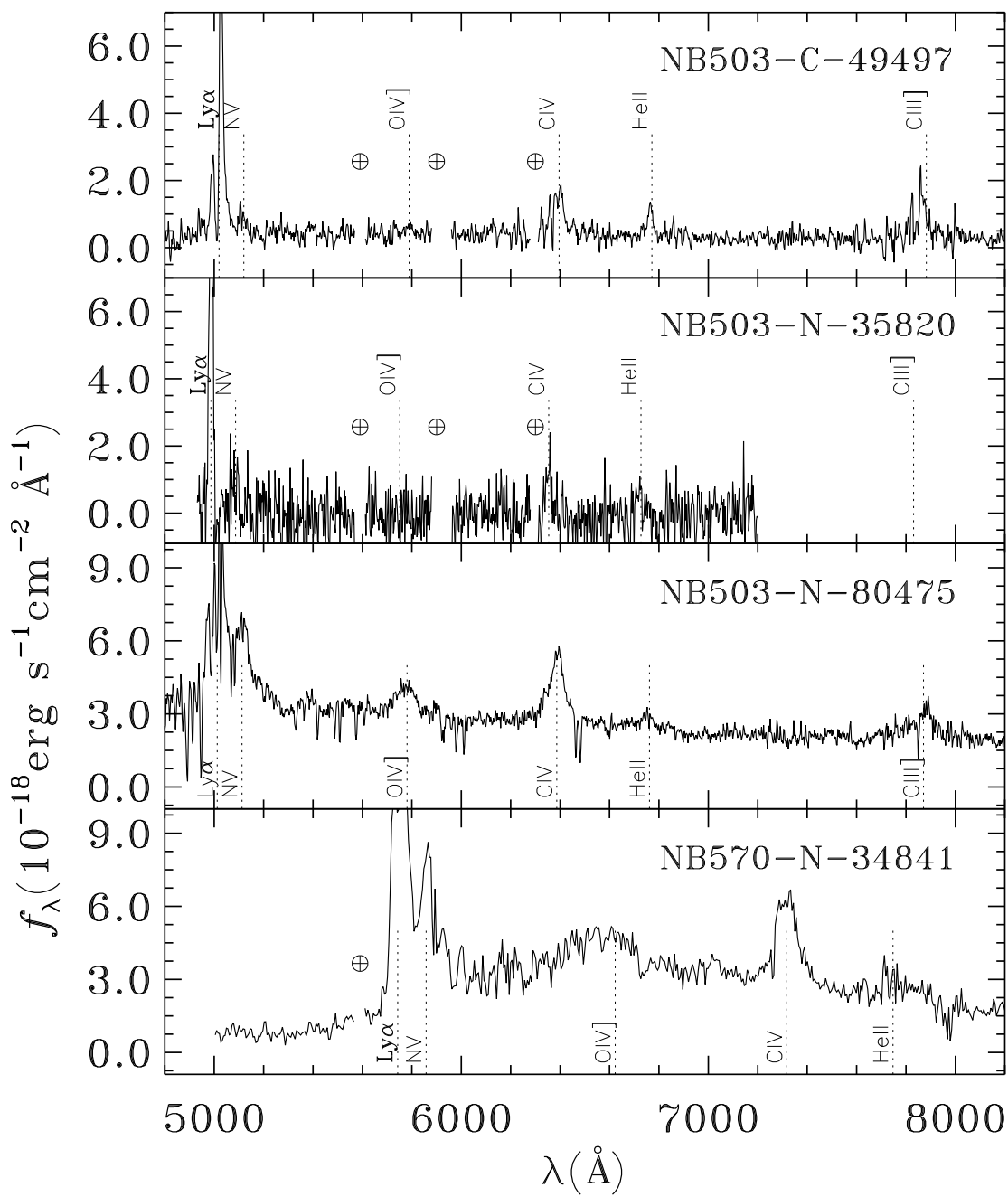}
\caption{
Spectra of LAEs with AGN activities. 
We show four AGN-LAEs that have spectroscopic data
(see Table \ref{tab:multi-wavelength_prop}).
Dotted lines
with the legend indicate wavelengths of
a typical emission from AGN. 
``$\oplus$'' marks the wavelengths
where significant residuals of sky subtraction
remain due to strong OH sky lines on a faint continuum.
\label{fig:disp_agn}}
\end{figure}

\begin{figure}
\epsscale{0.95}
\plotone{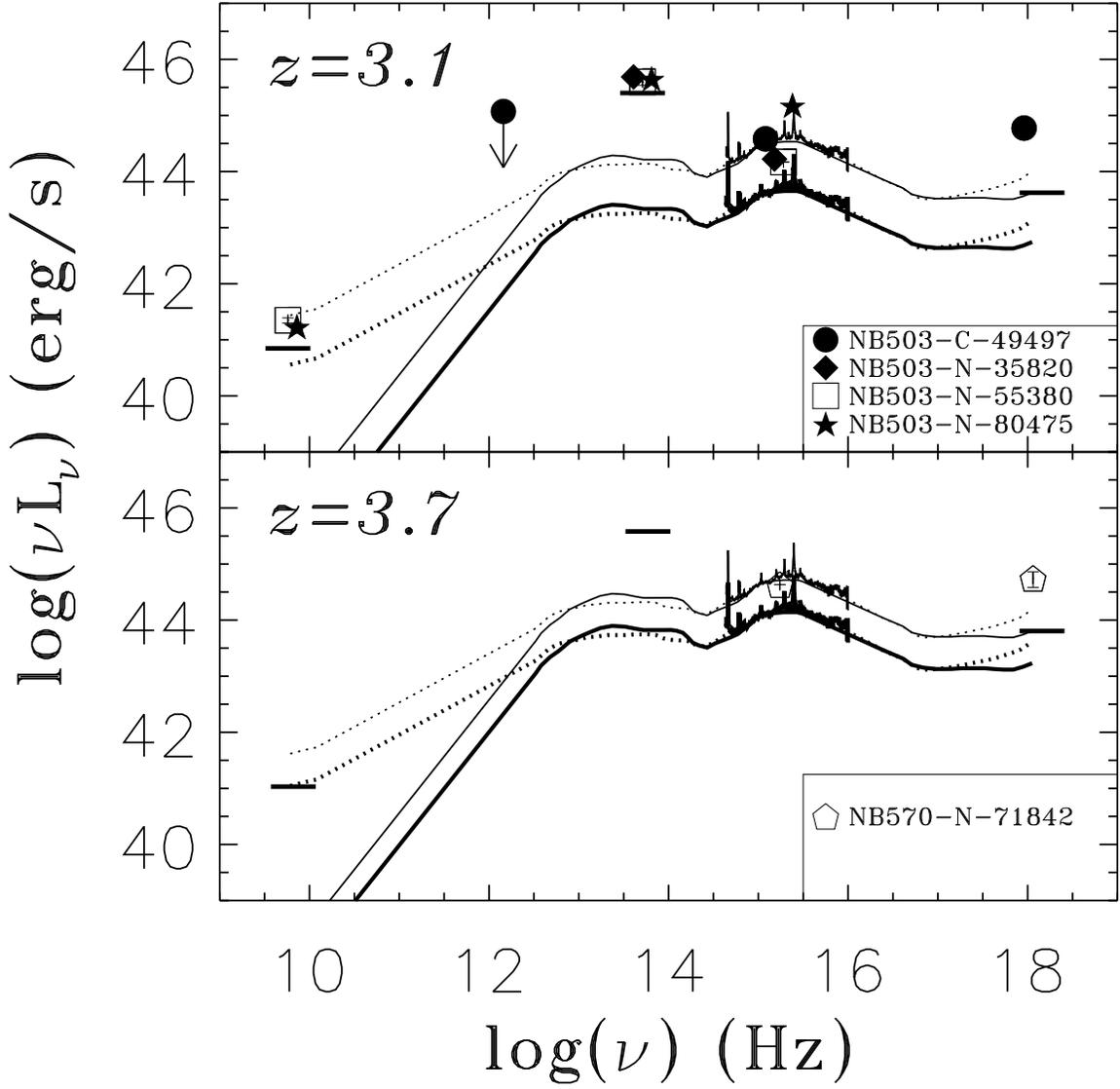}
\caption{
The spectral energy distribution (SED) of LAEs with
(an) X-ray, infrared, and/or radio counterpart(s).
The legend shows object names and the corresponding
symbols. Filled and open symbols represent an object with
and without a spectroscopic identification, respectively.
Solid and dotted curves are the templates of radio-quiet
and radio-loud AGNs taken from 
\citet{elvis1994,telfer2002,richards2003} (see text).
The thick and thin curves indicate the template SEDs
normalized with the detection limits of Ly$\alpha$ and
X-ray, respectively.
The horizontal solid bars present our detection limits
in X-ray, infrared, and radio bands.
\label{fig:multiband_sed}}
\end{figure}

\begin{figure}
\epsscale{1.0}
\plotone{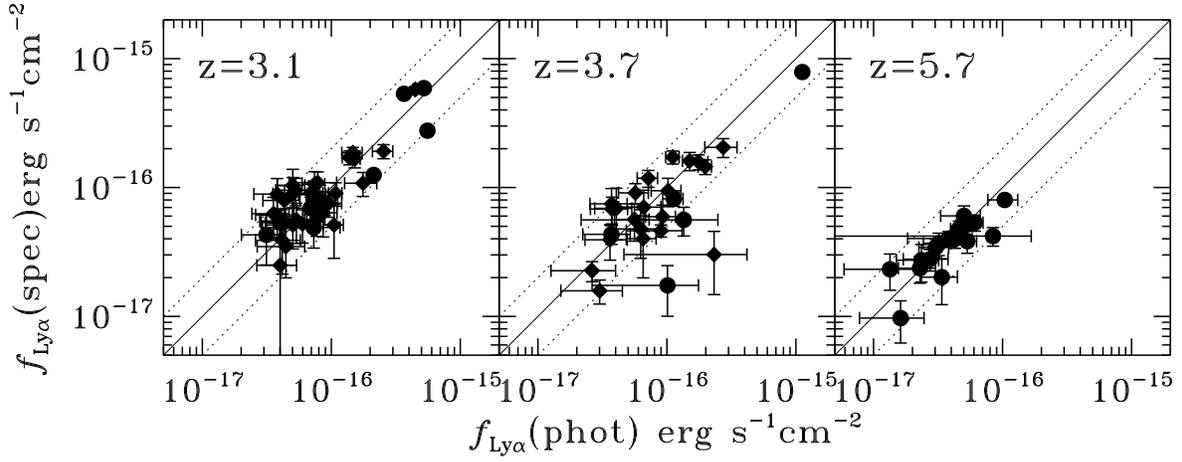}
\caption{
Comparison of Ly$\alpha$ fluxes measured with
our images, $f_{\rm Ly\alpha} {\rm (phot)}$, 
and spectra, $f_{\rm Ly\alpha} {\rm (spec)}$,
for our LAEs
at $z=3.1$ (left), $3.7$ (middle), and $5.7$ (right).
The circles and diamonds indicate LAEs with 
FOCAS and VIMOS spectra, respectively.
Since our FOCAS observations used a relatively 
narrow slit, we apply slit-loss corrections
for the Ly$\alpha$ fluxes of FOCAS. The error
bars of circles and diamonds show the ranges
of a 95\% confidence level.
The solid line at each panel
presents the equality of $f_{\rm Ly\alpha} {\rm (phot)}$
and $f_{\rm Ly\alpha} {\rm (spec)}$.
We plot dotted lines for the differences 
from the equality by a factor of 2.
\label{fig:comp_flya_all}}
\end{figure}

\begin{figure}
\epsscale{0.95}
\plotone{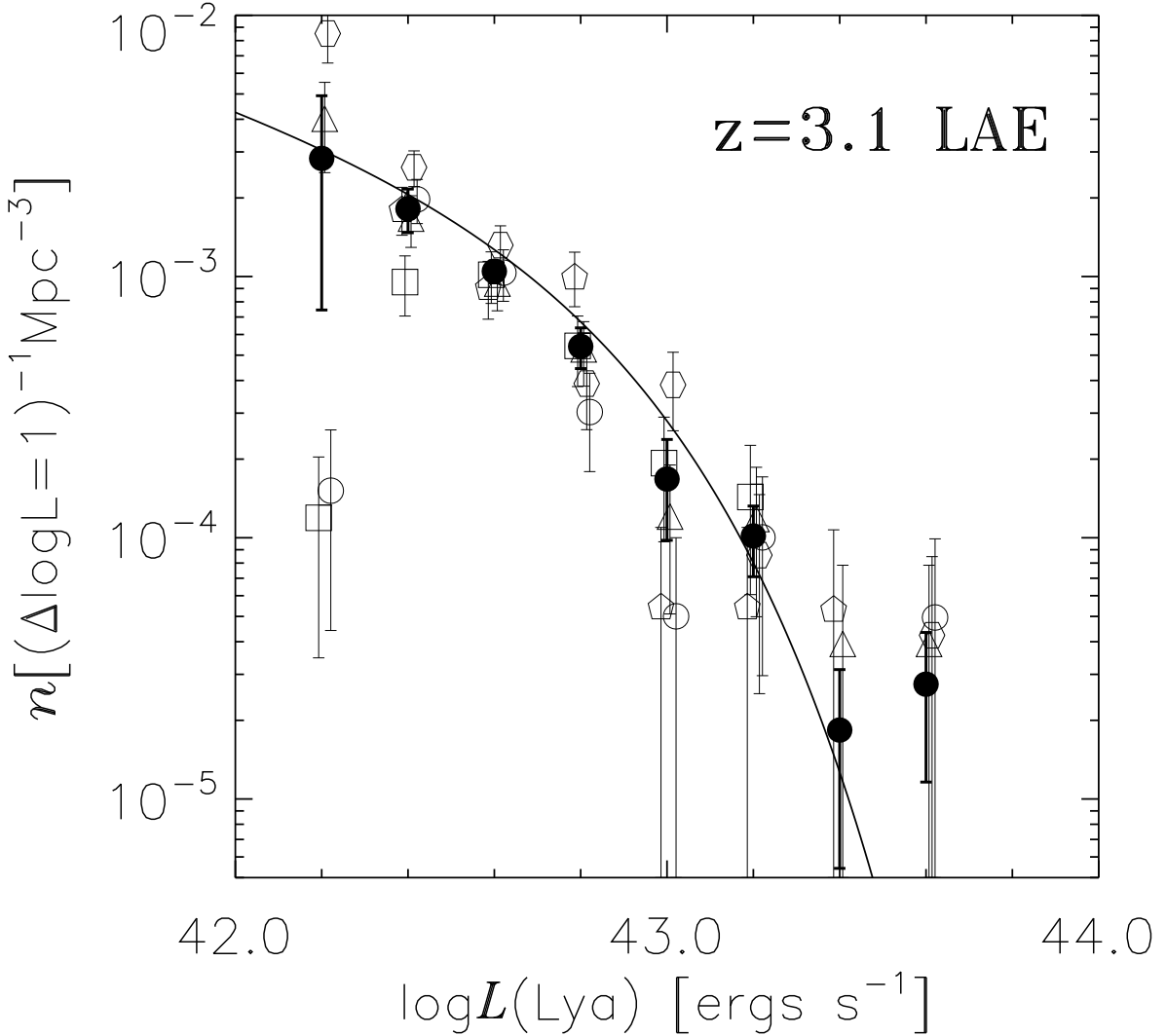}
\caption{
Luminosity functions (LFs) of LAEs at $z=3.1$.
The filled circles are LFs obtained by the classical method.
The error bars include the field-to-field variation
(see eq. \ref{eq:field_variation}).
The open circles, hexagons, triangles,
squares, and pentagons represent
the LFs of five $\sim 0.2$ deg fields, SXDS-C, SXDS-N,
SXDS-S, SXDS-E, and SXDS-W, respectively,
which are also derived by the classical method.
In order to avoid the overlaps of points, we
slightly shift all the open symbols along the abscissa.
The exact luminosity
of an open symbol is the same as a magnitude of a neighboring 
filled circle.
The solid line is the best-fit Schechter function with $\alpha=-1.5$
estimated by our simulations, but not the Schechter function fitted
to the LFs of the classical method.
\label{fig:lumifun_full_diff_nb503}}
\end{figure}

\begin{figure}
\epsscale{1.0}
\plotone{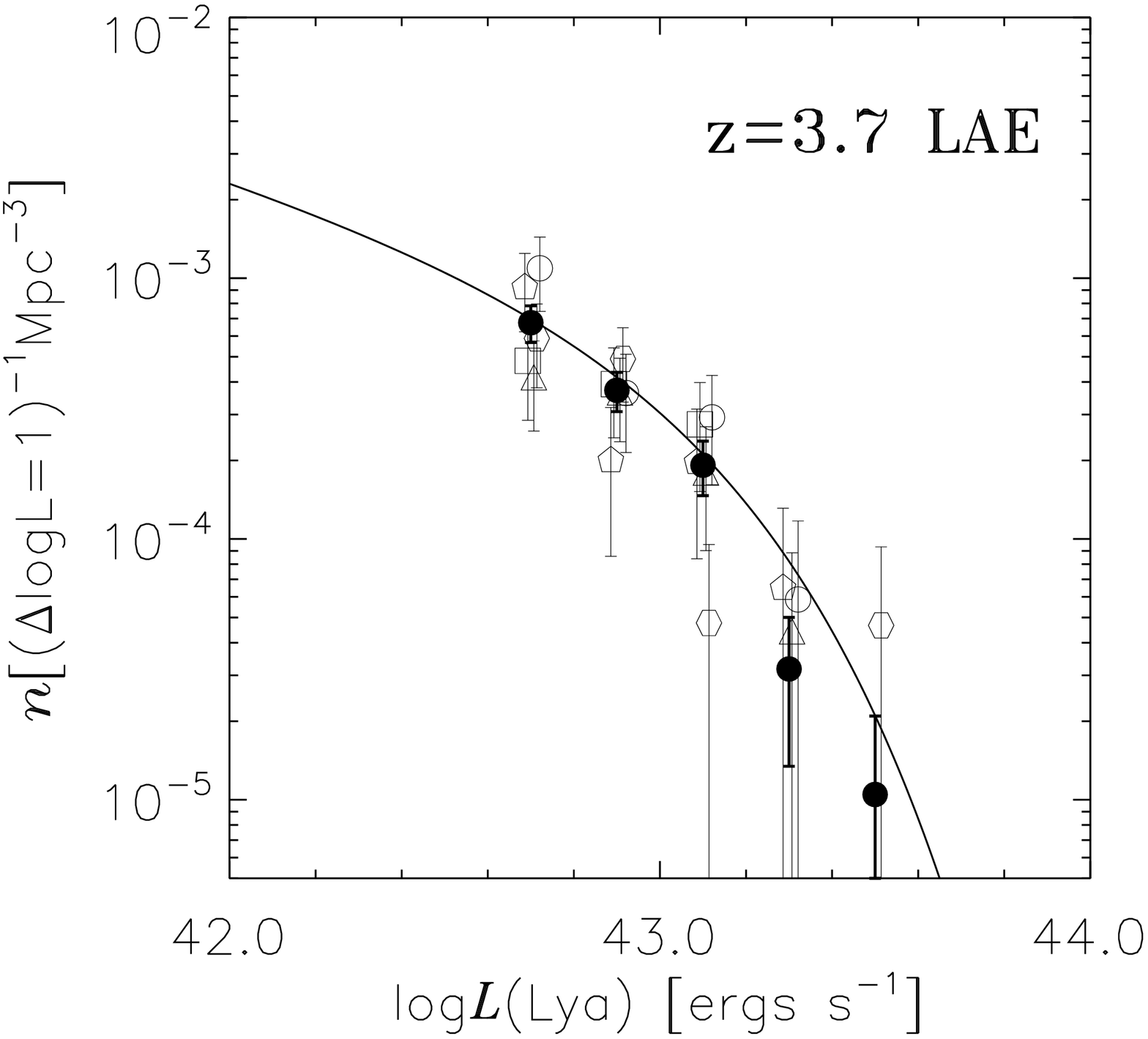}
\caption{
Same as Figure \ref{fig:lumifun_full_diff_nb503},
but for $z=3.7$.
\label{fig:lumifun_full_diff_nb570}}
\end{figure}

\begin{figure}
\epsscale{1.0}
\plotone{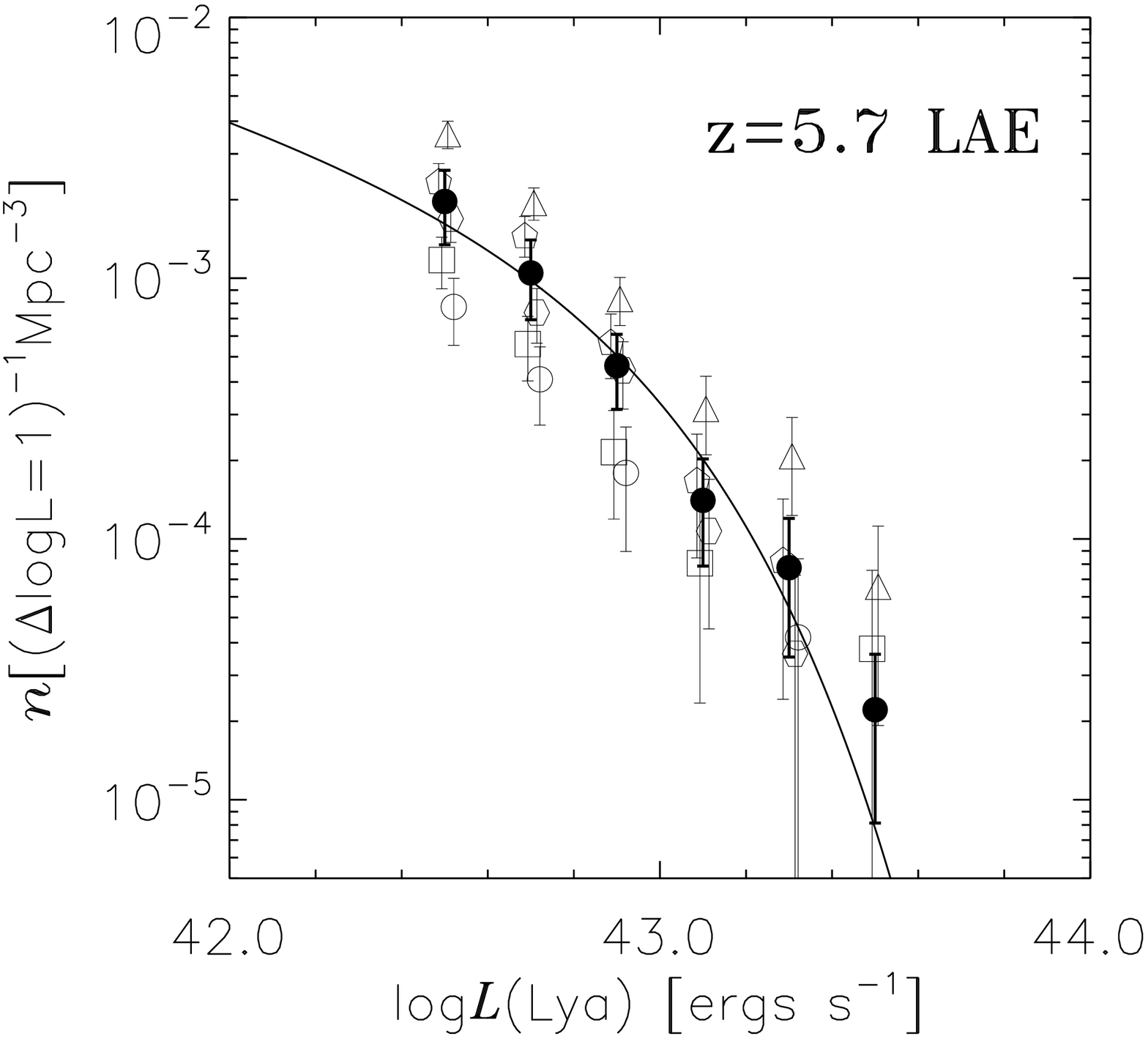}
\caption{
Same as Figure \ref{fig:lumifun_full_diff_nb503},
but for $z=5.7$.
\label{fig:lumifun_full_diff_nb816}}
\end{figure}

In order to avoid such uncertainties, we perform Monte Carlo 
simulations to find the best-fit Schechter parameters 
for the Ly$\alpha$ LF. These Monte Carlo simulations
are developed by \citet{shimasaku2006}, and the details
of simulations are described in their paper. We briefly
summarize the procedure of these simulations.
We generate a mock catalog of LAEs 
with a set of the Schechter 
parameters ($\alpha, \phi^\star, L^\star$) 
and a Gaussian sigma of probably distribution of EW, $\sigma_{\rm EW}$.
We uniformly distribute them in comoving space
over the redshift range of all LAEs that can be selected with our narrow-band filter.
We {\lq}observe{\rq} these LAEs with the narrow-
and broad-bands same to the real band responses 
(Figure \ref{fig:plot_sed_SXDSfilter_paper}), and
add to their flux densities photon noise corresponding 
to the actual observation. We select LAEs 
by the same criteria as for selecting the actual LAEs,
and derive the number densities and color distributions 
from the mock catalog. 
We compare these results
with the observational measurements of number densities and color distributions.
By performing this set of simulations over a wide range 
of the parameters, we find the best-fit Schechter parameters
by $\chi^2$ fitting.
We compare the results of these simulations with
the observational measurements corrected for
the detection completeness, since the simulations
are not affected by the detection completeness, $f_{\rm det}$.
In Figures \ref{fig:number_density_nb503}-\ref{fig:color_hist_all},
we plot the predicted 
surface densities and color distributions of LAEs
for some parameter sets.

The difference in $\chi^2$ for $\alpha$ values 
is found to be insignificant.
We fix the $\alpha$ value to
-1.0, -1.5, and -2.0, and carry out $\chi^2$ fitting, varying
$\phi^\star$, $L^\star$, and $\sigma_{\rm EW}$.
Figures \ref{fig:number_density_nb503}-\ref{fig:number_density_nb816}
present that all three models of $\alpha=-1.0$, $-1.5$, and $-2.0$ 
reproduce the observed counts well, although 
shallower $\alpha$ gives a slightly better fit. 
This suggests that the Ly$\alpha$ LF 
of LAEs is approximated well by the Schechter function.
The best-fit parameters, thus obtained, are summarized in
Table \ref{tab:lya_lumifun_schechter}.
We adopt the $\alpha=-1.5$ results as the fiducial set of 
the best-fit Schechter parameters (see \citealt{gronwall2007}).
Our best-fit Schechter parameters with $\alpha=-1.5$
are 
$\phi^*=(9.2_{-2.1}^{+2.5}, 3.4_{-0.9}^{+1.0}, 7.7_{-3.9}^{+7.4}) 
\times 10^{-4}$ Mpc$^{-3}$
and
$L^*_{\rm Ly\alpha}=(5.8_{-0.7}^{+0.9}, 10.2_{-1.5}^{+1.8}, 6.8_{-2.1}^{+3.0}) 
\times 10^{42}$ erg s$^{-1}$
at 
$z=(3.1, 3.7, 5.7)$.

The best-fit Schechter functions for $\alpha=-1.5$ obtained above 
are shown in 
Figures \ref{fig:lumifun_full_diff_nb503}-\ref{fig:lumifun_full_diff_nb816}
with the solid lines.
Note that the LFs from the classical method (filled circles) 
are consistent with the best-fit Schechter functions from
our simulations.
This means that the simple method gives a good approximation. 
This is probably because the uncertainties (a) and (b) in the classical
method are negligible and/or cancel with each other.
The number densities of the best-fit Schechter functions are 
slightly larger than those of the classical method.
This slight difference can be explained by
the underestimates of 
the number density and Ly$\alpha$ luminosity
with the classical method 
due to the large survey volume of the top-hat redshift distributions
and the small Ly$\alpha$ flux measured with 
a bandpass of the triangle-shaped narrow bands.

Note that these best-fit LFs are based on our LAE samples
selected with a limit of EW (Table \ref{tab:sample}).
If we extrapolate the EW distribution down to $EW=0$\AA\ with our simulations,
we can obtain estimates of the LFs for all LAEs with $EW>0$\AA.
We refer a $\phi^\star$ value of LFs for all ($EW>0$\AA) LAEs 
as for $\phi^{\star}_0$, and present the best-fit $\phi^{\star}_0$
in Table \ref{tab:lya_lumifun_schechter}.
These $\phi^{\star}_0$ values only differ from $\phi^\star$ by $\sim 10$\%.
There are two reasons for this small difference: (i) The typical
distribution of EW, $\sigma_{\rm EW}$, is much wider than that
of EW limit of our selection, $EW_{\rm lim}$ 
(see Tables \ref{tab:sample} and \ref{tab:lya_lumifun_schechter}).
(ii) the EW range missed by our selection ($0<EW<EW_{\rm lim}$) 
is significantly shorter than that of our selection ($EW>EW_{\rm lim}$).
The combination of these facts provide this small ($\sim 10$\%-level)
difference between $\phi^\star$ and $\phi^{\star}_0$.

\subsubsection{Comparison with Previous Measurements}
\label{sec:comparison_with_previous_measurements}

We overplot previous measurements of Ly$\alpha$ LFs
for $z\simeq 3$, $4$ and $6$ in Figure \ref{fig:lumifun_full_diff_evol}.
Our LFs of $z\sim 3$ and $4$ agree very well with
all of the previous measurements typically within the error bars.
It is noticeable that the best-fit Schechter function of 
\citet{gronwall2007} (from the 0.3 deg$^2$ survey) 
shows an extremely good agreement
with ours within a factor of $\sim 0.1$,
although the luminosity range of their measurements is
limited to the relatively faint luminosity
(i.e. $\log L\simeq 42-43$ erg s$^{-1}$).

In the bottom panel of Figure \ref{fig:lumifun_full_diff_evol},
the previous measurements of $z\sim 6$ LAEs show large scatters 
over Poisson errors. However, our measurements are consistent
with those of \citet{shimasaku2006} (red open circles) 
as well as the recalculated LF of \citet{hu2004} (squares), 
both of which are based on deep and moderately-wide 
($\simeq 0.2$ deg$^2$) field imaging and spectroscopy.
Moreover, $z=5.7$ LF of the recent 2-deg$^2$ imaging survey of 
\citet{murayama2007} (red open diamonds) agrees with ours
within $1 \sigma$ error bars at $\log L> 43.0$ erg s$^{-1}$ 
where the measurements of \citet{murayama2007} are complete.
(Note that the imaging data of \citet{murayama2007} are shallower
than ours by $\sim 1$ magnitude, and that the 
data points of \citet{murayama2007} are not corrected 
for incompleteness.)
The best-fit Schechter function of \citet{malhotra2004} (dotted line)
is significantly lower than our LFs. \citet{shimasaku2006}
and \citet{tapken2006}
have also pointed out that
the measurements of \citet{malhotra2004}
are lower than their LFs. 
Our $z=5.7$ LF is consistent with the measurements of
\citet{ajiki2006} (stars) and \citet{tapken2006} (hexagon) 
within the error bars, 
but our measurements would be systematically higher than theirs. 
This difference is probably originated from
i) the method of Ly$\alpha$ luminosity estimates from photometric data, 
ii) the detection-completeness correction
and
iiii) field-to-field variation raised by
smaller field surveys of \citet{ajiki2006} and \citet{tapken2006}.
The differences in selection criteria for LAEs between 
our sample and the others are probably not a main reason for 
this discrepancy, 
since, for our data, changing the $i'- NB816$ 
threshold over $1.2$ -- $1.5$ only makes a difference in
number density by 11\%.

\subsubsection{Evolution of Ly$\alpha$ Luminosity Function}
\label{sec:evolution_of_lya_LF}

\begin{figure}
\epsscale{0.5}
\plotone{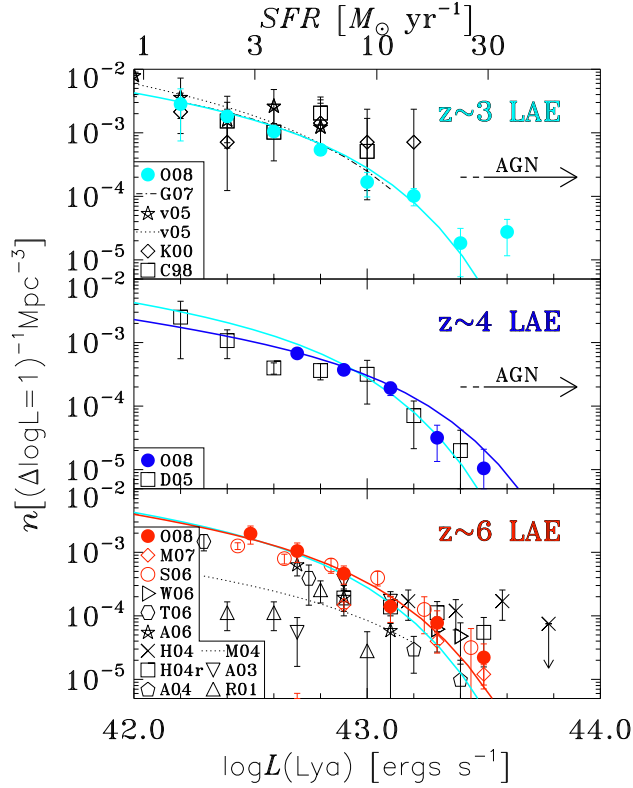}
\caption{
The best measurements of Ly$\alpha$ luminosity functions (LFs)
at $z\sim 3$ (top), $4$ (middle), and $6$ (bottom),
together with the previous measurements.
The cyan, blue, and red filled circles are the 
LFs calculated with the classical method
at $z=3.1$, $3.7$, and $5.7$, respectively,
that are labeled as O08 in the legend.
The corresponding best-fit LFs from the simulations
are plotted with colored solid lines.
These filled circles and solid lines are 
the same as those presented 
in Figures 
\ref{fig:lumifun_full_diff_nb503}-\ref{fig:lumifun_full_diff_nb816}.
Note that the $z=3.1$ LFs (cyan lines) are overplotted
on the middle and bottom panels for references.
The other symbols and dotted-lines are the previous
measurements and best-fit Schechter functions
whose references are indicated in the legend on each panel: 
$G07$---\citet{gronwall2007},
$v05$---\citet{vanbreukelen2005},
$K00$---\citet{kudritzki2000}, 
$C98$---\citet{cowie1998},
$D05$---\citet{dawson2005},
$M07$---\citet{murayama2007},
$S06$---\citet{shimasaku2006},
$W06$---\citet{westra2006},
$T06$---\citet{tapken2006},
$A06$---\citet{ajiki2006},
$H04$---\citet{hu2004},
$H04r$---recalculated LFs of \citet{hu2004},
$A04$---\citet{ajiki2004},
$M04$---\citet{malhotra2004},
$A03$---\citet{ajiki2003}, and
$R01$---\citet{rhoads2001}.
We recalculate the LFs of \citet{hu2004} data,
and obtain the points of $H04r$ whose values
are close to those of the similar re-estimation by
\citet{tapken2006}. 
The arrows in the top and middle panes
represent the luminosity range where LFs
are dominated by LAEs with AGN activities
(see Section \ref{sec:lae_hosting_agn}).
For the reader's eye guide,
we plot ticks of SFR obtained from eq. (\ref{eq:lya_sfr})
on the upper abscissa axis.
\label{fig:lumifun_full_diff_evol}}
\end{figure}

\begin{figure}
\epsscale{1.0}
\plotone{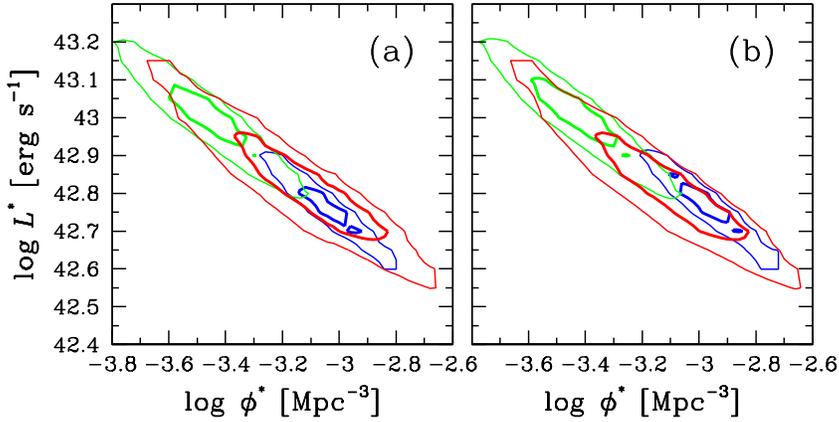}
\caption{
Error ellipses of our Schechter parameters,
$L^*_{\rm Ly\alpha}$ and $\phi^*$.
The left panel shows the ellipses for our
observational data, while the right panel
presents those for all LAEs
with a positive emission ($EW>0$) 
estimated by our simulations.
Blue, green, and red contours represent $z=3.1$,
$3.7$, and $5.7$ LAE-LFs with the fixed slope of 
$\alpha=-1.5$. Thick and thin lines indicate 
$1$ and $2\sigma$ confidence levels, respectively.
\label{fig:con_Lstar_phi_1.5_3NB_NB816EW500}}
\end{figure}

Our LFs are appropriate to identify an evolutionary trend of LF,
since our LFs from $z=3.1$ to $5.7$ are derived 
by the same procedure with similar data sets from the same instrument.
In Figure \ref{fig:lumifun_full_diff_evol}, we compare 
our Ly$\alpha$ LFs of LAEs at $z\sim 3$, $4$, and $6$.
The LFs do not change within error bars at these redshifts. 
The left panel of Figure \ref{fig:con_Lstar_phi_1.5_3NB_NB816EW500}
shows the error ellipses of Schechter parameters of our LFs at $z=3.1-5.7$.
All the $2\sigma$-error contours well overlap each other.
At the fixed slope of $\alpha=-1.5$, the best-fit parameters
of LFs change from $z=3.1-5.7$ only by factors of 1.8 and 2.7
in $L^*$ and $\phi^*$, respectively.
Since these error ellipses are obtained by our observational
data whose limiting Ly$\alpha$ EW is different at each redshift, 
we also plot in the right panel of 
Figure \ref{fig:con_Lstar_phi_1.5_3NB_NB816EW500}
the ellipses for all LAEs with a positive emission ($EW>0$) 
which are obtained from the extrapolation with our simulations.
In this case, the overlaps of $z=3.1$ and $z=3.7$
error ellipses become smaller. However, the error ellipses of $z=5.7$ LAEs
are much larger than those of $z=3.1$ and $3.7$. These large
$z=5.7$ error ellipses still appear to be consistent
with $z=3.1$ and $3.7$ error ellipses.
The results for LAEs with $EW>0$ are
similar to those for the observed LAEs.
This is because the number density of observed LAEs
($\phi^*$) differs from that of all LAEs ($\phi_0^*$)
only by $\sim 10$\%, as discussed in 
Section \ref{sec:ly_alpha_luminosity_functions}.
Thus, we conclude that Ly$\alpha$ LFs of LAEs do not evolve by more than
a factor of 2-3 either in luminosity or number density,
and that drastic evolution 
in the Ly$\alpha$ LF of LAEs from $z \sim 3.1$ to $z=5.7$ 
is ruled out.
It should be noted that the central positions
of $z=3.7$ contours appear to be shifted 
to brighter $L^*$ and lower $\phi^*$ than those of $z=3.1$ and $5.7$,
although this shift is not significant within $1-2\sigma$ levels
This small shift may indicate
a moderate evolution of LFs between these redshifts.
However, it is possible that AGNs at the bright-end LF
more strongly affect the Schechter fit of $z=3.7$ LF
than that of $z=3.1$ and $5.7$ LFs, due to the
shallow detection limit of our $z=3.7$ LAEs (see
Section \ref{sec:lae_hosting_agn}).

\begin{figure}
\epsscale{0.8}
\plotone{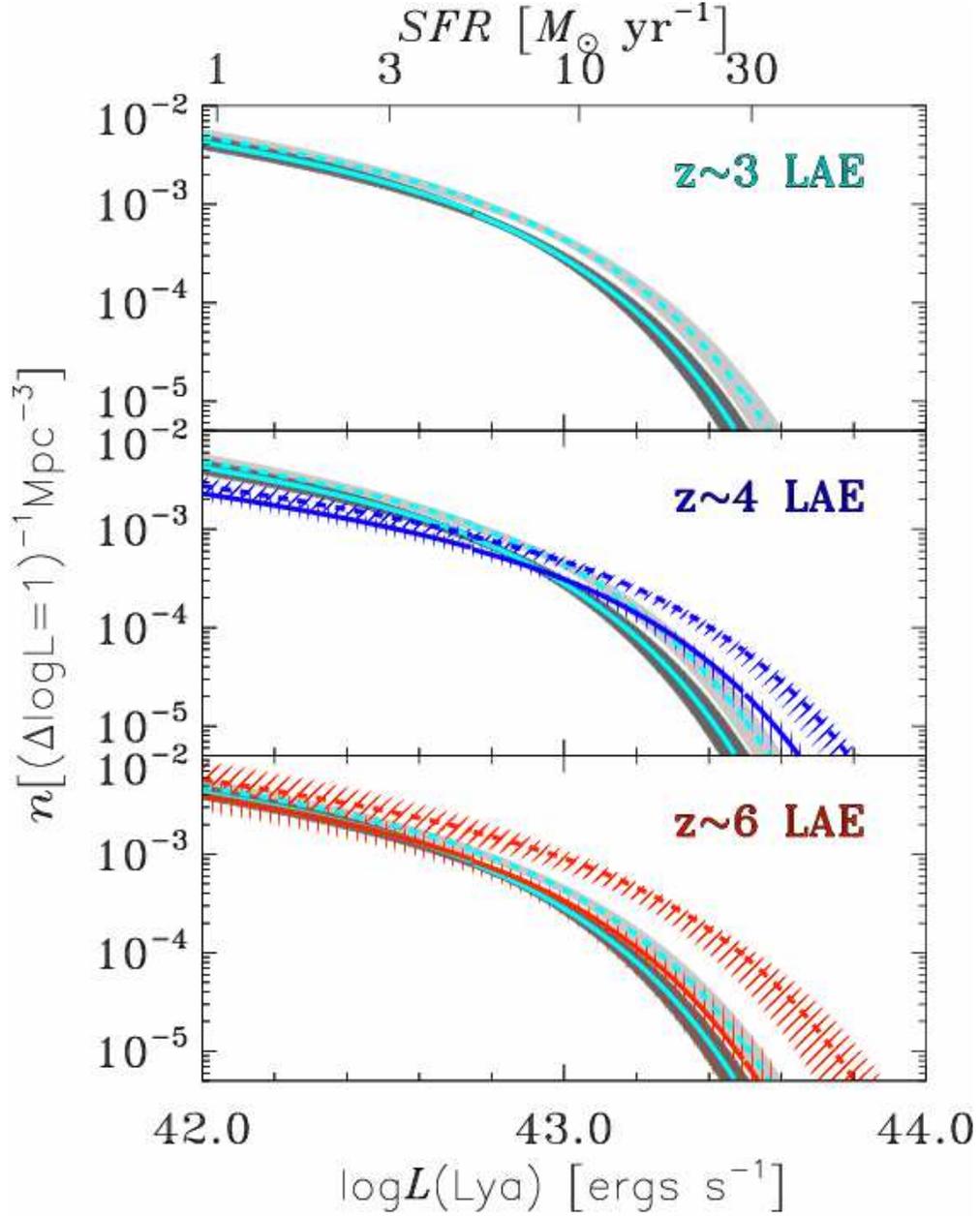}
\caption{
Inferred {\it intrinsic} Ly$\alpha$ luminosity functions (LFs),
together with the best-estimate {\it apparent} Ly$\alpha$ LFs.
Dashed- and solid-line curves are the {\it intrinsic} and {\it apparent} 
Ly$\alpha$ LFs, respectively.
Cyan, blue, and red curves denote for LFs at $z=3.1$ (top panel), 
$3.7$ (middle panel), and $5.7$ (bottom panel), respectively. 
Cyan lines are repeatedly plotted at each panel for comparison.
Light-gray and gray shades around cyan lines indicate
1 $\sigma$ errors of {\it intrinsic} and {\it apparent}
Ly$\alpha$ LFs at $z=3.1$. Areas of vertical and hatched lines
are the same, but for $z=3.7$ (blue) and $5.7$ (red) 
in the middle and bottom panels, respectively.
\label{fig:lumifun_full_diff_evol_implication}}
\end{figure}

Note that Ly$\alpha$ fluxes from high-$z$ objects are
generally attenuated by neutral hydrogen of 
intervening inter-galactic medium (IGM). It is
known that high-$z$ galaxies have an asymmetric
Ly$\alpha$ emission whose blue side of line is 
more strongly absorbed by neutral hydrogen of IGM
than low-$z$ galaxies.
(e.g. \citealt{hu2004,shimasaku2006}).
Thus, the Ly$\alpha$ LFs 
in Figure \ref{fig:lumifun_full_diff_evol} (solid lines) 
should be called {\it apparent}-Ly$\alpha$ LFs.
Due to the absorption of IGM, we cannot directly 
measure {\it intrinsic}-Ly$\alpha$ fluxes that are emitted from LAEs.
However, we can constrain the evolution of {\it intrinsic}-Ly$\alpha$ LFs
with {\it apparent}-Ly$\alpha$ LFs.
First, the lower limits of {\it intrinsic}-Ly$\alpha$ LFs
are obviously given by the {\it apparent}-Ly$\alpha$ LFs (solid lines
of Figure \ref{fig:lumifun_full_diff_evol}).
Second, we can estimate the {\it intrinsic}-Ly$\alpha$ luminosity
with the average Gunn-Peterson (GP) optical depths.
The GP optical depths are estimated in various studies.
Here, based on the results of \citet{fan2006} (Fan), 
\citet{madau1995} (Madau), and \citet{meiksin2006} (Meiksin),
we calculate the ratios of intrinsic Ly$\alpha$ flux, $f_{\rm Ly\alpha}^{int}$,
to apparent Ly$\alpha$ flux, $f_{\rm Ly\alpha}^{app}$.
Assuming that IGM absorbs a blue half of symmetric
Ly$\alpha$ emission line,
we estimate the ratios $f_{\rm Ly\alpha}^{app}/f_{\rm Ly\alpha}^{int}$
with the formulae of (Fan, Madau, Meiskin) to be
(0.85, 0.81, 0.84), (0.76, 0.73, 0.76), and
(0.53,0.54,0.52) at $z=3.1$, $3.7$, and $5.7$, respectively.
Since the $f_{\rm Ly\alpha}^{app}/f_{\rm Ly\alpha}^{int}$ ratios
change only by $3-5$\% between these three estimates,
we use the classic \citealt{madau1995}'s model.
The dashed lines in Figure \ref{fig:lumifun_full_diff_evol_implication}
correspond to the {\it intrinsic}-Ly$\alpha$ LFs that we estimate.
This figure indicates
that {\it intrinsic}-Ly$\alpha$ LFs may evolve significantly,
and that luminosities and/or number densities of LAEs
are intrinsically brighter/higher at $z=5.7$ than $z=3.1$.

Note that we estimate these {\it intrinsic}-Ly$\alpha$ LFs
with two assumptions. The first assumption is that 
the density of neutral hydrogen atoms in the intervening IGM 
is the cosmic average value 
at each epoch. However, hydrogen clouds around LAEs
are likely more ionized by the UV radiation from LAEs.
On the other hand, the chance of intervening galaxies 
with molecular hydrogen probably increases around LAEs, 
due to clustering of galaxies. 
The second assumption is that the {\it intrinsic}-Ly$\alpha$ emission of LAEs  
has a symmetric profile whose central wavelength is not
red- or blue-shifted. \citet{shapley2003} find
in a composite spectrum of $z=3$ LBGs that
Ly$\alpha$ emission is redshifted by 360 km s$^{-1}$ from 
the average of continuum absorption lines.
Because the typical Ly$\alpha$ FWHM of LAEs
is as small as $<300$ km s$^{-1}$ (e.g. \citealt{rhoads2003,hu2004,shimasaku2006}),
the IGM will absorb only a small fraction of Ly$\alpha$ photons
if they are redshifted by about 360 km s$^{-1}$.
However, this is true only if Ly$\alpha$ properties of LAEs 
are similar to those of $z=3$ LBGs. In fact, 
the typical Ly$\alpha$ FWHM of LAEs is smaller than
that of $z=3$ LBGs ($450\pm 150$ km s$^{-1}$ \citealt{shapley2003}). 
It is possible that the velocity offset of
Ly$\alpha$ emission may also be smaller in LAEs than in LBGs (360 km s$^{-1}$).
In this way, the inferred {\it intrinsic}-Ly$\alpha$ features
(such as those found in Figure \ref{fig:lumifun_full_diff_evol_implication})
are real only if these two assumptions are correct.
We need high-resolution spectroscopy for LAEs
and sophisticated numerical simulations to assess
the {\it intrinsic}-Ly$\alpha$ LF, and to derive a reliable 
conclusion on its evolution.

In summary, it is interesting that {\it apparent} LF of LAEs does 
not show a significant evolution over the long redshift
range of $z=3.1$ to $5.7$. We point out the possibility that
evolution of IGM and LAEs may be tuned to realize 
this no evolution at $z=3.1-5.7$, if the amount of external
absorption changes with redshift.

\subsection{UV Luminosity Functions}
\label{sec:uv_luminosity_functions}

We derive the UV LFs of our LAEs at $z=3.1$, $3.7$,
and $5.7$. We use our LAE catalogs, and obtain
UV-continuum luminosities from magnitudes of a broad band
whose bandpass does not include Ly$\alpha$ but only continuum emission.
We choose $R$ magnitudes for $z=3.1$ and $3.7$ LAEs, 
and $z'$ magnitudes for $z=5.7$ LAEs, where the corresponding
wavelengths are $\sim 1300-1600$\AA\ in rest frame.
We apply no k-corrections to the UV magnitudes,
since the difference of UV magnitudes in this narrow-wavelength
range is negligibly small for LAEs with a flat continuum (Section \ref{sec:uv_color}).
We calculate number densities from the survey volume
and the number counts as a function of UV magnitudes
by weighting with the detection completeness
(Section \ref{sec:photometric_lae_completeness}).
Note that we need a careful estimation of total magnitudes
of broad band. Since the aperture sizes and shapes
of MAG\_AUTO magnitudes are defined by the sources on the
narrow-band images, the MAG\_AUTO magnitudes of broad-band
images could be contaminated by the neighboring continuum sources.
Thus, we identify and deblend sources on our broad-band images with SExtractor,
and cross-match the broad-band sources with our LAEs.
If a LAE do not have a corresponding broad-band source detected 
over a $3\sigma$ level, we just use photometry of a $2''$-diameter aperture.
We refer to these magnitudes as a total magnitudes of broad band,
and derive UV LFs with these magnitudes. For comparison, we also calculate
the LFs from the simple $2''$-diameter aperture photometry with no
completeness correction. These simple estimates provide 
lower limits of UV LFs.

\begin{figure}
\epsscale{0.40}
\plotone{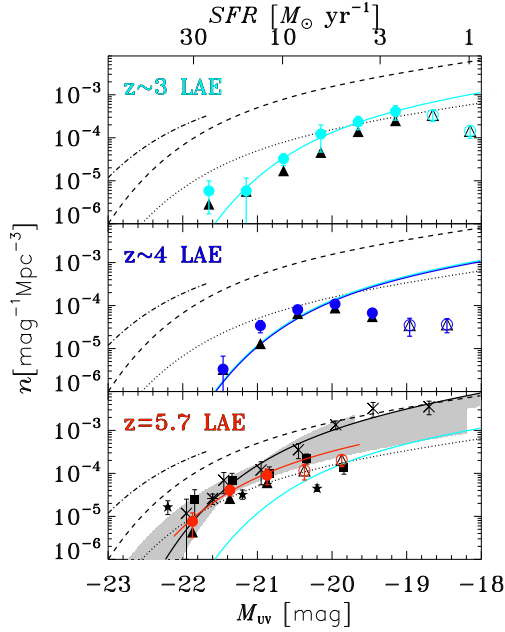}
\caption{
The UV luminosity functions (LFs) of LAEs at $z=3.1-5.7$
The cyan, blue, and red circles indicate
the best-estimates of LFs for our $z=3.1$ (top panel),
$3.7$ (middle panel), and $5.7$ (bottom panel) LAEs.
The solid lines represent the best-fit Schechter functions
for the best estimates. The best-fit Schechter function
of $z=3.1$ LAEs is presented on each panel for reference.
The triangles are the lower limits
of these LFs at each redshift 
that are derived with the aperture photometry of broad-band images 
(see text). 
For circles and triangles, we show the reliable and less-reliable measurements 
with filled and open symbols, which are obtained with UV magnitudes
at $>5\sigma$ and $2-5\sigma$ levels, respectively.
We only use these $>5\sigma$ data (filled circles) 
for our Schechter-function fit (see text).
The filled squares and stars on the bottom panel
are the UV LFs of LAEs at $z=5.7$ obtained by \citet{shimasaku2006}
and \citet{hu2006}.
On each panel, we also plot the UV LFs of dropout galaxies for comparison.
The dashed and dot-dashed lines are the $z=3$ LFs of \citet{steidel1999}
and \citep{paltani2006}, respectively. The dotted line indicates the
$z=3$ LF of \citet{steidel1999}, but $\phi^*$ of the LF is multiplied by 
$1/10$. Note that the LF of $z=4$ dropout galaxies are almost same as
that of $z=3$ \citep{steidel1999,ouchi2004a,beckwith2006,yoshida2006}.
On the bottom panel, we show the UV LFs of dropout galaxies at $z\sim 6$
with crosses \citep{bouwens2006} and asterisk \citep{shimasaku2006}.
The gray region on the bottom panel indicates the range of
the best-fit Schechter functions for $i$-dropouts obtained
by various studies \citep{bunker2004,yan2004,malhotra2005,bouwens2006},
which show the uncertainties of measurements for $z=6$ dropout LF.
For the reader's eye guide,
we plot ticks of SFR obtained from eq. (\ref{eq:uv_sfr})
on the upper abscissa axis.
\label{fig:lumifun_UV_diff_evol}}
\end{figure}

Figure \ref{fig:lumifun_UV_diff_evol} shows the UV-LFs of our LAEs,
together with the previous measurements of LAEs and
dropout galaxies (or LBGs) at each redshift. Our UV-LFs are
consistent with those of \citet{hu2004} and \citet{shimasaku2006} at $z=5.7$,
although their measurements show slightly higher number densities than ours.
To check the consistency, we calculate $z=5.7$ LF in the same manner as
\citet{shimasaku2006} who simply used $2''$-diameter aperture 
magnitudes with aperture corrections estimated with their $NB816$ image.
Our LF by the method of \citet{shimasaku2006} presents a very good
agreement with that of \citet{shimasaku2006}. However, we find that
the aperture corrections defined with $NB816$ band overestimate
the broad-band magnitudes. This is because the Ly$\alpha$
morphology in the narrow band is generally more extended 
than UV morphology on the broad band. Thus, we
take our best estimates based on the MAG\_AUTO magnitudes
with the source deblending on the broad-band images.

In Figure \ref{fig:lumifun_UV_diff_evol},
we find that UV-LFs of LAEs do not change at $z\simeq 3-4$, but
evolve from $z\sim 4$ to $5.7$. There exist an emergence
of UV-bright LAEs at $z=5.7$. We compare these UV LFs with those of dropout
galaxies. 
At $z=3-4$, number densities of our LAEs are as much as 10\% 
of dropout galaxies down to $M_{\rm UV}\simeq -20$,
which corresponds to $\simeq M_{\rm UV}^*$ or $\simeq M_{\rm UV}^*+1$
\citep{steidel1999,ouchi2004a,sawicki2006,beckwith2006,yoshida2006,vanzella2006b}.
Note that the EW limit of our $z=3-4$ LAE samples is $\simeq 40-60$\AA.
This ratio of Ly$\alpha$ emitting galaxies to dropout galaxies
is consistent with that reported by \citet{shapley2003} who find that
25\% and $\sim 2$\% of their $z=3$ dropout galaxies have Ly$\alpha$ emission
line with $EW_0\ge 20$\AA\ and $EW_0\ge 100$\AA, respectively.
On the other hand, at $z=5.7$, the number densities of our LAEs seems 
comparable to, or at least more than 50\% of, dropout galaxies at $z\sim 6$,
where the EW limit of our $z=5.7$ LAE is about $\simeq 30$\AA.
Spectroscopic studies of \citet{dow-hygelund2006} report that 6 Ly$\alpha$ emitting
galaxies with $EW_0\ge 20$\AA\ are found in their $i'$-dropout sample of CL1252
and HUDF parallel fields, and conclude that about $\sim 30$\% of $i'$-dropout galaxies
have a Ly$\alpha$ emission line. This result is consistent with
the measurements of previous studies \citep{stanway2004a,stanway2004b,vanzella2006a}.
More recently, \citet{stanway2007} show that 16 out of 24 $i'$-dropout galaxies
do not have a Ly$\alpha$ emission line with $EW_0\ge 25$\AA\ in the GLARE/HUDF field.
Most of the previous spectroscopic studies of $i'$-dropout galaxies
suggest that $\sim 30$\% of $i'$-dropout galaxies down to $\sim L*$
have a Ly$\alpha$ emission line with $EW_0\ge 20$\AA.
It is not clear why our ratio of dropout galaxies and LAEs is
higher than those of spectroscopic results. There are three possible explanations.
The first is the difference of the definition in the ratio.
We compare the ratio of LAEs to dropout galaxies at the same redshift, but
\citet{dow-hygelund2006} and \citet{stanway2007} obtain the ratio
of Ly$\alpha$ emitting dropouts to all the dropouts.
If all the LAEs are identified by dropout technique, these two ratios should
be comparable. However, LAEs with no significant continuum are likely missed 
by the dropout selection. 
Moreover, the $i$ dropout selection
systematically misses the strong Ly$\alpha$ emission lines 
at $z\lesssim 6$.
The selection differences may cause the discrepancy between
these two ratios estimated by the different methods.
In fact, the theoretical model of \citet{dijkstra2007} suggests 
that LAE selections identify more population III galaxies with a large EW
than dropout selections, and that the EW distribution
of LAEs should not be the same as that of dropouts.
The second explanation is that LFs of dropout galaxies have systematic
errors. Indeed, we compare the LFs of dropout galaxies of 
\citet{bouwens2006} to estimate the ratio. However, the measurements of dropouts' LFs
scatter by a factor of 2 or more in the different estimates by various authors.
We show the uncertainties of estimates with the gray shade 
in Figure \ref{fig:lumifun_UV_diff_evol}.
The third explanation is the evolution of UV LFs from $z=5.5$ to $6.5$
as claimed by \citet{dow-hygelund2006}.
The UV-LFs of dropout galaxies decrease in number density and/or luminosity 
from $z\sim 4$ to $6$ 
(e.g. \citealt{lehnert2003,ouchi2004a,shimasaku2005,bunker2004}),
and even more significantly evolve from $z\sim 6$ to higher redshifts
of $z\sim 7-8$ \citep{bouwens2006}. Since $i$-dropout galaxy samples
distribute over $z=5.5-6.5$, the LFs of $i$-dropout galaxies are
measured at higher redshifts in average than that of our LAEs (i.e. $z=5.7$). 
This would not allow us to obtain the accurate ratio of Ly$\alpha$ emitting
population to dropout galaxies.
For these reasons, it is likely that the simple comparison between 
LFs of dropout galaxies and LAEs
would not give a ratio between all $z\sim 6$ objects and Ly$\alpha$ emitting
population.

It is interesting to compare the shapes of UV LF of
LAEs and dropout galaxies.
At $z=3-4$, we find a significant deficit of 
UV-bright LAEs with respect to dropout galaxies.
This deficit of UV-bright LAEs are reported by \citet{ouchi2003}
at $z=4.9$ based on the comparison of dropout and LAE LFs. 
This tendency is also consistent with the results of
Section \ref{sec:lya_uv} and the claims of
\citet{ando2006,shimasaku2006,vanzella2006b} 
who find that UV-bright galaxies do not have
a large Ly$\alpha$ EW from spectroscopic samples at $z=5-6$.
On the other hand, this tendency cannot be clearly seen for $z\sim 6$
in Figure \ref{fig:lumifun_UV_diff_evol}, probably
due to the large uncertainties of LF of dropout galaxies.

We fit the Schechter function,
\begin{equation}
\psi(M)dM=C\phi^*\exp\left\{-C(\alpha+1)(M-M^*)-\exp[-C(M-M^*)]\right\} dM,
\label{eq:schechter_m}
\end{equation}
to the UV LFs, where $C\equiv 0.4 \ln(10)$:
$\alpha$ is the power-law slope, $\phi^*$ is the normalization factor,
and $M^*$ is the characteristic absolute magnitude. 
We fit the Schechter function with the
measurements over a $5\sigma$ level 
(filled circles in Figure \ref{fig:lumifun_UV_diff_evol}).
Since the faintest points of our $z=3.7$ LAEs appear to
be suffered from the incompleteness of source detection,
we do not use the point for the fitting. Since our LAE samples
are not enough to derive $\alpha$,
we fix the slope to $\alpha = -1.5$. For our $z=3.7$ sample,
we fix the $M^*$ to the best-fit $M^*$ value of our $z=3.1$ LAEs.
We summarize the results of the fitting in Table \ref{tab:muv_lumifun_schechter}.

We find no evolution at $z\sim 3-4$, even considering the error bars. 
The difference between $z=3.1$ and $3.7$ LFs
is within 50\% in number density or luminosity.
On the other hand, UV LFs of LAEs evolve significantly from $z\sim 3-4$ to $6$.
If it comes from the pure luminosity evolution, the UV LFs
brighten by a $\sim$ 1 magnitude from $z\sim3-4$ to $z=5.7$.
Note that the EW limits of our selections are different
between these redshifts. However, the difference of sample selection
only changes by $\sim 10$\% in number density, according the results of our simulations 
(Section \ref{sec:ly_alpha_luminosity_functions}). Thus, 
this evolutionary trend of UV LFs
is not affected by the selection effects.

\subsection{Ly$\alpha$- and UV-Luminosity Densities
Contributed from Ly$\alpha$ Emitters}
\label{sec:evolution}

We calculate Ly$\alpha$ and UV-luminosity densities of our LAEs
from the luminosity functions (LFs) 
derived in sections \ref{sec:ly_alpha_luminosity_functions} 
and \ref{sec:uv_luminosity_functions}.
We show Ly$\alpha$ and UV-luminosity densities
in Tables \ref{tab:lya_lumifun_schechter} and \ref{tab:muv_lumifun_schechter},
respectively. These tables present 
Ly$\alpha$ (UV) luminosity densities 
down to our observational detection limits, 
$\rho^{\rm obs}_{\rm Ly\alpha}$ ($\rho^{\rm obs}_{\rm UV}$),
and total luminosity densities down
to zero luminosity by extrapolation,
$\rho^{\rm tot}_{\rm Ly\alpha}$ ($\rho^{\rm tot}_{\rm UV}$). 
These observational and total luminosity densities
provide approximated lower and upper limits.
For a reference, 
we estimate total Ly$\alpha$ luminosity densities of LAEs with $EW>0$,
from our simulations, which are shown under 
$\rho^{\rm tot}_{\rm 0Ly\alpha}$ in Table \ref{tab:lya_lumifun_schechter}.

In Table \ref{tab:lya_lumifun_schechter},
the total Ly$\alpha$ luminosity density is 
$\sim 10^{40}$ erg$^{-1}$s$^{-1}$Mpc$^{-3}$
at $z=3.1-5.7$, and does not change within a $1\sigma$ level
from $z=3.1$ to $5.7$. Table \ref{tab:muv_lumifun_schechter}
shows that the total UV luminosity density 
would increase from 
$\sim 4\times10^{25}$ erg$^{-1}$s$^{-1}$Hz$^{-1}$Mpc$^{-3}$
to $\sim 8\times10^{25}$ erg$^{-1}$s$^{-1}$Hz$^{-1}$Mpc$^{-3}$ 
at $z=3-4$ to $5.7$, 
although the difference is only a 1 $\sigma$ level because of large errors.
This possible increase of UV luminosity density
could be due to the fact that UV bright LAEs increase at $z=5.7$
as shown in Figure \ref{fig:lumifun_UV_diff_evol}.

Note that the entire Ly$\alpha$ (and UV) emission is not only
originated from star-formation activities, but also from cooling radiation, 
shock winds, and AGN, and that Ly$\alpha$ (and UV) emission is attenuated 
by dust and IGM. However, we roughly estimate the contribution of star-formation
from LAEs with those luminosity densities to check the consistency
with the previous studies. We use the simple prescription 
of Ly$\alpha$ luminosity, $L_{\rm Ly\alpha}$, and 
star-formation rate, $SFR,$
\begin{equation}
SFR({\rm M_\odot yr^{-1}}) =
L_{\rm Ly\alpha}({\rm erg\ s^{-1}})/(1.1\times10^{42}),
\label{eq:lya_sfr}
\end{equation}
combining the relation of H$\alpha$ luminosity and star-formation rate
\citep{kennicutt1998} and the case B approximation \citep{brocklehurst1971}.
With the assumptions of Salpeter IMF and solar metallicity \citep{madau1998},
the relation between UV luminosity and star-formation rate,
is:
\begin{equation}
{\rm SFR(M_\odot yr^{-1})} = 
L_{UV}{\rm (erg\ s^{-1} Hz^{-1})}/(8\times 10^{27}),
\label{eq:uv_sfr}
\end{equation}
where $L_{\rm UV}$ is the UV luminosity
measured in 1500\AA.
In Figures \ref{fig:lumifun_full_diff_evol} and \ref{fig:lumifun_UV_diff_evol},
we plot ticks of star-formation rates corresponding to
Ly$\alpha$ and UV luminosities based on these formulae of 
eqs (\ref{eq:lya_sfr}) and (\ref{eq:uv_sfr}). Assuming these equations,
SFRs of a typical ($L^*$) LAE are
$SFR ({\rm M_\odot yr^{-1}}) = 
(5.3^{+0.8}_{-0.6},9.3^{+1.6}_{-1.4},6.2^{+2.7}_{-1.9})$
and
$(4.5 \pm 2.0, \sim 4.5 , 9.4 \pm 7.0)$
at $z=(3.1,3.7,5.7)$, which are estimated from Ly$\alpha$
and UV-continuum $L^*$, respectively. Thus, a typical LAE
has a SFR of $\simeq 5-10$ $M_\odot$ yr$^{-1}$
at $z=3.1-5.7$, although these SFRs are correct under 
the assumptions given by eqs. (\ref{eq:lya_sfr}) and (\ref{eq:uv_sfr}).

We estimate the nominal lower limits of cosmic SFR density (SFRD)
from the luminosity densities down to our detection limits, 
$\rho^{\rm obs}_{\rm Ly\alpha}$.
From Table \ref{tab:lya_lumifun_schechter},
the observed Ly$\alpha$ luminosity density, 
$\rho^{\rm obs}_{\rm Ly\alpha}$ = 
($4.8_{-1.0}^{+1.2}$,$2.4_{-0.6}^{+0.7}$,
$3.6_{-1.7}^{+3.1}$)$\times 10^{39}$ erg$^{-1}$s$^{-1}$Mpc$^{-3}$
for redshifts of (3.1, 3.7, 5.7).
If we naively estimate SFR densities (SFRD)
from these Ly$\alpha$ luminosity densities
with eq. (\ref{eq:lya_sfr}), we obtain
$SFRD =$
$(4.3_{-0.9}^{+1.1}, 2.1_{-0.5}^{+0.6}, 3.2_{-1.6}^{+2.8} ) \times 10^{-3}$ 
 $M_\odot$yr$^{-1}$Mpc$^{-3}$
for $z=(3.1, 3.7, 5.7)$.
Similarly, we estimate SFRDs from the observed UV luminosity
densities, $\rho^{\rm obs}_{\rm UV}$, which are
($1.2_{-0.7}^{+1.4}$, $0.5_{-0.1}^{+0.1}$,
$0.9_{-0.7}^{+2.5}$)$\times 10^{25}$ erg$^{-1}$s$^{-1}$Hz$^{-1}$Mpc$^{-3}$
for redshifts of (3.1, 3.7, 5.7).
Based on eq. (\ref{eq:uv_sfr}), we obtain
$SFRD =$
$(1.5_{-0.8}^{1.8}, 0.7_{-0.1}^{+0.1}, 1.1_{-0.8}^{+3.1}) \times 10^{-3}$ 
 $M_\odot$yr$^{-1}$Mpc$^{-3}$
for $z=(3.1, 3.7, 5.7)$.
These Ly$\alpha$ and UV SFRDs are consistent with those obtained by 
the previous studies (see \citealt{taniguchi2005} and references therein). 

If we estimate the nominal total SFRDs 
from the total luminosity densities of
$\rho^{\rm tot}_{\rm Ly\alpha}$ and 
$\rho^{\rm tot}_{\rm UV}$ that are given by the extrapolation
of luminosity functions, we obtain
$SFRD =$
$(8.5_{-1.5}^{+1.8}, 5.7_{-1.2}^{+1.3}, 8.3_{-3.4}^{+5.9}) \times 10^{-3}$ 
and
$(5.5_{-3.0}^{+6.5}, 5.1_{-0.7}^{+0.8}, 9.4_{-6.9}^{+26}) \times 10^{-3}$ 
 $M_\odot$yr$^{-1}$Mpc$^{-3}$
for $z=(3.1, 3.7, 5.7)$, respectively.
We compare the SFRD estimated from Ly$\alpha$ LF of $z=3.1$ LAEs
that is recently obtained by MUSYC. \citet{gronwall2007} have found that
SFRD contributed by $z=3.1$ LAEs is $\simeq 6.5-8.6\times 10^{-3}$ 
$M_\odot$yr$^{-1}$Mpc$^{-3}$. 
Our result of $8.5_{-1.5}^{+1.8}\times 10^{-3}$ $M_\odot$yr$^{-1}$Mpc$^{-3}$
is very consistent with their finding.

The SFRD from UV luminosity density may increase from $z=3.1$ to $5.7$
by a factor of 2,
but again the large errors do not distinguish between evolutionary effects
and uncertainties. If we take into account the large error bars,
$SFRD \simeq 5-9 \times 10^{-3}$ $M_\odot$yr$^{-1}$Mpc$^{-3}$
at $z=3.1-5.7$. 
We compare these SFRDs with those estimated from dropout galaxies.
The results of SFRD measurements with no dust correction
show that the SFRDs of dropouts
are $\simeq 2-3\times 10^{-2}$ $M_\odot$ yr$^{-1}$ Mpc$^{-3}$
at $z\sim 3-6$ \citep{steidel1999,bouwens2006,hopkins2006}. 
Thus, LAEs contribute 
roughly about $\simeq 20-40$\% of the entire cosmic SFRD
at $z=3-6$.

\section{Properties of Ly$\alpha$ Emitters}
\label{sec:properties_of_lya_lines}

\subsection{Statistics of Ly$\alpha$ Equivalent Width}
\label{sec:lya_ew}

\begin{figure}
\epsscale{0.65}
\plotone{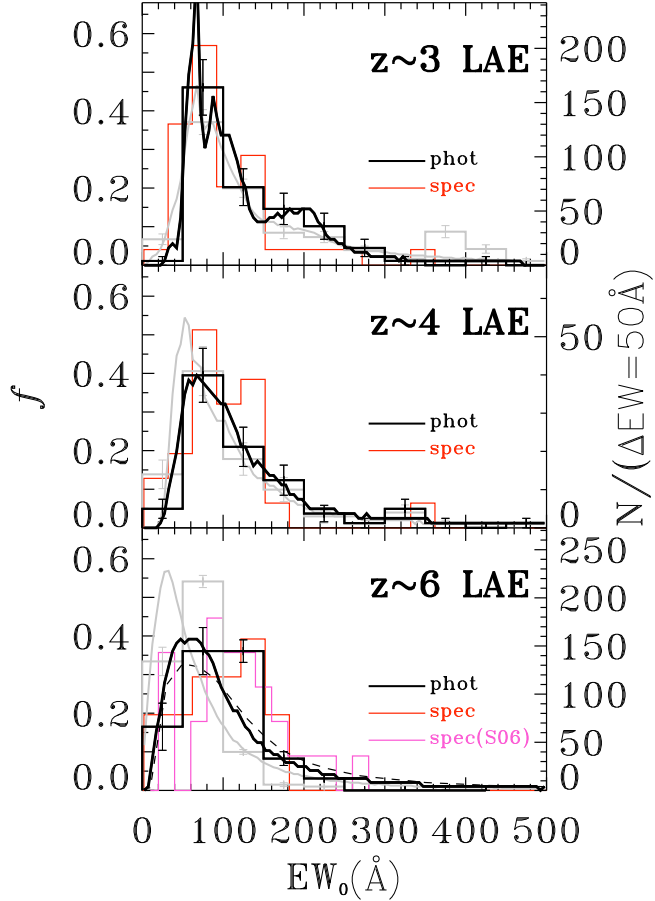}
\caption{
Histogram of the rest-frame Ly$\alpha$ equivalent widths (EW)
for our LAEs at $z=3.1$ (top), 
$3.7$ (middle), and $5.7$ (bottom).
The black histograms and curves
represent the best-estimate $EW_0$ and 
the $EW_0$ probability distribution estimated from errors of measurements, 
respectively, for photometrically selected LAEs
with $\log L({\rm Ly\alpha})\gtrsim 42.6$
and $EW_0^{\rm int}\gtrsim 70-80$. The gray histograms
and curves are the same, but for 
all the photometrically selected LAEs.
The ticks of vertical axes indicate 
a fraction (left-hand side) and the number (right-hand side)
in a bin size of $\Delta EW=50$\AA. The right-hand ticks
correspond to the number of all the photometrically selected LAEs.
The red histograms present the $EW_0$ distribution of our 
spectroscopically identified LAEs. 
On the bottom panel, the magenta histogram and black dashed
line show the best-estimate $EW_0$ and 
the $EW_0$ probability distribution for $z=5.7$ LAE sample
obtained by DEIMOS spectroscopy in the Subaru Deep Field \citep{shimasaku2006}.
\label{fig:ew_hist_all_data}}
\end{figure}

Figure \ref{fig:ew_hist_all_data} presents the histogram of 
rest-frame equivalent width ($EW_0$) for our 
LAEs. We calculate $EW_0$ of our LAEs from
our photometric measurements by the modeling described
in Section \ref{sec:ly_alpha_luminosity_functions}.
We then obtain two estimates of the $EW_0$ distribution by
(i) taking the best measurements of $EW_0$ for each LAE 
and (ii) summing the probability distribution of $EW_0$ that
is defined by errors of $EW_0$.
We show in Figure \ref{fig:ew_hist_all_data} 
these two estimates with (i) histograms (+Poisson errors) 
and (ii) solid curves. We find that the results of (i) and (ii)
agree fairly well for $z\sim 3$ and $4$ LAEs.
However, for our $z\sim 6$ LAEs, we see a significant difference
at $EW_0=100-150$\AA. This discrepancy between (i) and (ii)
for $z\sim 6$ LAEs is probably due to the large errors
of the relatively shallow off-band (i.e. $z'$-band) photometry.
Thus, we regard the difference of (i) and (ii) as the uncertainties
of our measurements.
The gray histograms and curves indicate
the $EW_0$ distribution for all of our photometrically-selected LAEs.
Since our three LAE samples at $z=3.1$, $3.7$, $5.7$ do not have 
the same limits of Ly$\alpha$ luminosity and $EW_0$, we make
subsamples of LAEs that share the same detection limits
for comparison. We set $\log L({\rm Ly\alpha})\gtrsim 42.6$ erg s$^{-1}$
and $EW_0^{\rm int}\gtrsim 70-80$\AA\ for the limits of our subsamples,
because our three LAE samples are complete in these limits.
The histograms of these subsamples are plotted with
black lines at each panel. We compare these histograms
with those for spectroscopically identified LAEs.
Although the statistical accuracy is limited for
the spectroscopic LAE samples, the $EW_0$ distributions of 
spectroscopic and photometric samples are consistent.

\begin{figure}
\epsscale{0.75}
\plotone{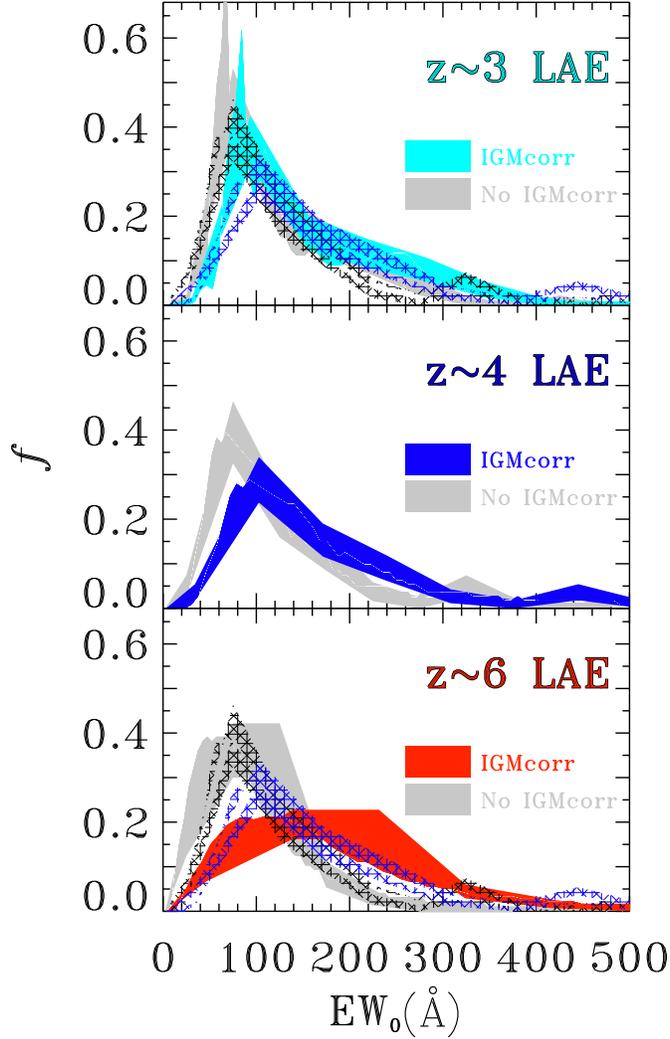}
\caption{
Same as Figure \ref{fig:ew_hist_all_data}, but 
for the comparison of histograms for 
apparent (``No IGMcorr'') 
and intrinsic (``IGMcorr'') $EW_0$
from $z=3.1$ to $5.7$. At each panel,
the gray region indicates 
the apparent $EW_0$ histogram with uncertainties 
whose area corresponds to the allowed parameters
of the black curves and lines (+error bars)
of Figure \ref{fig:ew_hist_all_data}.
The cyan, blue, and red areas are the same as
the gray regions, but for intrinsic $EW_0$
corrected for the IGM absorption,
assuming the average Ly$\alpha$ opacity (see text). 
For comparison, we repeatedly plot 
the $z\sim 4$ gray and blue areas 
with the black and blue meshes, respectively,
on the top and bottom panels.
\label{fig:ew_hist_all_implication}}
\end{figure}

In Figure \ref{fig:ew_hist_all_implication} 
we plot gray areas showing the $EW_0$ distributions 
for the subsamples with the uncertainties of (i) and (ii).
We find no clear evolution of $EW_0$ distribution from 
$z=3.1$ to $5.7$ with the uncertainties of our measurements. 
Since these $EW_0$ distributions are based on
the apparent EW, we correct for 
the absorption of IGM with the average attenuation at $z=3.1$,
$3.7$, $5.7$ (see Section \ref{sec:evolution_of_lya_LF}).
We present in Figure \ref{fig:ew_hist_all_implication} 
the distribution of {\it intrinsic} $EW_0$
(i.e. $EW_0$ corrected for IGM absorption)
with cyan, blue, and red regions for $z=3.1$, $3.7$, and $5.7$,
respectively. The IGM corrected histograms of $z=3.1$ and $3.7$ appear
to be comparable within the large uncertainties of our measurements.
The histogram of $z=5.7$ (red histogram)
would be systematically different from those of $z=3.1$ and $3.7$, 
and could imply that 
$z=5.7$ LAEs may tend to have a large {\it intrinsic} $EW_0$.
However, the uncertainty of $EW_0$ measurements is as large as
the difference of histograms between at $z=3.7$ and $z=5.7$.
We cannot reject the possibility of no evolution of {\it intrinsic} $EW_0$
from $z=3.7$ to $5.7$.

We investigate the fraction of large-EW LAEs with the subsamples.
We define a threshold of $EW_0^{\rm int}\ge 240$\AA\ with
an {\it intrinsic} EW,
following previous studies (e.g. \citealt{malhotra2002,dawson2004,
shimasaku2006,saito2006}). This threshold of $EW_0^{\rm int}$
corresponds to the {\it apparent} EW of $EW_0^{\rm app}=192$\AA,
$175$\AA, and $130$\AA, assuming the average attenuation of IGM
(Section \ref{sec:evolution_of_lya_LF}). We find that the fractions
and errors of the large-EW LAEs in our subsamples
are
($z$, fraction, fraction within 95th percentile)=
($3.1$, 0.21, 0.11-0.29),
($3.7$, 0.26, 0.16-0.40),
and
($5.7$, 0.25, 0.15-0.33)
by calculations of (i).
\footnote{
The ranges of 95th percentile are obtained by bootstrap sampling.
}
From the calculations of (ii),
we obtain the similar fractions of 0.20, 0.20, and 0.27 
for $z=3.1$, $3.7$, and $5.7$ LAEs.
There is a possible but slight implication of the evolution
of the fraction of large-EW LAEs from $z=3.7$ to $5.7$ 
in the results of (ii).
However, again, this possible evolution is very uncertain
due to the large errors of EW measurements.
The best estimates of the fraction are $20-30$\% for z=3.1-5.7. 
Thus, we conclude with these uncertainties
that the fraction of large-EW LAEs 
is 10-40\% at $z=3.1-5.7$, which does not significantly 
change from $z=3.1-5.7$ within this percent range.

We compare our fraction of large-EW LAEs with those of
\citet{shimasaku2006} for $z=5.7$ LAEs. Since \citet{shimasaku2006}
calculate this fraction with the threshold of {\it apparent} $EW_0$
of $\ge 120$\AA\ (c.f. 130\AA\ in above calculations), we use
this threshold of {\it apparent} $EW_0\ge 120$\AA\ here.
With this threshold, we find the fraction and 95th-percentile range
of 0.34 and 0.24-0.42 for our $z=5.7$ LAEs. These numbers are
vary consistent with those claimed by \citet{shimasaku2006} who obtain
30-40\%.

\subsection{Relation between Ly$\alpha$ and UV-continuum emission}
\label{sec:lya_uv}

\begin{figure}
\epsscale{0.8}
\plotone{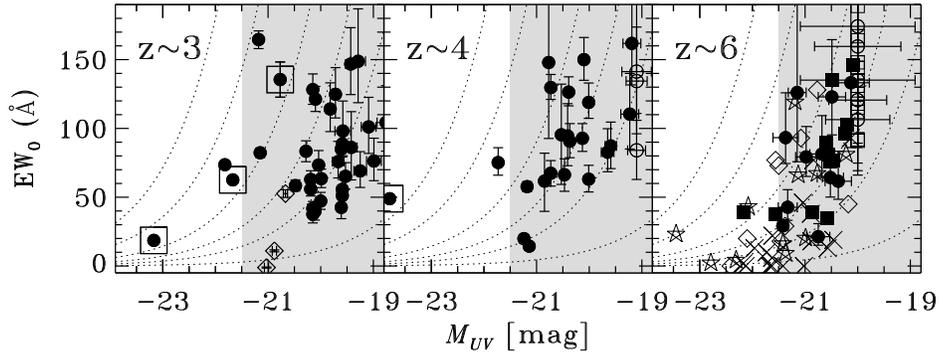}
\caption{
Rest-frame equivalent width ($EW_0$) as a function
of UV magnitude for $z\sim 3$ (left), $4$ (middle),
and $6$ (right) objects. Filled and open circles plot
our spectroscopically identified LAEs
with a UV magnitude brighter and fainter
than our $3\sigma$ limits, respectively.
The filled circles marked with an open square
indicate LAEs with AGN activities.
On the left panel, we show $EW_0$ values of $z\sim 3$ dropout galaxies
obtained by \citet{shapley2003}, which are the
average of the four subsamples of \citet{steidel2003}.
On the right panel, the filled squares represent
$z=5.7$ LAEs of \citet{shimasaku2006}.
The stars and diamonds present $EW_0$ of $z\sim 6$
dropout galaxies from \citet{stanway2007}
and the compilation of \citet{ando2006}.
The crosses plot $EW_0$ of $z\sim 5$ dropout galaxies
presented in \citet{ando2006}.
The dotted lines are loci of the constant Ly$\alpha$ luminosity
for $10^{44}$, $5\times 10^{43}$, $2\times 10^{43}$, $10^{43}$,
$5\times 10^{42}$, and $1\times 10^{42}$ erg s$^{-1}$ from
top left to the bottom. The shaded regions present the
magnitude range with EW-large emitters ($M_{\rm UV}\gtrsim -21.5$),
which is claimed by \citet{ando2006}.
\label{fig:corr_Muv_EW0_all}}
\end{figure}

Figure \ref{fig:corr_Muv_EW0_all} presents rest-frame Ly$\alpha$ $EW$ ($EW_0$)
as a function of UV-continuum magnitude for our spectroscopic samples. 
The effective wavelength of UV magnitude is $\sim 1300-1600$\AA.
The details of UV-magnitude measurements are presented in
Section \ref{sec:uv_luminosity_functions}.
Figure \ref{fig:corr_Muv_EW0_all} also plots $EW_0$ from
previous studies of dropout galaxies and LAEs.
On the right panel of Figure \ref{fig:corr_Muv_EW0_all}, 
$EW_0$ of $z\sim 6$ objects show a clear deficit of 
large $EW_0$ objects in the bright magnitude ($M_{\rm UV}\lesssim -21.5$).
This trend is found by \citet{ando2006} for $z=5-6$ galaxies, and discussed 
in \citet{shimasaku2006,stanway2007} for $z\sim 6$ objects. 
We confirm their findings on our plot of $z\sim 6$ objects.
For $z\sim 3$ and $4$ objects shown in Figure \ref{fig:corr_Muv_EW0_all},
we find the trend very similar to that of $z\sim 6$. We see no
large $EW_0$ objects with $EW_0\gtrsim 80$\AA\ for UV-bright 
($M_{\rm UV}\lesssim -21.5$) objects. At $z=3$, 
this trend is also reported by \citet{shapley2003}
who find that the Ly$\alpha$ emission strength increases 
toward fainter magnitudes
in their spectra of LBGs.

\subsection{UV-Continuum Color and Extinction}
\label{sec:uv_color}

We present UV-continuum colors as a function of 
UV magnitude for our spectroscopically identified LAEs 
at $z=3.1$ and $3.7$
in Figure \ref{fig:corr_Muv_contcolor_all}.
This figure compares those of dropout galaxies
at the similar redshift. Except for AGN, typical
$z=3-4$ LAEs have $i'-z'$ (or $R-i'$) colors of 
$\lesssim 0.05$. The UV-continuum colors of our LAEs
are consistent with those of $z=4.5$ LAEs \citep{finkelstein2007}.
On the other hand, dropout galaxies (or LBGs) 
generally have colors of $\simeq 0.1$.
Thus, LAEs appear to be bluer than dropout galaxies
in a UV-continuum color. 
This color difference indicates
that LAEs are younger or less-dusty than dropout-galaxy population.
We discuss more details in Section \ref{sec:implications_blue_continuum}.

\subsection{Composite Spectra}
\label{sec:composite_spectra}

We make a composite spectrum each for
our $z=3.1$, $3.7$, and $5.7$ LAEs.
From our spectroscopic samples,
we exclude AGNs (Table \ref{tab:multi-wavelength_prop}) and 
LAEs with uncertain identification (Table \ref{tab:laes_with_redshifts}). 
We also do not include the spectrum of NB503-N-42377 that has a marginal
{\sc Civ} line, so as to obtain the average of typical LAE spectra
(see Section \ref{sec:agn_spectra}).
Thus, we use 36, 25, and 15 spectra of $z=3.1$, $3.7$, and $5.7$, respectively.
Since some of spectra have systematic differences originated from
a faint AGN and/or errors of cosmic rays and subtraction of sky etc., 
we calculate a mean flux with rejections
of two largest and smallest values at each wavelength.
Figure \ref{fig:compspec_all} shows the composite spectra
of our LAEs whose average redshifts are $\left <z\right > = 3.13$,
$3.68$, and $5.68$. We have also performed the median stacking, 
and find no qualitative differences
between the rejected-mean and median results.
The Ly$\alpha$ luminosity of the composite spectrum is
$L(Ly\alpha)=(6.3\pm 0.1, 8.5\pm 0.2, 9.5\pm 0.3) \times 10^{42}$ erg s$^{-1}$
at $\left< z\right >=(3.13,3.68,5.68)$.
We have a signal-to-noise ratio high enough to detect
UV continua in $\left<z \right>=3.13$ and $3.68$ spectra,
but not in $\left< z\right >=5.68$ spectrum.

The composite spectrum of $\left< z\right >=5.68$ LAEs has Ly$\alpha$ emission line
with a clear asymmetry, which is a typical emission feature of
high-$z$ objects. The spectra of $\left<z \right>=3.13$ and $3.68$
do not show a significant asymmetry in Ly$\alpha$. This is because
a typical spectral resolution of our 
$z=3.1$ and $3.7$ LAEs is not high enough 
to resolve the asymmetry of Ly$\alpha$
($R\sim 500$; c.f. $R\sim 1000$ for $z=5.7$ LAEs).
Moreover, the IGM absorption
at $z=3-4$ is not as strong as $z=5.7$, which provides a weak asymmetric feature
in Ly$\alpha$ emission at $z=3-4$.

We find no significant absorption lines
in these composite spectra, probably due to the combination
of the weak metal absorption of LAEs and 
the relatively poor signal-to-noise ratios of
our composite spectra. 
\footnote{
The Si{\sc ii} absorption of $\left<z \right>=3.13$ LAE 
may exist, but the significance level is not high.
}
The composite spectra of $\left<z \right>=3.13$ and $3.68$
cover the wavelength of He{\sc ii}$\lambda1640$ emission.
The inset panels of Figure \ref{fig:compspec_all} 
magnify the spectra around this wavelength. No significant
feature is found for He{\sc ii} at $\left<z \right>=3.13$ and $3.68$.
We calculate upper limits of He{\sc ii} flux, $f_{\rm HeII}$, and
estimate the $3\sigma$ upper limits of the ratio, $f_{\rm HeII}/f_{\rm Ly\alpha}$,
where $f_{\rm Ly\alpha}$ is Ly$\alpha$ flux.
We obtain meaningful $3\sigma$ upper limits of the ratio,
$f_{\rm HeII}/f_{\rm Ly\alpha}=0.02$ and $0.06$
for $\left<z \right>=3.13$ and $3.68$ LAEs, respectively
Since He{\sc ii} emission is an indicator of primordial populations,
i.e. population III stars and cooling radiation, 
our upper limits place constraints on primordial populations at $z=3-4$.
Although theoretical models predict the existence of
primordial populations at these redshifts 
\citep{scannapieco2003,jimenez2006},
we have found no signature
at the level of the upper limits of the present observations.
Since the ratio of $f_{\rm HeII}/f_{\rm Ly\alpha}$ is
predicted as small as $\sim 0.1-0.001$ for population III star formation
and cooling radiation (e.g. \citealt{schaerer2003,yang2006}), 
we need future surveys whose
upper limits reach at the level of $f_{\rm HeII}/f_{\rm Ly\alpha}\simeq 0.001$.

\section{Discussion}
\label{sec:discussion}

\subsection{Evolution of Luminosity Functions}
\label{sec:evolution_of_luminosity_functions}

We have found four pieces of evidence
that provide a general picture in the evolution
of LAEs: 
(1) {\it apparent}-Ly$\alpha$ LFs of LAEs
show no evolution from $z=3.1$ to $5.7$. 
(2) These {\it apparent}-Ly$\alpha$ LFs
and the average Gunn-Peterson optical depth indicate that
{\it intrinsic}-Ly$\alpha$ LFs of LAEs must have
brighten in the luminosity and/or increased in the number density
from $z=3.1$ to $5.7$.
(3) The UV LFs of LAEs 
show the increase of luminosity and/or 
number densities from $z=3.1$ to $5.7$.
(4) The rest-frame {\it intrinsic}-Ly$\alpha$ EW, $EW_0^{\rm int}$ ,
distribution of LAEs does not change or positively evolves from $z=3.1$ to $5.7$.

The evolution of {\it intrinsic}-Ly$\alpha$ LFs is caused by
the emergence of Ly$\alpha$ emitting objects with a bright UV continuum
or the increasing trend of Ly$\alpha$ $EW_0$ for LAEs at $z\sim 5.7$.
There are three possible explanations for the evolution of LAEs.
First, the amount of dust mixed in HII regions may decrease in average
from $z=3.1$ to $5.7$, and the extinction of Ly$\alpha$ luminosity
through the resonance scattering 
decreases selectively. This implies that at $z=5.7$
even UV bright galaxies have a detectable Ly$\alpha$ emission line.
The second explanation is that
the young Universe at $z=5.7$ has numerous
exotic Ly$\alpha$ populations such as 
metal-poor population III and/or cooling clouds.
The third explanation is that the fraction of Ly$\alpha$-absorbed 
galaxies is lower at higher redshifts, due to an decrease in the 
amount of HI gas in and around galaxies with redshift.
In fact, some $z\sim 3$ LAEs show such an HI absorption originating 
from an inflow and/or outflow in their spectra (e.g. \citealt{tapken2007}), 
although the evolution of such HI clouds has not been 
observationally understood.
The CDM-based hierarchical model predicts that 
galaxies at higher redshifts are generally more compact 
(e.g. \citealt{mo1998,somerville2006}). It is possible that
the star formation in such compact high-$z$ galaxies is 
so efficient that cold gas is immediately converted into stars.
If high-$z$ LAEs are such compact galaxies, they will have little 
HI gas and thus their Ly$\alpha$ emission from HII regions can 
escape without absorption.
If this hypothesis is true, we may be able to 
explain the evolutionary connection between LAEs and dropouts:
the internal mass density of LAEs decreases with cosmic time by mergers 
and thus they evolve into dropout galaxies that are more massive and 
have more HI gas.

It should be noted that the evolutionary trend of UV-LF of LAEs
is opposite to that of dropout galaxies.
The UV-LFs of dropout
galaxies show a significant decrease in luminosity and/or number
density from $z\sim4$ to $z\sim 6$ (e.g. \citealt{lehnert2003,ouchi2004a,
shimasaku2005,bunker2004}). In case of luminosity evolution,
UV-LFs of dropouts get faint from $z\sim 3$ to $6$ 
by a factor of $1.7^{+0.4}_{-0.3}$ \citep{bouwens2006}.
On the other hand, UV-LFs of LAEs brighten by $\sim 1$  magnitude,
if it comes from the pure luminosity evolution.
This difference of LFs implies that Ly$\alpha$ emitting
population is more dominant at $z=6$ than $z=3$
at the given UV luminosity range.

\subsection{AGN fraction and Evolution of Faint AGN}
\label{sec:evolution_of_agn_fraction}

We identify 6 AGN-LAEs in total. In our photometric samples of LAEs (satisfying
our color-selection criteria), we find that 3 out of 356 $z=3.1$ LAEs and 1 out of 101
$z=3.7$ LAEs host AGNs (Section \ref{sec:lae_hosting_agn}).
Thus, the fraction of AGNs is about 1\% at $z=3.1$ and $3.7$, 
which is consistent with the result of \citet{gawiser2006}.
Note that the detection limits of our LAEs reach 
$(1-4)\times 10^{42}$ ergs s$^{-1}$ ($\sim 0.2-0.4 L^*_{\rm Ly\alpha}$). 
This AGN fraction is
the lower limit, since our multi-wavelength data are probably 
not deep enough to identify faint AGNs. 
We estimate the relation between Ly$\alpha$ luminosity and
the detection limits of our multi-wavelength data.
By the comparison of our detection limits and 
the QSO templates in Figure \ref{fig:multiband_sed} 
(see Section \ref{sec:lae_hosting_agn} for the definition of QSO templates), 
our X-ray data likely reach the deepest detection limit for AGNs 
among our multi-wavelength data.
We normalize the QSO templates to the detection limit
of our X-ray data, and measure Ly$\alpha$ luminosities of the QSO templates.
We find Ly$\alpha$-luminosity limits of
$L(Ly\alpha)=9.5\times 10^{42}$ and $1.5\times 10^{43}$
erg s$^{-1}$ for $z=3.1$ and $3.7$ objects, respectively,
assuming that the SED of AGN-LAEs is the same
as the QSO templates.
Down to these Ly$\alpha$-luminosity limits,
we find 3 out of 25 (1 out of 8) photometrically selected LAEs have an AGN
at $z=3.1$ ($z=3.7$).
Thus, the AGN fraction is estimated to be $\sim 10$\%
down to $L({\rm Ly\alpha})\simeq 1\times 10^{43}$ erg s$^{-1}$
for $z=3.1$ and $3.7$ LAEs. This Ly$\alpha$-luminosity limit corresponds to
$\simeq 2 L^*_{\rm Ly\alpha}$. 
Note that
the AGN fraction of LAEs is as small as 10\%
even with this bright detection limit of Ly$\alpha$ luminosity.

If we take a Ly$\alpha$ limit for the very bright
luminosity of $\log L({\rm Ly\alpha})>43.6$ and $>43.4$ erg s$^{-1}$ 
in our $z=3.1$ and $3.7$ samples, respectively, 
all of LAEs (2 out of 2 LAEs and 1 out of 1 LAE at $z=3.1$ and 3.7)
have AGN activities. These very bright limits correspond
to the brightest bins of our LF 
in Figure \ref{fig:lumifun_full_diff_evol} 
(see Section \ref{sec:ly_alpha_luminosity_functions}).
There appear possible humps of LF at $z=3.1$ and $3.7$
which are produced by the AGN-LAEs. 
The number density in these brightest bins
is about $\sim 10^{-5}$ Mpc$^{-3}$. Thus, a 
shallow wide-field LAE survey generally
finds this AGN-LAE population.

Steidel et al. (2002) have obtained the fraction of AGNs in
their $z=3$ LBG samples down to $\simeq M^*+1$ with optical spectra.
They have found the AGN fraction of 3\% for their LBG sample
with a relatively faint continuum limit of $\simeq M^*+1$. Since the
detection limits of this study is different from ours,
we cannot directly compare the AGN fraction of their LBGs and our LAEs.
Since we find the AGN fraction (and its lower limit) of our LAEs is 
10\% (and $\gtrsim 1$\%)
for our $\simeq 2 L^*_{\rm Ly\alpha}$ 
(and the entire [$\sim 0.2-0.4 L^*_{\rm Ly\alpha}$]) samples,
the AGN fraction of LAEs and LBGs would be comparable.

For $z=5.7$ LAEs, we identify no AGN with our spectra or multi-wavelength data.
In the same manner as the calculations above,
we estimate a Ly$\alpha$ detection limit of $z=5.7$ LAEs
that corresponds to the X-ray luminosity limit with the QSO templates.
We find that the Ly$\alpha$ detection limit is 
$L(Ly\alpha)\sim 4\times 10^{43}$ erg s$^{-1}$, but 
we have no LAEs with $L(Ly\alpha)>4\times 10^{43}$ erg s$^{-1}$.
On the other hand, there are two AGN-LAEs 
that are brighter than $L(Ly\alpha)\sim 4\times 10^{43}$ erg s$^{-1}$
(i.e. $\log L({\rm Ly\alpha})>43.6$ erg s$^{-1}$) at $z=3.1$. 
Therefore, it may imply that LAEs with a bright AGN
such found at $z=3$ would disappear at $z=5.7$. 
In fact, no AGN signatures such as N{\sc v} emission
have been confirmed in spectra of the brightest $z=5.7$ LAEs
that are found in other wide-field studies \citep{westra2006}.
No detection of AGN in $z=5.7$ LAEs implies that 
the number density of LAEs with AGN activities
would drop from $z=3$ to $5.7$.
Recent QSO surveys have measured QSO luminosity functions
accurately, and found that the space density of bright QSOs
with $M_{\rm UV}<-26.7$
decreases from $z=3$ to 6 
by more than an order of magnitude (e.g. \citealt{croom2004,fan2001,fan2004}).
This trend is the same as that of our finding in our LAE samples.
Because our AGN-LAEs are as faint as $M_{\rm UV}\simeq -23- -21$,
our AGN-LAEs probe for evolution of relatively faint AGN.
Thus, evolution of faint AGNs appears to follow the evolution of QSOs. 
This faint-AGN evolution implies that not only bright QSOs but also 
faint AGNs can provide less significant contribution than star-forming galaxies 
for reionizing the Universe at $z\gtrsim 6$. In other words,
UV radiation from star formation likely becomes more dominant
than that from faint AGNs at $z\sim 6$.

\subsection{Star-Formation Rates Indicated from Multi-Wavelength Data}
\label{sec:sfr_multiwavelength}
In Section \ref{sec:multi-wavelength_properties}, we have identified
AGNs with our multi-wavelength data. The multi-wavelength data
are too shallow to identify star-forming activities at high redshifts. 
The limiting SFRs of our multi-wavelength are as large as
$\sim 10^4 M_\odot {\rm yr^{-1}}$ for our X-ray and $24\mu$m data
and $\sim 1000 M_\odot {\rm yr^{-1}}$ for our radio data, respectively.
However, we find that our radio data provide 
a meaningful limit on SFR of our LAEs by stacking analysis.
We exclude LAEs with an individual radio detection, and 
stack radio images of our photometrically selected LAEs
in the same manner as \citet{simpson2006b}.
The stacked images show no detection in the radio band.
The $3\sigma$-upper limits of the stacked images are
$\left < f_{\rm 1.4GHz} \right > = (1.8, 7.0, 2.7)$ $\mu$Jy 
for our $z=(3.1,3.7,5.7)$ LAEs.

The relationship between SFR
and radio luminosity ($L_{\rm 1.4 GHz}$) is given by
\begin{eqnarray}
\label{eq:sfr_multiwavelength}
SFR({\rm M_\odot yr^{-1}}) & = & \frac{L_{\rm 1.4 GHz}({\rm W\ Hz^{-1}})}{5.3\times 10^{21} (\nu/GHz)^{-\alpha}},
\end{eqnarray}
for high-mass stars ($M>5M_\odot$; \citealt{condon1992}; 
see also \citealt{serjeant2002}), 
where $\nu$ is the frequency and $\alpha$ is the non-thermal ratio spectral index
($\alpha \simeq 0.8$). Assuming this SFR v.s. radio-luminosity relation,
we obtain $3\sigma$ upper limits of
$\left < SFR \right > =(29, 162, 160)$ $M_\odot {\rm yr^{-1}}$
for our $z=(3.1,3.7,5.7)$ LAEs.
Thus, the average SFRs of LAEs are smaller than $\simeq 30$ and $160$
$M_\odot {\rm yr^{-1}}$ at $z=3.1$ and $3.7/5.7$, respectively.
The result of our $z=5.7$ LAEs is similar to that of
\citet{carilli2006} who report
the $2\sigma$ upper limit of $\left < f_{\rm 1.4GHz} \right > = 2.5$ $\mu$Jy 
(i.e. $\sim 100$ $M_\odot {\rm yr^{-1}}$) 
for their stacked radio image of $z=5.7$ LAEs in COSMOS field.
The upper limits of radio SFR are consistent with
the SFRs of typical ($L^*$) LAEs,
$\simeq 5-10 {\rm\ M_\odot yr^{-1}}$ (Section \ref{sec:evolution}),
estimated from UV and Ly$\alpha$ luminosities.
This agreement between radio and optical SFRs
implies that majority of LAEs 
are not associated with dusty starbursts with significant ($\times 10-100$) 
hidden star-formation.
It is consistent with
the fact that no LAEs are detected in submm bands (Section \ref{sec:sub-mm_properties}).
These results support the small-dust extinction of LAEs from
the blue UV-continuum color (Section \ref{sec:implications_blue_continuum}).
On the other hand, no significant excess of radio SFR
may reject the possibility of an extremely-top heavy 
initial mass function for star-formation of LAEs, 
since radio fluxes are originated from 
non-thermal radiation of supernovae at the final stage of massive stars.

No radio detections in the stacked images also suggest that typical LAEs do not harbor
low-luminosity radio AGNs at $z=3.1-5.7$ (see also \citealt{carilli2006} 
for $z=5.7$ LAEs). The radio fluxes of our AGN-LAEs are
$f_{\rm 1.4 GHz} \simeq 100-200 \mu$Jy (Table \ref{tab:multi-wavelength_prop})
at $z=3.1$ and $3.7$, while our $3\sigma$ upper limits of stacked images
are $1.8-7.0 \mu$Jy. Thus, typical LAEs do not have a low-luminosity radio AGN
whose flux is brighter than $1/10-1/100$ of typical radio AGN-LAEs at $z=3-4$.

\subsection{Ly$\alpha$ Emissivity}
\label{sec:lya_emissivity}

Figure \ref{fig:corr_Muv_EW0_all} shows
that UV-bright ($M_{\rm UV}\lesssim -21.5$) galaxies
do not have a large $EW_0$ ($EW_0\gtrsim 80$\AA)
at $z=3-6$.
\citet{ando2006} suggest that
UV-bright galaxies are a
dusty and chemically-evolved population, and that such UV-bright
galaxies do not have a large EW.
There are two other possibilities to explain this trend.
First, the average stellar population of UV-bright galaxies 
would be older than that of UV-faint galaxies. Since Ly$\alpha$ emission
can be efficiently produced by hard-UV continuum from
early-type stars, galaxies can emit a strong Ly$\alpha$ emission of 
$EW_0\gtrsim 100$\AA\ within $\sim 10$ Myr after a starburst
\citep{schaerer2003}. 
Second, HI clouds of inter-stellar medium or inflows may be rich
in UV bright population, and absorb a Ly$\alpha$ emission line. 
It is likely that the combination of
age, neutral hydrogen, dust, and metallicity makes 
the deficit of large EW galaxies at bright UV magnitude.

\subsection{Implications from the Blue UV continuum}
\label{sec:implications_blue_continuum}

\begin{figure}
\epsscale{0.8}
\plotone{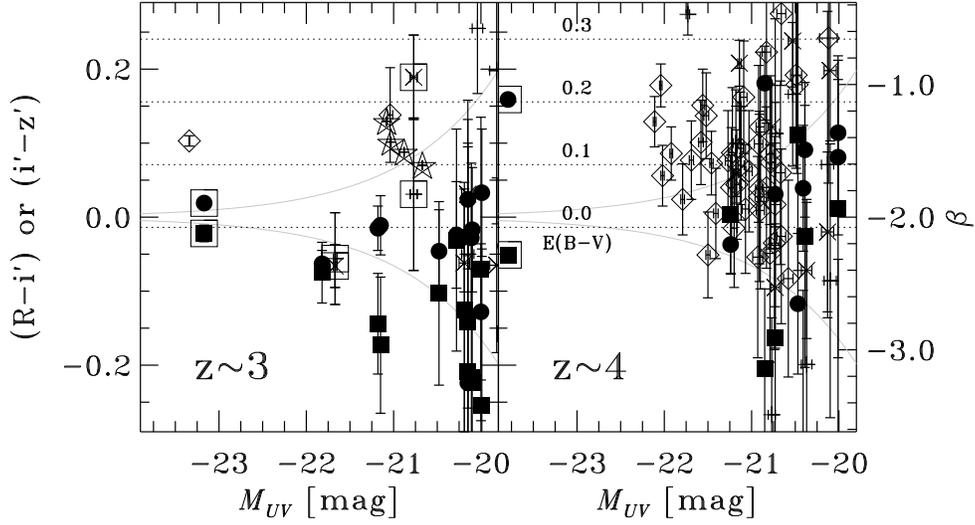}
\caption{
UV-continuum color as a function of UV magnitude
at $z\sim 3$ (left panel) and $z\sim 4$ (right panel).
The filled squares indicate a $i'-z'$ color of 
our LAEs at $z=3.1$ and $3.7$ from our spectroscopic samples.
The filled circles are the same, but for a $R-i'$ color.
For confused LAEs whose photometry appears to be contaminated 
by neighboring objects on a broad-band image, 
we use plus signs and crosses instead of filled squares and circles,
respectively.
We mark LAEs with AGN activities with a large open square.
The diamonds plot $i'-z'$ colors of spectroscopically identified
LBGs \citep{yoshida2006}. The star marks on the left panel
represent the average colors of LBGs obtained by 
\citet{shapley2003}. We estimate $i'-z'$ colors of Shapley et al.'s 
data with $E(B-V)$ and the eq. \ref{eq:ebv_iz}. 
The vertical axis on the right-hand side shows the UV-slope index, $\beta$,
that corresponds to $i'-z'$ colors at $z\sim 4$.
Note that this $\beta$ is based on the original definition
(See \citealt{calzetti2001}), which is different from $\beta_{\rm iz}$
defined by \citet{ouchi2004a}.
The difference of $\beta$ values between $R-i'$ and $i'-z'$ colors is
only $\lesssim 7$\% in the color range of $-0.3-0.3$.
The $E(B-V)$ values corresponding to the UV colors are 
indicated with dotted lines. The gray curves represent 
the size of $\pm 1\sigma$ error for a continuum-flat ($f_\nu={\rm const}$)
object.
\label{fig:corr_Muv_contcolor_all}}
\end{figure}

\begin{figure}
\epsscale{0.8}
\plotone{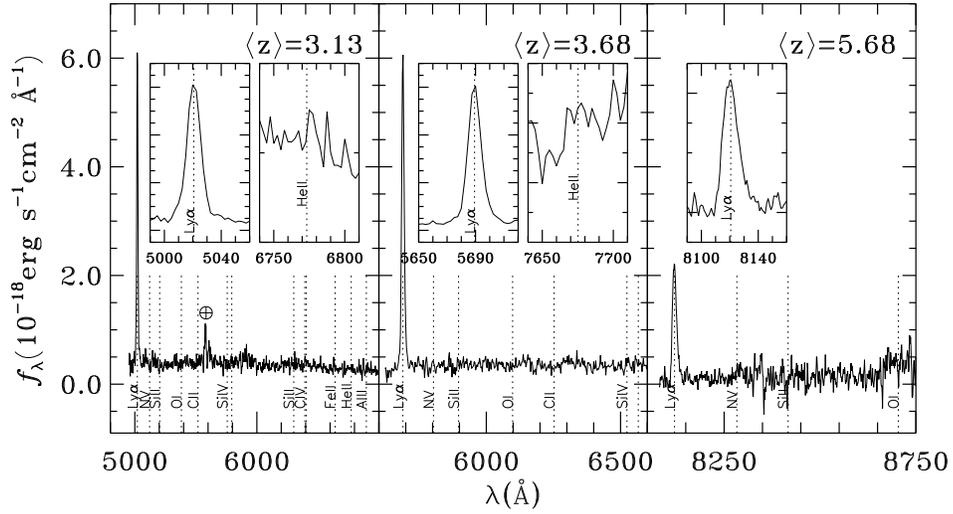}
\caption{
Composite spectra of our LAEs at $\left < z \right >=3.13$ (left),
$3.68$ (middle), and $5.68$ (right). The dotted lines
indicate the wavelengths of interstellar absorption
from star formation, or high ionized emission lines 
from AGN activities.
The plots of spectrum are magnified in the inset boxes
for wavelength ranges of Ly$\alpha$ (left) and He{\sc ii} (right).
For the $\left < z \right > =5.68$ LAE, we only show the inset box of Ly$\alpha$,
since our spectrum does not cover the wavelength of He{\sc ii}
emission for $z=5.7$ objects.
\label{fig:compspec_all}}
\end{figure}

We find that UV-continuum colors of LAEs are bluer than
those of dropout galaxies in Section \ref{sec:uv_color}.
This implies that LAEs are less-dusty and/or younger population
than dropout galaxies. Although more detailed analysis with
near-infrared data and population synthesis models
will be presented in Ono et al. (in preparation),
we compare the UV-continuum colors of LAEs with a simple model.
\citet{ouchi2004a} have obtained the relation between $i'-z'$
color and dust extinction, $E(B-V)$, with 
the stellar population synthesis model \citep{bruzual2003}
and the extinction law \citep{calzetti2000}
which reproduce a typical UV-optical SED of dropout galaxies at $z\sim 3$.
The relation of color and dust extinction is,
\begin{equation}
E(B-V)\simeq 0.0162+1.18(i'-z'),
\label{eq:ebv_iz}
\end{equation}
\citet{ouchi2004a} also confirm that this relation
agrees with the one of local starbursts.
This relation assumes that
the difference of UV color (or UV slope) is caused by dust extinction,
since the model fixes the age, metallicity, initial mass function,
and star-formation history.
We apply this simple model to our LAEs in Figure \ref{fig:corr_Muv_contcolor_all}.
Figure \ref{fig:corr_Muv_contcolor_all} implies that
LAEs have an extinction of $E(B-V)\lesssim 0.05$, while that
dropout galaxies typically have $E(B-V)\sim 0.1-0.2$.
There are some very blue LAEs whose UV colors scatter
around $-0.2-0.0$. The colors of these objects are still 
comparable to the color of $E(B-V)=0.0$ within a $\simeq 2\sigma$ level. 
However, if this very blue UV color is true, the colors of LAEs can be 
explained not only by extinction, but also by stellar population that is bluer than 
that of typical dropout galaxies.
\citet{gronwall2007} estimate $E(B-V)$ of $z=3.1$ LAEs with the
independent approach, i.e. the comparison of UV and Ly$\alpha$ luminosity.
They have obtained $E(B-V)<0.05$, which is consistent with
ours.

\section{Conclusions}
\label{sec:conclusions}

We have made a photometric sample of 858 LAE candidates and
a spectroscopic sample of 84 confirmed LAEs 
at $z=3.1$, $3.7$, and $5.7$
down to $L({\rm Ly}\alpha)\gtrsim 3\times 10^{42}$
erg s$^{-1}$ in a 1 deg$^2$ sky of the Subaru XMM-Newton
Deep Survey Field. Based on these samples, we have
studied photometric and spectroscopic properties
with the aid of multi-wavelength data from X-ray to radio.
The major findings of our study are summarized below.\\

1. We derive the LF of Ly$\alpha$ emission for each redshift and 
provide the best-fit Schechter parameters as summarized in Table 5, 
by carefully taking into account the statistical error 
and the field-to-field variation.
We find that the {\it apparent} Ly$\alpha$ LF shows no significant 
evolution between $z=3.1$ and $5.7$. Evolution, if any, is limited 
within factors of 1.8 and 2.7 in $L^*$ and $\phi^*$, respectively.
Although no evolution of the Ly$\alpha$ LF 
over $3\lesssim z\lesssim 5.7$ has already been suggested by previous 
studies,
we now have much stronger and more reliable constraints 
with smaller systematic and statistical 
errors. 
Since the average Ly$\alpha$ opacity of the IGM is larger
at earlier epochs, the absence of evolution in the apparent 
Ly$\alpha$ LF implies an intrinsic brightening of Ly$\alpha$ 
luminosity with increasing redshift being canceled out by 
increasing absorption of the IGM. 
We note, however, that this claim
is true only if the two assumptions made 
in Section \ref{sec:evolution_of_lya_LF} are correct.

2. The LF of UV continuum for LAEs shows an increase 
in number density and/or UV luminosity from $z\sim 3-4$ to $5.7$.
This evolution of the UV LF is consistent with 
the plausible evolution in the {\it intrinsic} Ly$\alpha$ LF, 
since we find that the distribution of the rest-frame {\it intrinsic} 
EW of LAEs does not change or positively evolves with redshift.
On the other hand, the UV LF of dropout galaxies 
is known to show an opposite evolutionary trend to 
that of LAEs, namely, a decrease in number
density and/or UV luminosity with increasing redshift.
This implies that the ratio in number density of LAEs 
to dropout galaxies
increases from $z\sim 3-4$ to $5.7$, and that galaxies with 
Ly$\alpha$ emission are more common at earlier epochs.

3. We identify six LAEs (in total) with AGN activities. In our 
photometric sample, 
three out of 356 $z=3.1$ LAEs and one out of 101 $z=3.7$ LAEs 
host AGNs. Thus, the lower limit of the AGN fraction down to 
$(1-4)\times 10^{42}$ erg s$^{-1}$ ($\sim 0.2-0.4 L^*_{\rm Ly\alpha}$)  
is $\simeq 1$\% at $z=3.1$ and $3.7$.
We use QSO templates given in the literature to find 
that $\sim 10\%$ of LAEs are AGNs 
above the Ly$\alpha$ luminosity where our sample is complete, 
$L({\rm Ly\alpha})\simeq 1\times 10^{43}$ erg s$^{-1}$
($\simeq 2 L^*_{\rm Ly\alpha}$).
It is notable that
100\% of LAEs (2 out of 2 LAEs and 1 out of 1 LAE at $z=3.1$ and 3.7)
host AGNs at the bright end of the LF with 
$\log L({\rm Ly\alpha})>43.6$ and $>43.4$ erg s$^{-1}$ at $z=3.1$ and $3.7$, 
respectively.
Shallow and wide-field narrow-band imaging would be an efficient 
way to search for LAEs hosting AGNs.

4. We obtain the $3\sigma$ upper limits of stacked radio images,
$\left < f_{\rm 1.4GHz} \right > = (1.8, 7.0, 2.7)$ $\mu$Jy 
for our $z=(3.1,3.7,5.7)$ LAEs. These upper limits indicate
$\left < SFR \right > =(29, 162, 160)$ $M_\odot {\rm yr^{-1}}$
at $z=(3.1,3.7,5.7)$ based on the SFR-radio luminosity relation of 
\citet{condon1992}. These upper limits are consistent with
the SFRs estimated from Ly$\alpha$ and UV luminosities. 
Thus, the majority of LAEs are not associated with dusty starbursts 
with significant ($\times 10-100$) hidden star-forming activities.
No radio detections in the stacked images also indicate that 
typical LAEs do not harbor a low-luminosity radio AGN at $z=3.1-5.7$.
At $z=3-4$, typical LAEs do not have a low-luminosity radio AGN
whose flux is brighter than $1/10-1/100$ of typical radio AGN-LAEs
of our SXDS field.

5. The fraction of LAEs that have very large Ly$\alpha$ EWs 
does not change significantly over $3.1\le z\le 5.7$;
the fraction of objects with $EW_0^{\rm int}\ge 240$\AA\ 
is 10--40 \%.

6. If we naively assume that all of the Ly$\alpha$ or UV emission 
in our LAEs originate from star-forming activities, 
LAEs contribute about $\simeq 20-40$\% of the entire cosmic SFRDs
at $z=3-6$.

7. We find a deficit of large-EW galaxies 
with bright UV continuum emission at all three redshifts.
This finding not only confirms the previous results of
\citet{shapley2003,ando2006,shimasaku2006} for $z=3$ and $5-6$, 
but also claims that this trend is common in the early universe.
This deficit is probably caused by a combination of 
old ages and large amount of neutral hydrogen, dust, and metallicity
in UV-bright galaxies.

8. The UV-continuum colors of LAEs are generally bluer than
those of dropout galaxies at the same UV luminosity. If we assume 
that LAEs have the same stellar population as dropout galaxies, 
then the UV spectral slope of LAEs
indicates $E(B-V)\lesssim 0.05$. 
The very blue LAEs probably have extremely low dust extinction 
and/or very young stellar populations.

9. We stack the spectra of the confirmed LAEs to estimate 
the $3\sigma$ upper limit of $f_{\rm HeII}/f_{\rm Ly\alpha}$ 
to be $0.02$ for $z=3.1$ and $0.06$ for $z=3.7$.
These low $f_{\rm HeII}/f_{\rm Ly\alpha}$ values are consistent 
with no or little primordial population being included in our sample.

\acknowledgments

We are grateful to Mark Dijkstra, Eric Gawiser, Kim Nilsson, and Edward Westra
for their helpful comments. We thank the anonymous referee 
for his/her constructive suggestions and comments that improved this article.
M.O. has been supported by the Hubble Fellowship program
through grant HF-01176.01-A awarded by the Space Telescope Science
Institute, which is operated by the Association of Universities
for Research in Astronomy, Inc. under NASA contract NAS 5-26555.
M.O. acknowledge support via Carnegie Fellowship.

\clearpage

\begin{deluxetable}{ccrcccl}
\tablecolumns{7}
\tabletypesize{\scriptsize}
\tablecaption{Summary of Imaging Observations and Data 
\label{tab:obs}}
\tablewidth{0pt}
\tablehead{
\colhead{Band} & 
\colhead{Field Name(s)} & 
\colhead{Exposure Time} &
\colhead{PSF size\tablenotemark{\ddag}} &
\colhead{Area} &
\colhead{$m_{\rm lim}$\tablenotemark{\dagger}}  &
\colhead{Date of Observations} \\
\colhead{} & 
\colhead{} & 
\colhead{(sec)} &
\colhead{(arcsec)} &
\colhead{(arcmin$^2$)} &
\colhead{(5$\sigma$ AB mag)}  &
\colhead{}
}
\startdata
$NB503$ & SXDS-C & 4200 & 0.79 (0.81) & 657 & 25.3 & 2003 Oct 23   \\
$NB503$ & SXDS-N & 4200 & 1.05 (1.05) & 766 & 25.1 & 2003 Oct 23   \\
$NB503$ & SXDS-S & 4200 & 0.91 (0.91) & 827 & 25.3 & 2003 Oct 23   \\
$NB503$ & SXDS-E & 4200 & 0.69 (0.83) & 685 & 25.4 & 2003 Oct 24   \\
$NB503$ & SXDS-W & 5400 & 0.67 (0.83) & 603 & 25.5 & 2003 Oct 24   \\
$NB570$ & SXDS-C & 4200 & 0.91 (0.91) & 627 & 24.6 & 2003 Oct 22,26   \\
$NB570$ & SXDS-N & 4200 & 0.91 (0.91) & 781 & 24.9 & 2003 Oct 26   \\
$NB570$ & SXDS-S & 5400 & 0.69 (0.81) & 833 & 25.2 & 2003 Oct 22   \\
$NB570$ & SXDS-E & 4200 & 0.75 (0.83) & 674 & 24.9 & 2003 Oct 26   \\
$NB570$ & SXDS-W & 4200 & 0.87 (0.87) & 559 & 24.8 & 2003 Oct 26   \\
$NB816$ & SXDS-C & 17182 & 0.65 (0.81) & 676 & 26.0 & 2003 Sep 28-30, Oct 22 \\
$NB816$ & SXDS-N & 14400 & 0.73 (0.85) & 810 & 26.0 & 2003 Sep 29, Oct 22 \\
$NB816$ & SXDS-S & 14400 & 0.65 (0.81) & 835 & 26.1 & 2003 Sep 28,30 Oct 22 \\
$NB816$ & SXDS-E & 14255 & 0.67 (0.83) & 722 & 26.0 & 2003 Sep 29-30, Oct 22 \\
$NB816$ & SXDS-W & 20400 & 0.69 (0.83) & 678 & 25.9 & 2003 Sep 30, Oct 22\phn \\
\cutinhead{Archival broad-band data\tablenotemark{\dagger\dagger}.}
$B$ & SXDS-C,N,S,E,W & $19800-20700$ & $0.78-0.84$ & $915-979$ & $27.5-27.8$ & \nodata\phn \\
$V$ & SXDS-C,N,S,E,W & $17460-19260$ & $0.72-0.82$ & $915-979$ & $27.1-27.2$ & \nodata\phn \\
$R$ & SXDS-C,N,S,E,W & $13920-14880$ & $0.74-0.82$ & $915-979$ & $27.0-27.2$ & \nodata\phn \\
$i'$ & SXDS-C,N,S,E,W & $18540-38820$ & $0.68-0.82$ & $915-979$ & $26.9-27.1$ & \nodata\phn \\
$z'$ & SXDS-C,N,S,E,W & $11040-18660$ & $0.70-0.76$ & $915-979$ & $25.8-26.1$ & \nodata\phn
\enddata
\tablenotetext{\ddag}{
The FWHM of PSFs in the reduced image. The value in parenthesis
indicates the FWHM of PSF that is matched with broad-band images 
in each field.
}
\tablenotetext{\dagger}{
The limiting magnitude defined by 
a $5\sigma$ sky noise 
in a $2''$-diameter circular aperture.
}
\tablenotetext{\dagger\dagger}{
The archival broad-band data of SXDS presented in Furusawa et al. (in preparation).
We summarize the properties of the 5-field images on a single line.  Note that the  
exposure time is {\it not} a total of the 5 fields, but 1 field.
More details are presented in Table 2 of Furusawa et al. in preparation.
}
\end{deluxetable}

\clearpage

\begin{deluxetable}{ccccccccccccc}
\tablecolumns{13}
\tabletypesize{\scriptsize}
\tablecaption{Ly$\alpha$ Emitters with spectroscopic redshifts
\label{tab:laes_with_redshifts}}
\setlength{\tabcolsep}{0.03in}
\tablewidth{0pt}
\tablehead{
\colhead{Object Name} & 
\colhead{$\alpha$(J2000)$^a$} &
\colhead{$\delta$(J2000)$^a$} & 
\colhead{$z$\tablenotemark{\ddag}} &
\colhead{$m_{\rm NB}$} &
\colhead{$m_{\rm BB}$} &
\colhead{$S_{\rm NB}$\tablenotemark{\dagger}} &
\colhead{$S_{\rm BB}$\tablenotemark{\dagger}} &
\colhead{$L({\rm Ly \alpha})$} &
\colhead{$EW_0^{\rm app}$} &
\colhead{$M_{\rm UV}$} &
\colhead{SFR} &
\colhead{Note} \\
\colhead{} & 
\colhead{} &
\colhead{} & 
\colhead{} &
\colhead{(mag)} &
\colhead{(mag)} &
\colhead{(kpc)} &
\colhead{(kpc)} &
\colhead{($10^{42}$erg s$^{-1}$)} &
\colhead{(\AA)} &
\colhead{(mag)} &
\colhead{($M_\odot$yr$^{-1}$)} &
\colhead{} \\
\colhead{(1)} & 
\colhead{(2)} &
\colhead{(3)} & 
\colhead{(4)} &
\colhead{(5)} &
\colhead{(6)} &
\colhead{(7)} &
\colhead{(8)} &
\colhead{(9)} &
\colhead{(10)} &
\colhead{(11)} &
\colhead{(12)} &
\colhead{(13)}
}
\startdata
\cutinhead{$z=3.1$ LAEs}
NB503-C-112924 & 02:18:50 & -04:50:54 & 3.134 & $24.9$ & $27.8$ & 4.1 & \nodata & $3.0\pm 0.5$ & $264.6^{+171.2}_{-84.3}$ & -18.0 & 0.9 & \nodata \\ 
NB503-C-114646 & 02:17:44 & -04:50:32 & 3.135 & $23.9$ & $25.4$ & 4.1 & $\lesssim 4$ & $6.6\pm 0.5$ & $55.9^{+5.1}_{-5.0}$ & -20.2 & 6.5 & \nodata \\ 
NB503-C-15140 & 02:18:58 & -05:12:26 & 3.147 & $23.6$ & $25.6$ & 4.4 & $\lesssim 4$ & $9.0\pm 0.8$ & $62.9^{+6.3}_{-6.1}$ & -20.2 & 6.5 & \nodata \\ 
NB503-C-29891 & 02:17:21 & -05:09:12 & 3.105 & $23.7$ & $25.9$ & 7.1 & $\lesssim 4$ & $14.8\pm 2.1$ & $124.8^{+19.4}_{-17.2}$ & -19.7 & 4.2 & \nodata \\ 
NB503-C-33149 & 02:18:39 & -05:08:27 & 3.141 & $24.6$ & $26.6$ & 6.9 & 4.5 & $3.5\pm 0.6$ & $76.3^{+16.6}_{-14.3}$ & -19.0 & 2.1 & \nodata \\ 
NB503-C-46996 & 02:18:33 & -05:04:57 & 3.129 & $23.9$ & $25.6$ & 4.3 & 4.3 & $5.9\pm 0.7$ & $47.2^{+6.3}_{-6.0}$ & -20.0 & 5.4 & \nodata \\ 
NB503-C-49497 & 02:18:06 & -05:04:08 & 3.129 & $21.8$ & $24.4$ & 8.3 & 6.2 & $44.4\pm 1.3$ & $62.6^{+2.1}_{-1.9}$ & -21.7 & (25.3) & AGN \\ 
NB503-C-51702 & 02:18:07 & -05:03:49 & 3.125 & $24.4$ & $25.8$ & 5.4 & $\lesssim 4$ & $3.5\pm 0.7$ & $37.6^{+7.4}_{-6.3}$ & -20.1 & 6.2 & \nodata \\ 
NB503-C-52511 & 02:17:12 & -05:03:37 & 3.125 & $24.7$ & $26.8$ & 5.6 & $\lesssim 4$ & $3.2\pm 0.5$ & $104.3^{+26.8}_{-20.8}$ & -18.8 & 1.7 & \nodata \\ 
NB503-C-53382 & 02:17:05 & -05:03:25 & 3.152: & $24.5$ & $25.5$ & 4.7 & 4.6 & $4.4\pm 0.5$ & $41.8^{+5.1}_{-5.5}$ & -20.1 & 6.2 & \nodata \\ 
NB503-C-57309 & 02:17:08 & -05:02:26 & 3.136 & $24.6$ & $25.8$ & 7.3 & 4.3 & $3.3\pm 0.5$ & $39.7^{+6.5}_{-6.3}$ & -20.1 & 6.0 & \nodata \\ 
NB503-C-66311 & 02:17:27 & -05:00:21 & 3.144 & $24.6$ & $26.4$ & 5.1 & $\lesssim 4$ & $3.5\pm 0.5$ & $69.1^{+14.0}_{-12.0}$ & -19.2 & 2.7 & \nodata \\ 
NB503-C-88741 & 02:17:00 & -04:55:40 & 3.113 & $23.6$ & $25.6$ & $\lesssim 4$ & $\lesssim 4$ & $11.7\pm 0.7$ & $128.2^{+11.4}_{-10.7}$ & -20.1 & 6.2 & \nodata \\ 
NB503-C-90464 & 02:17:43 & -04:55:21 & 3.111 & $24.8$ & $27.1$ & 4.1 & \nodata & $4.2\pm 0.8$ & $204.2^{+72.1}_{-53.8}$ & -18.6 & 1.6 & \nodata \\ 
NB503-C-99469 & 02:18:03 & -04:53:31 & 3.133 & $24.9$ & $27.2$ & 4.5 & \nodata & $2.7\pm 0.5$ & $137.2^{+51.3}_{-34.5}$ & -18.4 & 1.2 & \nodata \\ 
NB503-E-161419 & 02:20:24 & -04:52:05 & 3.105 & $24.6$ & $26.3$ & 4.6 & 4.2 & $6.1\pm 1.8$ & $86.2^{+26.0}_{-23.6}$ & -19.4 & 3.2 & \nodata \\ 
NB503-E-49946 & 02:19:09 & -04:47:22 & 3.152 & $24.3$ & $26.0$ & 4.7 & 4.6 & $5.2\pm 0.7$ & $65.2^{+9.9}_{-9.3}$ & -19.5 & 3.5 & \nodata \\ 
NB503-E-60750 & 02:19:16 & -04:58:38 & 3.129 & $23.5$ & $25.1$ & $\lesssim 4$ & $\lesssim 4$ & $9.2\pm 0.6$ & $58.4^{+4.2}_{-3.9}$ & -20.5 & 8.4 & \nodata \\ 
NB503-N-32683 & 02:17:41 & -04:43:47 & 3.106 & $24.4$ & $25.9$ & 5.0 & 7.4 & $7.3\pm 1.0$ & $114.1^{+19.0}_{-16.4}$ & -19.8 & 4.6 & \nodata \\ 
NB503-N-33460 & 02:17:27 & -04:43:36 & 3.119 & $23.3$ & $25.6$ & 5.3 & 4.1 & $12.9\pm 0.6$ & $121.3^{+9.2}_{-9.0}$ & -20.1 & 6.0 & \nodata \\ 
NB503-N-34067 & 02:17:44 & -04:43:29 & 3.152 & $24.6$ & $>28.0$ & $\lesssim 4$ & \nodata & $4.4\pm 0.7$ & $343.0^{+292.7}_{-120.6}$ & -17.6 & 0.6 & \nodata \\ 
NB503-N-35820 & 02:17:35 & -04:42:59 & 3.102 & $23.6$ & $25.3$ & 6.6 & 6.1 & $21.1\pm 1.9$ & $135.5^{+12.8}_{-12.8}$ & -20.8 & (11.0) & AGN \\ 
NB503-N-42377 & 02:17:42 & -04:41:08 & 3.154 & $22.2$ & $23.9$ & 4.2 & 4.4 & $38.8\pm 0.8$ & $73.6^{+1.5}_{-1.5}$ & -21.8 & 29.0 & C{\sc iv}?\tablenotemark{\P} \\ 
NB503-N-72492 & 02:17:42 & -04:33:20 & 3.108 & $24.3$ & $26.2$ & 7.2 & 6.3 & $8.0\pm 1.0$ & $146.8^{+26.4}_{-22.2}$ & -19.4 & 3.2 & \nodata \\ 
NB503-N-80475 & 02:17:41 & -04:31:30 & 3.123 & $21.8$ & $22.6$ & $\lesssim 4$ & 4.3 & $31.0\pm 0.6$ & $18.6^{+0.4}_{-0.3}$ & -23.2 & (100.7) & AGN \\ 
NB503-N-87126 & 02:17:54 & -04:30:06 & 3.112 & $23.5$ & $25.4$ & 6.1 & 4.6 & $12.4\pm 1.1$ & $83.4^{+7.6}_{-7.6}$ & -20.3 & 7.0 & \nodata \\ 
NB503-N-92276 & 02:17:42 & -04:28:58 & 3.133 & $24.3$ & $26.4$ & 4.2 & 5.6 & $4.5\pm 0.7$ & $75.9^{+14.7}_{-11.9}$ & -19.7 & 4.0 & \nodata \\ 
NB503-S-105705 & 02:17:06 & -05:16:34 & 3.106 & $24.6$ & $26.5$ & 7.1 & 5.3 & $6.6\pm 1.3$ & $148.8^{+38.1}_{-30.1}$ & -19.3 & 2.8 & \nodata \\ 
NB503-S-35558 & 02:18:18 & -05:31:57 & 3.132 & $24.7$ & $26.6$ & 6.6 & 5.4 & $3.4\pm 0.5$ & $98.1^{+21.8}_{-18.0}$ & -19.6 & 3.7 & \nodata \\ 
NB503-S-45244 & 02:18:26 & -05:29:45 & 3.156 & $23.1$ & $24.8$ & $\lesssim 4$ & $\lesssim 4$ & $18.5\pm 0.6$ & $82.3^{+3.4}_{-3.4}$ & -21.1 & 15.7 & \nodata \\ 
NB503-S-54416 & 02:17:05 & -05:27:35 & 3.133 & $23.9$ & $26.0$ & 5.9 & 9.3 & $6.7\pm 0.6$ & $88.7^{+12.0}_{-9.5}$ & -19.6 & 3.7 & \nodata \\ 
NB503-S-56809 & 02:17:57 & -05:27:00 & 3.123 & $24.2$ & $27.3$ & 5.9 & \nodata & $5.9\pm 0.8$ & $233.7^{+95.9}_{-50.7}$ & -18.3 & 1.2 & \nodata \\ 
NB503-S-66012 & 02:17:23 & -05:24:49 & 3.167 & $24.6$ & $25.6$ & 4.3 & 6.4 & $7.8\pm 1.0$ & $73.4^{+10.5}_{-10.1}$ & -20.0 & 5.6 & \nodata \\ 
NB503-S-89393 & 02:19:04 & -05:19:58 & 3.125 & $24.4$ & $26.0$ & 5.9 & 5.9 & $3.8\pm 0.8$ & $42.6^{+9.4}_{-8.2}$ & -19.6 & 3.8 & \nodata \\ 
NB503-S-94275 & 02:18:41 & -05:18:49 & 3.102 & $22.8$ & $24.5$ & 4.1 & $\lesssim 4$ & $46.5\pm 1.9$ & $164.6^{+6.3}_{-6.6}$ & -21.2 & 16.1 & \nodata \\ 
NB503-W-100818 & 02:16:04 & -04:49:43 & 3.125 & $24.0$ & $26.0$ & 4.1 & $\lesssim 4$ & $6.3\pm 0.6$ & $85.0^{+11.3}_{-10.3}$ & -19.6 & 3.8 & \nodata \\ 
NB503-W-31855 & 02:16:52 & -04:56:43 & 3.130 & $24.4$ & $26.9$ & 5.1 & 9.1 & $4.6\pm 0.6$ & $137.3^{+39.2}_{-27.6}$ & -18.6 & 1.5 & \nodata \\ 
NB503-W-67066 & 02:16:27 & -05:02:43 & 3.139 & $23.9$ & $25.8$ & 7.3 & 11.9 & $6.3\pm 0.8$ & $55.7^{+7.2}_{-7.2}$ & -19.6 & 3.7 & \nodata \\ 
NB503-W-85295 & 02:16:14 & -05:02:15 & 3.141 & $24.5$ & $26.0$ & $\lesssim 4$ & 4.3 & $3.8\pm 0.6$ & $51.2^{+7.6}_{-8.2}$ & -19.6 & 3.7 & \nodata \\ 
NB503-W-97848 & 02:16:06 & -04:48:12 & 3.121 & $24.6$ & $26.6$ & 4.1 & $\lesssim 4$ & $3.7\pm 0.6$ & $101.1^{+21.5}_{-19.3}$ & -19.1 & 2.3 & \nodata \\ 
NB503-W-98167 & 02:16:06 & -04:50:57 & 3.124 & $23.8$ & $25.6$ & 4.1 & $\lesssim 4$ & $6.9\pm 0.7$ & $63.5^{+6.9}_{-6.5}$ & -20.0 & 5.4 & \nodata \\ 
\cutinhead{$z=3.7$ LAEs}
NB570-C-107443 & 02:17:47 & -04:52:30 & 3.648 & $23.4$ & $24.3$ & 4.7 & $\lesssim 4$ & $33.6\pm 4.7$ & $75.2^{+10.8}_{-9.2}$ & -21.7 & 26.7 & \nodata \\ 
NB570-C-116152 & 02:18:56 & -04:50:59 & 3.683 & $24.5$ & $26.4$ & 4.1 & $\lesssim 4$ & $4.7\pm 0.8$ & $87.3^{+17.3}_{-16.5}$ & -19.6 & 3.7 & \nodata \\ 
NB570-C-13685 & 02:17:09 & -05:12:29 & 3.691 & $23.9$ & $27.6$ & 5.7 & \nodata & $8.3\pm 1.1$ & $335.6^{+182.9}_{-93.9}$ & -18.3 & 1.1 & \nodata \\ 
NB570-C-30225 & 02:18:26 & -05:10:03 & 3.699 & $23.0$ & $24.8$ & 5.7 & 4.7 & $19.0\pm 1.1$ & $57.7^{+3.6}_{-3.7}$ & -21.2 & 16.1 & \nodata \\ 
NB570-C-40856 & 02:17:01 & -05:07:28 & 3.686 & $24.0$ & $26.8$ & 4.8 & 4.5 & $7.2\pm 0.9$ & $161.8^{+36.1}_{-30.2}$ & -19.2 & 2.6 & \nodata \\ 
NB570-C-78067 & 02:17:13 & -04:58:53 & 3.684 & $23.5$ & $24.6$ & 5.7 & 4.5 & $7.9\pm 1.6$ & $14.4^{+2.5}_{-2.9}$ & -21.1 & 15.5 & \nodata \\ 
NB570-E-103151 & 02:19:44 & -04:50:55 & 3.669 & $23.2$ & $25.8$ & 4.6 & $\lesssim 4$ & $21.9\pm 2.0$ & $150.0^{+16.1}_{-15.0}$ & -20.1 & 6.0 & \nodata \\ 
NB570-E-142687 & 02:20:11 & -04:50:08 & 3.644 & $24.4$ & $25.4$ & 5.6 & 4.2 & $16.6\pm 7.0$ & $95.4^{+36.8}_{-36.1}$ & -20.5 & 8.8 & \nodata \\ 
NB570-E-169608 & 02:20:30 & -04:48:15 & 3.702 & $24.6$ & $26.8$ & 5.3 & 5.0 & $4.7\pm 0.9$ & $110.4^{+29.6}_{-25.4}$ & -19.2 & 2.7 & \nodata \\ 
NB570-E-58808 & 02:19:13 & -04:51:59 & 3.692 & $23.3$ & $25.7$ & $\lesssim 4$ & $\lesssim 4$ & $14.2\pm 0.9$ & $126.4^{+11.1}_{-10.0}$ & -20.4 & 7.8 & \nodata \\ 
NB570-E-62593 & 02:19:15 & -04:55:11 & 3.672 & $23.6$ & $25.5$ & 6.0 & $\lesssim 4$ & $12.8\pm 1.7$ & $67.4^{+9.2}_{-9.3}$ & -20.7 & 10.6 & \nodata \\ 
NB570-E-65332 & 02:19:17 & -05:07:39 & 3.693 & $23.8$ & $25.9$ & $\lesssim 4$ & $\lesssim 4$ & $9.1\pm 0.8$ & $92.8^{+10.7}_{-9.5}$ & -20.1 & 6.1 & \nodata \\ 
NB570-N-33608 & 02:17:27 & -04:44:14 & 3.683 & $24.0$ & $25.9$ & 5.7 & $\lesssim 4$ & $7.1\pm 1.1$ & $63.1^{+10.1}_{-9.3}$ & -20.0 & 5.5 & \nodata \\ 
NB570-N-34841 & 02:18:31 & -04:43:54 & 3.724 & $21.8$ & $22.2$ & $\lesssim 4$ & $\lesssim 4$ & $145.0\pm 1.5$ & $48.9^{+0.7}_{-0.7}$ & -23.8 & (178.2) & AGN \\ 
NB570-N-42463 & 02:18:40 & -04:42:11 & 3.692 & $24.4$ & $26.4$ & 4.4 & $\lesssim 4$ & $4.7\pm 0.8$ & $82.7^{+17.1}_{-14.2}$ & -19.6 & 3.9 & \nodata \\ 
NB570-N-86993 & 02:18:13 & -04:30:57 & 3.668 & $23.8$ & $25.7$ & 5.2 & $\lesssim 4$ & $11.5\pm 1.5$ & $90.7^{+12.8}_{-12.0}$ & -20.4 & 7.6 & \nodata \\ 
NB570-S-125887 & 02:18:07 & -05:20:48 & 3.638 & $24.3$ & $25.2$ & 5.9 & 4.4 & $28.4\pm 11.4$ & $147.9^{+48.7}_{-55.2}$ & -20.8 & 11.0 & \nodata \\ 
NB570-S-132143 & 02:18:27 & -05:19:47 & 3.677 & $23.4$ & $25.9$ & 4.1 & $\lesssim 4$ & $14.2\pm 1.3$ & $118.9^{+14.1}_{-11.6}$ & -20.0 & 5.5 & \nodata \\ 
NB570-S-141871 & 02:18:10 & -05:18:13 & 3.684 & $24.5$ & $26.9$ & $\lesssim 4$ & $\lesssim 4$ & $4.6\pm 0.8$ & $134.4^{+39.3}_{-28.5}$ & -18.9 & 2.1 & \nodata \\ 
NB570-S-154087 & 02:17:43 & -05:16:11 & 3.682 & $24.9$ & $27.3$ & \nodata & \nodata & $3.3\pm 0.9$ & $141.3^{+72.6}_{-45.2}$ & -18.4 & 1.2 & \nodata \\ 
NB570-S-175068 & 02:18:28 & -05:13:06 & 3.701 & $24.8$ & $26.8$ & 5.7 & $\lesssim 4$ & $3.8\pm 1.0$ & $84.1^{+26.8}_{-21.2}$ & -19.1 & 2.3 & \nodata \\ 
NB570-S-32766 & 02:18:35 & -05:35:50 & 3.671 & $23.0$ & $25.4$ & $\lesssim 4$ & $\lesssim 4$ & $24.8\pm 1.5$ & $129.7^{+9.5}_{-8.6}$ & -20.7 & 10.6 & \nodata \\ 
NB570-S-84321 & 02:17:45 & -05:27:35 & 3.648 & $24.4$ & $25.1$ & 7.2 & $\lesssim 4$ & $12.4\pm 4.6$ & $61.7^{+20.3}_{-22.0}$ & -20.9 & 11.9 & \nodata \\ 
NB570-S-99194 & 02:17:14 & -05:25:16 & 3.729 & $24.9$ & $25.6$ & \nodata & $\lesssim 4$ & $11.6\pm 2.7$ & $94.5^{+23.5}_{-21.6}$ & -20.4 & 7.9 & \nodata \\ 
NB570-W-53415 & 02:16:42 & -04:58:55 & 3.667 & $24.2$ & $25.7$ & $\lesssim 4$ & $\lesssim 4$ & $8.2\pm 1.5$ & $66.3^{+11.6}_{-12.0}$ & -20.5 & 8.4 & \nodata \\ 
NB570-W-55371 & 02:16:40 & -05:01:29 & 3.699 & $24.2$ & $24.7$ & 4.7 & $\lesssim 4$ & $5.2\pm 0.8$ & $19.9^{+3.0}_{-3.3}$ & -21.2 & 17.0 & \nodata \\ 
\cutinhead{$z=5.7$ LAEs}
NB816-E-127266 & 02:20:12 & -04:49:50 & 5.681 & $24.9$ & $>27.0$ & $\lesssim 3$ & \nodata & $8.0\pm 1.3$ & $159.6^{+222.1}_{-73.4}$ & -19.2 & 2.6 & \nodata \\ 
NB816-E-129103 & 02:20:13 & -04:51:09 & 5.744 & $24.8$ & $26.1$ & $\lesssim 3$ & \nodata & $11.5\pm 1.4$ & $64.3^{+19.9}_{-14.2}$ & -20.5 & 8.8 & \nodata \\ 
NB816-E-141288 & 02:20:21 & -04:53:14 & 5.671 & $24.8$ & $>27.0$ & $\lesssim 3$ & \nodata & $9.6\pm 1.8$ & $120.6^{+124.3}_{-45.8}$ & -19.6 & 3.9 & \nodata \\ 
NB816-E-147538 & 02:20:26 & -04:52:34 & 5.718 & $23.7$ & $25.1$ & conf. & conf. & $18.9\pm 1.7$ & $29.2^{+4.2}_{-3.5}$ & -21.4 & (20.1) & \nodata \\ 
NB816-S-36496 & 02:18:22 & -05:33:37 & 5.650 & $25.1$ & $26.5$ & $\lesssim 3$ & \nodata & $13.2\pm 3.4$ & $133.2^{+92.5}_{-47.1}$ & -20.1 & 6.1 & \nodata \\ 
NB816-S-39206 & 02:18:19 & -05:33:11 & 5.676 & $25.3$ & $>27.0$ & 4.5 & \nodata & $5.7\pm 1.5$ & $135.1^{+300.8}_{-65.8}$ & -18.9 & 2.0 & \nodata \\ 
NB816-S-41408 & 02:18:14 & -05:32:49 & 5.673 & $23.9$ & $25.7$ & conf. & conf. & $17.5\pm 3.0$ & $42.7^{+12.8}_{-9.5}$ & -21.3 & (18.5) & \nodata \\ 
NB816-S-44568 & 02:18:17 & -05:32:22 & 5.644 & $24.6$ & $>27.0$ & $\lesssim 3$ & \nodata & $36.0\pm 4.7$ & $855.6^{+1327.0}_{-402.0}$ & -19.0 & 2.2 & \nodata \\ 
NB816-S-46148 & 02:18:23 & -05:32:05 & 5.680 & $24.9$ & $>27.0$ & 3.6 & \nodata & $8.2\pm 1.5$ & $174.2^{+311.2}_{-84.9}$ & -18.9 & 2.0 & \nodata \\ 
NB816-S-49611 & 02:17:43 & -05:31:35 & 5.629 & $25.0$ & $25.7$ & $\lesssim 3$ & $\lesssim 3$ & $28.9\pm 14.2$ & $126.1^{+69.6}_{-50.3}$ & -21.1 & 15.7 & \nodata \\ 
NB816-S-50308 & 02:17:48 & -05:31:27 & 5.690 & $24.0$ & $26.0$ & 3.2 & \nodata & $15.0\pm 1.2$ & $81.4^{+27.9}_{-16.3}$ & -20.7 & 10.2 & \nodata \\ 
NB816-S-59282 & 02:17:51 & -05:30:03 & 5.712: & $25.1$ & $26.0$ & 3.8 & \nodata & $4.8\pm 1.3$ & $21.3^{+9.5}_{-6.2}$ & -20.8 & 10.8 & \nodata \\ 
NB816-S-61269 & 02:17:45 & -05:29:36 & 5.688 & $23.7$ & $26.1$ & 3.5 & \nodata & $21.3\pm 1.7$ & $93.3^{+32.6}_{-18.8}$ & -21.4 & 19.2 & \nodata \\ 
NB816-S-66352 & 02:17:49 & -05:28:54 & 5.696 & $24.0$ & $25.9$ & $\lesssim 3$ & conf. & $16.2\pm 1.1$ & $79.3^{+22.1}_{-14.2}$ & -21.0 & (13.4) & \nodata \\ 
NB816-S-67673 & 02:17:45 & -05:28:42 & 5.751 & $25.0$ & $26.1$ & 3.9 & \nodata & $10.9\pm 1.4$ & $61.6^{+21.8}_{-13.1}$ & -20.4 & 7.6 & \nodata \\ 
NB816-S-70769 & 02:17:43 & -05:28:07 & 5.685: & $23.9$ & $26.1$ & $\lesssim 3$ & \nodata & $17.8\pm 1.2$ & $122.9^{+41.6}_{-27.6}$ & -20.5 & 8.5 & \nodata \\ 
NB816-S-77389 & 02:17:50 & -05:27:08 & 5.693 & $24.3$ & $>27.0$ & 6.1 & \nodata & $11.9\pm 1.9$ & $106.4^{+107.0}_{-40.3}$ & -19.6 & 3.6 & \nodata \\ 
\enddata

\tablecomments{
(1): Object name. (2)-(3): RA and Dec. (4): Redshift. (5)-(6): Magnitudes in narrow 
and broad bands, ($NB503$,$R$), ($NB570$,$R$), and ($NB816$,$z'$),
for $z=3.1$, 3.7, and 5.7 LAEs, respectively. (7)-(8) Half of FWHMs in physical kpc in
the narrow and broad bands. We only show reliable FWHMs 
for sources with a $>5\sigma$ significance.
 (9): Ly$\alpha$ luminosity in $10^{42}$ erg s$^{-1}$.
(10): Rest-frame {\it apparent} equivalent width of Ly$\alpha$ emission line. 
(11): UV total magnitude defined in Section \ref{sec:uv_luminosity_functions}.
(12): Star-formation rate estimated from the UV total magnitude. 
The values in parenthesis present formal SFRs which are estimated
for AGNs or confused objects.
(13): AGN classification and notes. 
}
\tablenotetext{a}{
See the published version for the exact coordinates with decimals.
}
\tablenotetext{\ddag}{
In column (4), ``:'' marks objects with an uncertain spectroscopic identification. 
}
\tablenotetext{\dagger}{
``conf'' in the columns (7) and (8) indicates 
the sources confused by (a) neighboring object(s),
which do not give an accurate FWHM.
}
\tablenotetext{\P}{
The object with a marginal C{\sc iv} detection, NB503-N-42377. 
See Section \ref{sec:agn_spectra}.
}
\end{deluxetable}

\clearpage 

\begin{deluxetable}{cccccccc}
\tabletypesize{\scriptsize}
\tablecaption{Samples of Ly$\alpha$ Emitters
\label{tab:sample}}
\tablewidth{0pt}
\tablehead{
\colhead{Redshift} &
\colhead{Survey Area} &
\colhead{Magnitude Range\tablenotemark{\ddag}} &
\colhead{$L_{Ly\alpha}$\tablenotemark{\dagger}} &
\colhead{$EW_0^{app}$\tablenotemark{\dagger\dagger}} &
\colhead{$N_{\rm phot}$} & 
\colhead{$N_{\rm spec}$} & 
\colhead{Selection Criteria} \\
\colhead{} &
\colhead{(arcmin$^2$)} &
\colhead{(AB mag)} & 
\colhead{(erg s$^{-1}$)} & 
\colhead{(\AA)} & 
\colhead{} & 
\colhead{} &
\colhead{} 
}
\startdata
$z=3.1\pm0.03$ & 3538 & $NB503=22.4-25.3$ & $\gtrsim 1\times 10^{42}$ & $\gtrsim 64$ & 356 & 41 & eq.(\ref{eq:laeselection_nb503})\\
$z=3.7\pm0.03$ & 3474 & $NB570=22.7-24.7$ & $\gtrsim 4\times 10^{42}$ & $\gtrsim 44$ & 101 & 26 & eq.(\ref{eq:laeselection_nb570})\\
$z=5.7\pm0.05$ & 3722 & $NB816=23.6-26.0$ & $\gtrsim 3\times 10^{42}$ & $\gtrsim 27$ & 401 & 17 & eq.(\ref{eq:laeselection_nb816})

\enddata

\tablenotetext{\ddag}{$2''$-diameter aperture magnitudes.
}
\tablenotetext{\dagger}{The approximate limit of Ly$\alpha$ luminosity.}
\tablenotetext{\dagger\dagger}{The approximate limit of rest-frame {\it apparent} 
equivalent width of Ly$\alpha$ emission. 
The corresponding {\it intrinsic} equivalent widths
corrected for the IGM absorption
are estimated to be $EW_0^{int}=80$, $60$, and $50$\AA,
for $z=3.1$, $3.7$, and $5.7$ samples, respectively.
}
\end{deluxetable}

\clearpage
\thispagestyle{empty}

\begin{deluxetable}{cccccccccccccccccc}
\tabletypesize{\scriptsize}
\rotate
\tablecaption{Ly$\alpha$ Emitters with AGN activities
\label{tab:multi-wavelength_prop}}
\tablewidth{0pt}
\setlength{\tabcolsep}{0.005in}
\tablehead{
\colhead{ID(opt)} & 
\colhead{RA} &
\colhead{Dec} & 
\colhead{$z$} &
\colhead{Det} &
\colhead{ID(X)} &
\colhead{sep(X)} &
\colhead{$f_{2-10{\rm keV}}$} &
\colhead{sep(IR)} &
\colhead{$f_{24\mu {\rm m}}$} &
\colhead{$f_{850\mu {\rm m}}$} &
\colhead{ID(R)} &
\colhead{sep(R)} &
\colhead{$f_{1.4{\rm GHz}}$} &
\colhead{$L_{{\rm Ly}\alpha}$} &
\colhead{$M_{{\rm UV}}$} &
\colhead{FWHM(NB)}  &
\colhead{Type} \\
\colhead{} & 
\colhead{(J2000)} &
\colhead{(J2000)} & 
\colhead{} &
\colhead{} &
\colhead{} &
\colhead{($``$)} &
\colhead{($10^{-15}$cgs)} &
\colhead{($``$)} &
\colhead{($\mu$Jy)} &
\colhead{(mJy)} &
\colhead{} &
\colhead{($``$)} &
\colhead{($\mu$Jy)} &
\colhead{($10^{43}$erg s$^{-1}$)} &
\colhead{(mag)} &
\colhead{(kpc)} &
\colhead{} \\
\colhead{(1)} & 
\colhead{(2)} &
\colhead{(3)} & 
\colhead{(4)} &
\colhead{(5)} &
\colhead{(6)} &
\colhead{(7)} &
\colhead{(8)} &
\colhead{(9)} &
\colhead{(10)} &
\colhead{(11)} &
\colhead{(12)} &
\colhead{(13)} &
\colhead{(14)} &
\colhead{(15)} &
\colhead{(16)} &
\colhead{(17)}  &
\colhead{(18)} 
}
\startdata
NB503-C-49497 & 02:18:06.826 & -05:04:08.80 & $3.129$ & SPXooo   & X0700 & 2.6 & $7.1\pm 0.3$   & \nodata & $<240$   & $\lesssim 4$   & \nodata & \nodata & $<60$ & 4.4 & -21.67 & $16.5$ & BL\\ 
NB503-N-35820 & 02:17:35.593 & -04:42:59.83 & $3.102$ & SPoIoo   & \nodata & \nodata & $<0.7$   & 0.9 & $463\pm 48$   & \nodata   & \nodata & \nodata & $<60$ & 2.1 & -20.77 & $13.2$ & BL?\\ 
NB503-N-55380 & 02:17:01.676 & -04:37:20.24 & $(3.1)$ & oPoIoR   & \nodata & \nodata & $<0.6$   & 1.6 & $378\pm 48$   & \nodata   & VLA0267 & 1.8 & $190\pm20$ & 4.2 & -20.64 & $21.9$ & \nodata \\ 
NB503-N-80475 & 02:17:41.987 & -04:31:30.55 & $3.123$ & SooIoR   & \nodata & \nodata & $<0.9$   & 1.2 & $431\pm 48$   & \nodata   & VLA0351 & 0.7 & $147\pm20$ & 3.1 & -23.17 & $7.8$ & BL \\
NB570-N-71842 & 02:18:57.776 & -04:34:31.53 & $(3.7)$ & oPXooo   & X0950 & 1.7 & $4.4\pm 1.4$   & \nodata & $<240$   & \nodata   & \nodata & \nodata & $<60$ & 3.0 & -21.65 & $8.5$ & \nodata \\ 
NB570-N-34841 & 02:18:31.379 & -04:43:54.77 & $3.724$ & Sooooo   & \nodata & \nodata & $<1.3$   & \nodata & $<240$   & \nodata   & \nodata & \nodata & $<60$ & 14.5 & -23.79 & $<6.1$ & BL
\enddata

\tablecomments{
(1): Object name from the optical data. (2)-(3): RA and Dec in the optical data.
(4): Redshift. The values in parenthesis are photometric redshifts.
(5) The identification of AGN-LAEs in our spectroscopy and
multi-wavelength data. ``X'', ``I'', ``M'', or ``R'' means that the LAE has
either an X-ray, infrared, submm, or radio detection. The flags of ``S'' and ``P'' 
indicate a spectroscopically identified AGN and an object satisfying our color-selection
criteria of LAEs, respectively.
(``o'' means no detection or identification in the corresponding condition.)
(6) and (12): ID names in the X-ray and radio data presented in
Ueda et al. (in preparation) and \citet{simpson2006a}, respectively.
(7),(9), and (13): Separation between positions of the optical center
and a counterpart of X-ray, infrared, and radio. (8), (10), (11), and (14):
Fluxes of X-ray 2-10 keV (in $10^{-15}$ erg s$^{-1}$ cm$^{-2}$), infrared $24\mu$m
(in $\mu$Jy), submm $850\mu$m (in mJy), and radio 1.4GHz (in $\mu$Jy),
respectively. For the upper limits of X-ray fluxes, we assume 
an absorption of neutral hydrogen with a column
density of $10^{22}$ cm$^{-2}$ and a spectral power of $\gamma=1.8$. 
We omit the effects of X-ray reflection, which 
provides the most conservative upper limit in column (8).
(15) Ly$\alpha$ luminosity in $10^{43}$ erg s$^{-1}$ derived from our photometric data.
(16) Total UV magnitude in AB mag. 
(17) FWHM in the narrow band in physical kpc.
(18) Type of object. ``BL'' indicates an AGN with a broad-line emission 
($v_{\rm FWHM} \gtrsim 1000$ km s$^{-1}$) in high ionized lines 
(i.e. N{\sc v}, Si{\sc iv}, C{\sc iv}, He{\sc ii}, and/or {\sc Ciii]}).
NB503-N-35820 shows a possible broad line with $1321\pm 362$ km s$^{-1}$
in {\sc Civ}, although the signal-to-noise ratio is as small as 4.
}

\end{deluxetable}

\clearpage

\begin{deluxetable}{ccccccccccc}
\tablecolumns{11}
\tabletypesize{\scriptsize}
\tablecaption{Summary of the Ly$\alpha$ luminosity functions.
\label{tab:lya_lumifun_schechter}}
\tablewidth{0pt}
\setlength{\tabcolsep}{0.0in}
\tablehead{
\colhead{$z$} &
\colhead{$\phi^*$} &
\colhead{$L_{\rm Ly\alpha}^*$\tablenotemark{\dagger}} &
\colhead{$\alpha$} &
\colhead{$\sigma_{\rm EW}$} &
\colhead{$\chi_{\rm r}^2$} &
\colhead{$n^{\rm obs}$} &
\colhead{$\rho_{\rm Ly \alpha}^{\rm obs}$} &
\colhead{$\rho_{\rm Ly \alpha}^{\rm tot}$} &
\colhead{$\phi_0^*$} &
\colhead{$\rho_{0 \rm Ly \alpha}^{\rm tot}$} \\
\colhead{} &
\colhead{($10^{-4}$Mpc$^{-3}$)} &
\colhead{($10^{42}$erg s$^{-1}$)} &
\colhead{} &
\colhead{($10^{2}$\AA)} &
\colhead{} &
\colhead{($10^{-4}$Mpc$^{-3}$)} &
\colhead{($10^{39}$erg s$^{-1}$Mpc$^{-3}$)} &
\colhead{($10^{39}$erg s$^{-1}$Mpc$^{-3}$)} &
\colhead{($10^{-4}$Mpc$^{-3}$)} &
\colhead{($10^{39}$erg s$^{-1}$Mpc$^{-3}$)} \\
\colhead{(1)} &
\colhead{(2)} &
\colhead{(3)} &
\colhead{(4)} &
\colhead{(5)} &
\colhead{(6)} &
\colhead{(7)} &
\colhead{(8)} &
\colhead{(9)} &
\colhead{(10)} &
\colhead{(11)} 
}
\startdata
\cutinhead{$\alpha=-1.5$ (fix)}
3.1 &  $9.2_{-2.1}^{+2.5}$ &  $5.8_{-0.7}^{+0.9}$ & $-1.5\ $ & $1.3_{-0.1}^{+0.1}$ & $1.57$ & $15.0_{-2.8}^{+3.2}$ & $4.8_{-1.0}^{+1.2}$ & $9.4_{-1.7}^{+2.0}$ & $10.8_{-2.5}^{+3.0}$ & $11.0_{-2.0}^{+2.3}$ \\ 
3.7 &  $3.4_{-0.9}^{+1.0}$ &  $10.2_{-1.5}^{+1.8}$ & $-1.5\ $ & $1.5_{-0.4}^{+0.1}$ & $0.70$ & $2.9_{-0.7}^{+0.8}$ & $2.4_{-0.6}^{+0.7}$ & $6.2_{-1.3}^{+1.5}$ & $3.7_{-1.0}^{+1.1}$ & $6.7_{-1.4}^{+1.6}$ \\ 
5.7 &  $7.7_{-3.9}^{+7.4}$ &  $6.8_{-2.1}^{+3.0}$ & $-1.5\ $ & $2.7$(fix) & $1.04$ & $6.8_{-3.1}^{+5.2}$ & $3.6_{-1.7}^{+3.1}$ & $9.2_{-3.7}^{+6.5}$ & $7.7_{-3.9}^{+7.5}$ & $9.2_{-3.7}^{+6.6}$ \\ 
\cutinhead{$\alpha=-1.0$ (fix)}
3.1 &  $14.9_{-3.2}^{+4.2}$ &  $4.1_{-0.5}^{+0.6}$ & $-1.0\ $ & $1.3_{-0.1}^{+0.2}$ & $1.72$ & $13.1_{-2.1}^{+2.6}$ & $4.5_{-0.8}^{+1.1}$ & $6.1_{-1.0}^{+1.3}$ & $17.5_{-3.7}^{+5.0}$ & $7.1_{-1.2}^{+1.5}$ \\ 
3.7 &  $5.7_{-1.5}^{+2.6}$ &  $7.2_{-1.6}^{+1.3}$ & $-1.0\ $ & $1.4_{-0.2}^{+0.2}$ & $0.70$ & $2.9_{-0.8}^{+0.9}$ & $2.4_{-0.7}^{+0.9}$ & $4.1_{-1.0}^{+1.3}$ & $6.2_{-1.6}^{+2.9}$ & $4.5_{-1.1}^{+1.4}$ \\ 
5.7 &  $10.4_{-2.5}^{+12.0}$ &  $5.4_{-1.7}^{+0.7}$ & $-1.0\ $ & $2.7$(fix) & $1.10$ & $6.3_{-1.9}^{+4.0}$ & $3.5_{-1.2}^{+2.3}$ & $5.6_{-1.6}^{+3.6}$ & $10.5_{-2.5}^{+12.1}$ & $5.7_{-1.6}^{+3.6}$ \\ 
\cutinhead{$\alpha=-2.0$ (fix)}
3.1 &  $3.9_{-1.5}^{+1.1}$ &  $9.1_{-1.2}^{+2.6}$ & $-2.0\ $ & $1.3_{-0.1}^{+0.1}$ & $1.51$ & $18.5_{-5.1}^{+6.3}$ & $5.4_{-1.6}^{+2.1}$ & \nodata & $4.5_{-1.7}^{+1.3}$ & \nodata \\ 
3.7 &  $1.3_{-0.3}^{+0.5}$ &  $16.2_{-2.4}^{+2.4}$ & $-2.0\ $ & $1.5_{-0.3}^{+0.1}$ & $0.71$ & $2.9_{-0.6}^{+0.9}$ & $2.3_{-0.5}^{+0.8}$ & \nodata & $1.5_{-0.3}^{+0.6}$ & \nodata \\ 
5.7 &  $3.6_{-2.5}^{+3.5}$ &  $9.5_{-3.1}^{+8.2}$ & $-2.0\ $ & $2.7$(fix) & $0.97$ & $6.9_{-4.0}^{+8.2}$ & $3.4_{-2.1}^{+4.8}$ & \nodata & $3.6_{-2.5}^{+3.5}$ & \nodata 
\enddata
\tablecomments{
(1): Redshift.
(2)-(4): Best-fit Schechter parameters in units of
$10^{-4}$Mpc$^{-3}$ and $10^{42}$erg s$^{-1}$ for
$\phi^*$ and $L^*_{\rm Ly\alpha}$, respectively. 
$\alpha$ is fixed to $-1.5$, $-1.0$, and $-2.0$ in
1-3, 4-6, 7-9 lines, respectively.
(5): Best-fit $\sigma_{\rm EW}$ in units of $10^2$\AA.
For $z=5.7$ LAEs, we fix $\sigma_{\rm EW}=270$\AA., 
(6): Reduced $\chi^2$ of the fitting.
(7)-(8): Number densities (in $10^{-4}$Mpc$^{-3}$)
and Ly$\alpha$ luminosity densities (in $10^{39}$erg s$^{-1}$ Mpc$^{-3}$)
calculated with the best-fit Schechter parameters down to
the observed limit of Ly$\alpha$ luminosity, i.e. 
$\log L_{\rm Ly\alpha}=42.1$, $42.6$, and $42.4$ (erg s$^{-1}$), for 
$z=3.1$, $3.7$, $5,7$, respectively.
(9): Inferred total Ly$\alpha$ luminosity densities 
integrated down to $L_{\rm Ly\alpha}=0$ 
with the best-fit Schechter parameters.
(10): Same as (2) but for all LAEs with $EW>0$\AA\ estimated from the simulations.
(11): Same as (9) but for all LAEs with $EW>0$\AA\ estimated from the simulations.\\
We refer to the results of $\alpha=-1.5$
as the best estimates of LFs. 
}
\tablenotetext{\dagger}{
$L^*_{\rm Ly\alpha}$ is the apparent value, i.e.
observed Ly$\alpha$ luminosity with no correction for IGM absorption.
If we assume the simple IGM absorption model in
Section \ref{sec:evolution_of_lya_LF},
the apparent $L^*_{\rm Ly\alpha}$ is divided
by a factor of (0.81, 0.73, 0.54)
to obtain the intrinsic $L^*_{\rm Ly\alpha}$ at $z=(3.1, 3.7, 5.7)$.
For examples, in the case of $\alpha=-1.5$,
the intrinsic $L^*_{\rm Ly\alpha}$ is estimated to be 
$(7.1_{-0.9}^{+1.0}, 14.1_{-2.1}^{+2.5}, 12.6_{-3.9}^{+5.6}) \times
10^{42}$erg s$^{-1}$ at $z=(3.1, 3.7, 5.7)$.
}
\end{deluxetable}

\clearpage

\begin{deluxetable}{ccccccccc}
\tablecolumns{9}
\tabletypesize{\scriptsize}
\tablecaption{Summary of the UV luminosity functions.
\label{tab:muv_lumifun_schechter}}
\tablewidth{0pt}
\setlength{\tabcolsep}{0.02in}
\tablehead{
\colhead{$z$} &
\colhead{$\phi^*$} &
\colhead{$M_{\rm 1500}^*$} &
\colhead{$\alpha$} &
\colhead{$\chi_{\rm r}^2$} &
\colhead{Mag. Range} &
\colhead{$n^{\rm obs}$} &
\colhead{$\rho_{\rm UV}^{\rm obs}$} &
\colhead{$\rho_{\rm UV}^{\rm tot}$} \\
\colhead{} &
\colhead{($10^{-4}$Mpc$^{-3}$)} &
\colhead{(mag)} &
\colhead{} &
\colhead{} &
\colhead{(mag)} &
\colhead{($10^{-4}$Mpc$^{-3}$)} &
\colhead{($10^{25}$erg s$^{-1}$ Hz$^{-1}$Mpc$^{-3}$)} &
\colhead{($10^{25}$erg s$^{-1}$ Hz$^{-1}$Mpc$^{-3}$)} \\
\colhead{(1)} &
\colhead{(2)} &
\colhead{(3)} &
\colhead{(4)} &
\colhead{(5)} &
\colhead{(6)} &
\colhead{(7)} &
\colhead{(8)} &
\colhead{(9)} 
}
\startdata
3.1 &  $5.6_{-3.1}^{+6.7}$ &  $-19.8\pm 0.4$ & $-1.6$ & $0.43$ & $-21.9<M<-18.9 $ & $4.0_{-2.2}^{+4.7}$ & $1.2_{-0.7}^{+1.4}$ & $4.4_{-2.4}^{+5.2}$ \\ 
3.7 &  $5.2_{-0.7}^{+0.8}$ &  $-19.8$(fix)$^{\dagger}$ & $-1.6$ & $2.74$ & $-21.7<M<-19.7 $ & $1.0_{-0.1}^{+0.1}$ & $0.5_{-0.1}^{+0.1}$ & $4.1_{-0.5}^{+0.6}$ \\ 
5.7 &  $4.4_{-3.2}^{+11.9}$ &  $-20.6\pm 0.6$ & $-1.6$ & $0.29$ & $-22.1<M<-20.6 $ & $0.8_{-0.6}^{+2.1}$ & $0.9_{-0.7}^{+2.5}$ & $7.5_{-5.5}^{+20.4}$
\enddata
\tablecomments{
(1): Redshift.
(2)-(4): Best-fit Schechter parameters in units of
$10^{-4}$Mpc$^{-3}$ and AB magnitude for
$\phi^*$ and $M^*_{\rm 1500}$. $\alpha$ is fixed to $-1.6$.
(5): Reduced $\chi^2$ of the fitting.
(6): Magnitude range of UV LFs that are used for the fitting. 
(7)-(8): Number densities (in $10^{-4}$Mpc$^{-3}$)
and UV luminosity densities (in $10^{25}$erg s$^{-1}$ Hz$^{-1}$ Mpc$^{-3}$)
calculated with the best-fit Schechter parameters down to
the observed limit of UV luminosity, i.e. 
$M_{\rm 1500}=-18.9$, $-19.7$, and $-20.6$, for 
$z=3.1$, $3.7$, $5.7$, respectively.
(9): Inferred total UV luminosity densities 
integrated down to $M_{\rm 1500}=\infty$ 
with the best-fit Schechter parameters.
}
\tablenotetext{\dagger}{
For the $z=3.7$ LF, we fix $M^*_{1500}$ to -19.8
that is the best estimate of our $z=3.1$ LF.
}
\end{deluxetable}

\clearpage 

\clearpage 

\clearpage

\clearpage

\clearpage

\clearpage 

\clearpage

\clearpage 

\clearpage 

\clearpage 

\clearpage 

\clearpage 

\clearpage

\clearpage 

\clearpage

\clearpage 

\clearpage

\clearpage

\clearpage

\clearpage

\clearpage

\clearpage

\clearpage

\clearpage

\clearpage

\clearpage

\clearpage

\end{document}